\documentclass[12pt,a4paper]{article}
\usepackage{jheppub}  
\usepackage{amssymb} 
\usepackage{amsmath}
\usepackage{mathtools}
\usepackage{amsfonts}    
\usepackage{dsfont}
\usepackage{pdfpages}
\usepackage{verbatim}
\usepackage{tensor}
\usepackage{mathrsfs}
\usepackage[mathscr]{euscript}
\usepackage{tikz}
\usetikzlibrary{patterns,decorations.pathreplacing,calc,shapes.misc,decorations.pathmorphing}

\newcommand{\ba}{\begin{align}}

\newcommand{\be}{\begin{equation}}
\newcommand{\ee}{\end{equation}}
\def\bd{\begin{tikzpicture}}
\def\ed{\end{tikzpicture}}

\tikzset{rubout/.style={/utils/exec=\tikzset{rubout/.cd,#1},
 decoration={show path construction,
      curveto code={
       \draw [white,line width=\pgfkeysvalueof{/tikz/rubout/line width}+2*\pgfkeysvalueof{/tikz/rubout/halo}] 
        (\tikzinputsegmentfirst) .. controls
        (\tikzinputsegmentsupporta) and (\tikzinputsegmentsupportb)  ..(\tikzinputsegmentlast); 
       \draw [line width=\pgfkeysvalueof{/tikz/rubout/line width},shorten <=-0.1pt,shorten >=-0.1pt] (\tikzinputsegmentfirst) .. controls
        (\tikzinputsegmentsupporta) and (\tikzinputsegmentsupportb) ..(\tikzinputsegmentlast);  
      }}},rubout/.cd,line width/.initial=2pt,halo/.initial=0.5pt}

\tikzset{cross/.style={cross out, draw=black, minimum size=2*(#1-\pgflinewidth), inner sep=0pt, outer sep=0pt},
cross/.default={1pt}}
      
\tikzset{
  pics/torus/.style n args={3}{
    code = {
      \providecolor{pgffillcolor}{rgb}{1,1,1}
      \begin{scope}[
          yscale=cos(#3),
          outer torus/.style = {draw,line width/.expanded={\the\dimexpr2\pgflinewidth+#2*2},line join=round},
          inner torus/.style = {draw=pgffillcolor,line width={#2*2}}
        ]
        \draw[outer torus] circle(#1);\draw[inner torus] circle(#1);
        \draw[outer torus] (180:#1) arc (180:360:#1);\draw[inner torus,line cap=round] (180:#1) arc (180:360:#1);
      \end{scope}
    }
  }
}
      
\tikzset{
    partial ellipse/.style args={#1:#2:#3}{
        insert path={+ (#1:#3) arc (#1:#2:#3)}
    }
}

\newcommand{\circlecovering}[1]{\begin{tikzpicture}[declare function={f(\x)=0.2*sin(\x)+\x/1000;}]
\draw[rubout={line width=2pt,halo=1pt},decorate] 
   plot[variable=\x,domain=-50:{#1*360-110},samples=55,smooth] ({f(\x)},{cos(\x)}) to[out=90,in=270] cycle;
\end{tikzpicture}}

\newcommand{\ket}[1]{\left| #1\right\rangle}
\newcommand{\abs}[1]{\left| #1 \right|}
\newcommand*{\tran}{^{\mkern-1.5mu\mathsf{T}}}
\DeclareMathOperator\tr{tr}

\renewcommand\sl{\mathfrak{sl}}
\newcommand\su{\mathfrak{su}}

\newcommand\PSL{\text{PSL}}
\newcommand\GL{\text{GL}}
\newcommand\Sp{\text{Sp}}
\newcommand\PSU{\text{PSU}}
\newcommand\U{\text{U}}
\newcommand\SL{\text{SL}}
\newcommand\SU{\text{SU}}

\newcommand\CC{\mathbb{C}}
\newcommand\ZZ{\mathbb{Z}}
\newcommand\RR{\mathbb{R}}
\newcommand\HH{\mathbb{H}}
\newcommand\CP{\mathbb{CP}}
\newcommand\TT{\mathbb{T}}

\let\S\relax
\DeclareMathOperator\S{S}
\DeclareMathOperator\AdS{AdS}
\DeclareMathOperator\Sym{Sym}

\newcommand\id{\mathds{1}}
\DeclareMathOperator\Hom{Hom}
\DeclareMathOperator\Aut{Aut}
\DeclareMathOperator\Out{Out}

\DeclareMathOperator\Jac{Jac}
\DeclareMathOperator\Tor{Tor}
\DeclareMathOperator\MCG{MCG}
\DeclareMathOperator\vol{vol}

\DeclareMathOperator\Stab{Stab}

\let\Im\relax
\DeclareMathOperator\Im{Im}
\DeclareMathOperator*{\Res}{Res}
\DeclareMathOperator*{\QRes}{QRes}

\allowdisplaybreaks[1]

\title{Summing over Geometries in String Theory}

\author{Lorenz Eberhardt} 
\affiliation{School of Natural Sciences, Institute for Advanced Study, \\
\hspace*{0.3cm}Einstein Drive 1, Princeton,  NJ 08540, USA}
\emailAdd{elorenz@ias.edu}

\abstract{We examine the question how string theory achieves a sum over bulk geometries with fixed asymptotic boundary conditions. We discuss this problem with the help of the tensionless string on $\mathcal{M}_3 \times \mathrm{S}^3 \times \TT^4$ (with one unit of NS-NS flux) that was recently understood to be dual to the symmetric orbifold $\text{Sym}^N(\TT^4)$. We strengthen the analysis of \cite{Eberhardt:2020bgq} and show that the perturbative string partition function around a fixed bulk background already includes a sum over semi-classical geometries and large stringy corrections can be interpreted as various semi-classical geometries. We argue in particular that the string partition function on a Euclidean wormhole geometry factorizes completely into factors associated to the two boundaries of spacetime.
Central to this is the remarkable property of the moduli space integral of string theory to localize on covering spaces of the conformal boundary of $\mathcal{M}_3$. We also emphasize the fact that string perturbation theory computes the grand canonical partition function of the family of theories $\bigoplus_N\text{Sym}^N(\TT^4)$.
The boundary partition function is naturally expressed as a sum over winding worldsheets, each of which we interpret as a `stringy geometry'. We argue that the semi-classical bulk geometry can be understood as a condensate of such stringy geometries. We also briefly discuss the effect of ensemble averaging over the Narain moduli space of $\TT^4$ and of deforming away from the orbifold by the marginal deformation.

}

\begin{document}

\maketitle

\makeatletter
\g@addto@macro\bfseries{\boldmath}
\makeatother

\section{Introduction}
The AdS/CFT correspondence has provided us with a unique glimpse into the properties of quantum gravity and consistency of the correspondence is still surprising from a variety of angles.

There are essentially two classes of proposals that seem to have qualitatively different properties. On the one hand, there are `top down' constructions derived from string theory, such as the duality between type IIB string theory on $\AdS_5 \times \S^5$ and $\mathcal{N}=4$ SYM or the duality between string theory on $\AdS_3 \times \S^3 \times \TT^4$ and a deformation of the symmetric orbifold CFT $\Sym^N(\TT^4)$ \cite{Maldacena:1997re, Witten:1998qj}. 
On the other hand, there are `bottom up' constructions of dual pairs, which start from a semiclassical gravity theory. Since it is not known how to quantize gravity in higher dimensions, these examples are all low-dimensional. The prime example is JT-gravity \cite{Jackiw:1984je, Teitelboim:1983ux}, whose boundary dynamics is given by the Schwarzian theory \cite{Jensen:2016pah, Engelsoy:2016xyb, Maldacena:2016upp}. The Schwarzian theory in turn describes the infrared dynamics of the SYK model \cite{Maldacena:2016hyu, Kitaev:2017awl} that is also described by a double-scaled matrix model \cite{Saad:2019lba, Stanford:2019vob}. The main difference is that the duality in these cases is between a gravity theory and an \emph{ensemble} of CFTs. 
There has been recently also a proposal for an exotic $\U(1)$-gravity theory in three dimensions, that is described holographically by free bosons averaged over the moduli space of Narain lattices \cite{Afkhami-Jeddi:2020ezh, Maloney:2020nni}. 

\begin{figure}
\begin{equation*}
Z(\S^1 \sqcup \S^1)= \ \raisebox{-0.5\height}{\includegraphics[width=.3\textwidth]{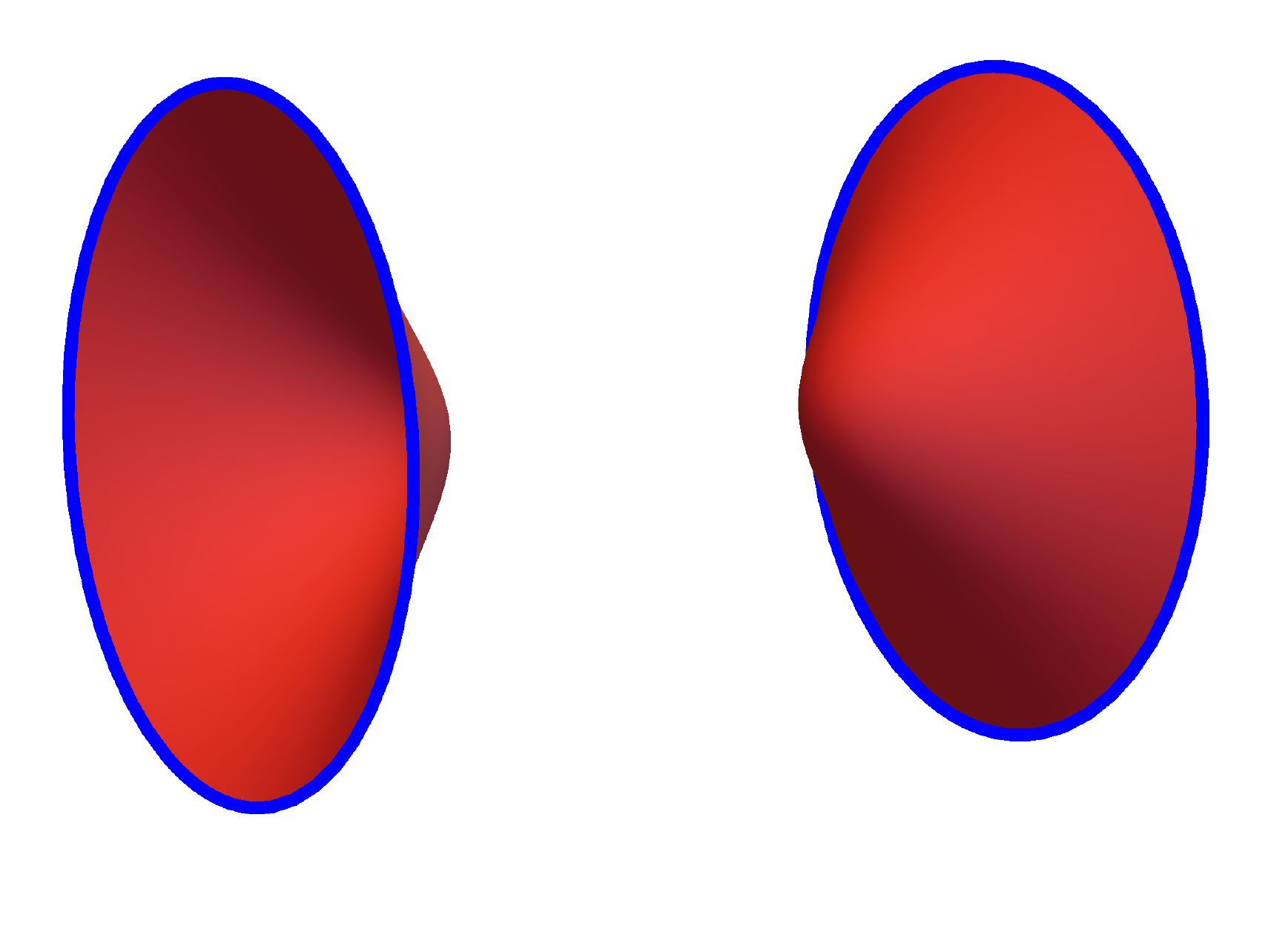}} \ +\  
\raisebox{-0.5\height}{\includegraphics[width=.3\textwidth]{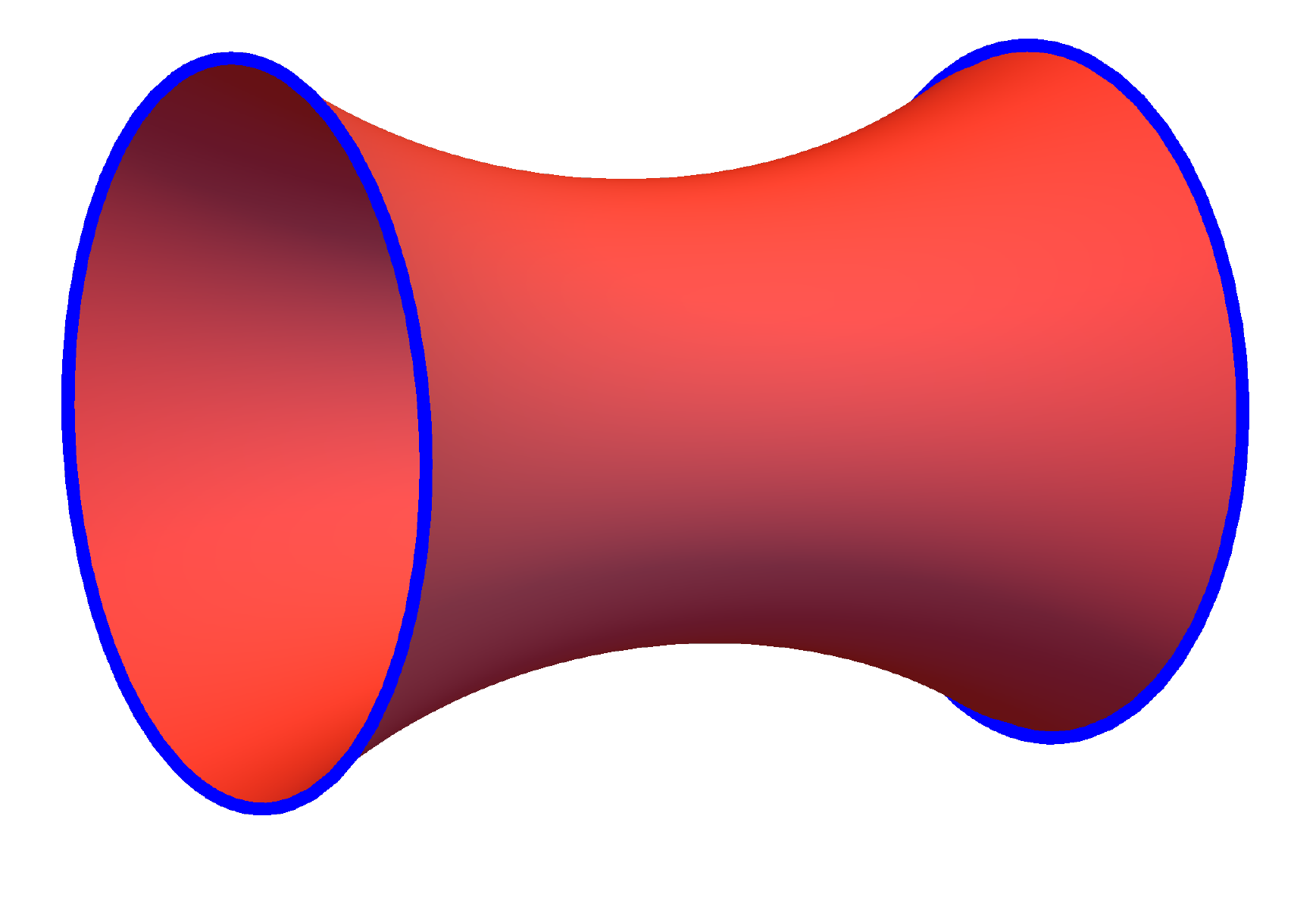}} \ + \dots
\end{equation*}
\caption{The gravitational computation of the partition function with two boundary circles in a two-dimensional theory of gravity. The presence of the wormhole geometry generically destroys factorization of the answer and leads to an ensemble dual on the blue boundary.} \label{fig:wormhole sum}
\end{figure}

The averaged examples behave more intuitively from a gravity point of view. They naturally involve a sum over \emph{all} bulk geometries, including also Euclidean wormhole geometries that are responsible for much of the recent progress on the information paradox \cite{Penington:2019kki, Almheiri:2020cfm}. Inclusion of wormhole geometries spoils factorization of the boundary partition function on disconnected boundaries and leads to the ensemble interpretation \cite{Cotler:2016fpe, Harlow:2018tqv}. See Figure~\ref{fig:wormhole sum}. The typical member of the ensemble in those dualities also generically seems to exhibit chaos. The dual gravitational theory captures only the averaged signal and it is an important open problem to explain where the random noise comes from in the gravitational description. See \cite{Saad:2018bqo, Bousso:2020kmy, Cotler:2020ugk, Belin:2020hea, Stanford:2020wkf, Marolf:2020xie, Blommaert:2020seb, Mertens:2020hbs} for recent progress on this.

It has been a subject of debate how to relate these two classes of proposals. The bulk properties of the stringy examples of the correspondence away from their supergravity regime are much more alien to us, which is mainly due to our lack of understanding and computational power of string theory (or M-theory). In particular, for such stringy examples of holography, there cannot be a non-zero wormhole correction to the partition function, since it would be inconsistent with a single boundary theory.\footnote{By a wormhole solution we mean here and in the following a connected bulk manifold whose conformal boundary is disconnected.} In the supergravity approximation, the stringy examples $\AdS_5 \times \S^5$ and $\AdS_3 \times \S^3 \times \TT^4$  admit Euclidean wormhole solutions \cite{Maldacena:2004rf, ArkaniHamed:2007js, Marolf:2021kjc}, in tension with a single local boundary CFT. This already indicates that string theory modifies the `sum over geometries' prescription in a non-trivial way.
\medskip

In this paper, we revisit the question directly within string theory. Our example is the symmetric orbifold CFT $\Sym^N(\TT^4)$. This precise theory is conjectured to be dual to `tensionless' string theory on $\AdS_3 \times \S^3 \times \TT^4$ \cite{Eberhardt:2018ouy, Eberhardt:2019ywk, Dei:2020zui},  see also \cite{Gaberdiel:2018rqv, Giribet:2018ada} for earlier work. The string background is supported only by one unit of NS-NS flux (and no R-R flux). This duality has been subjected to very stringent tests: not only has the full spectrum been matched with the symmetric product orbifold \cite{Eberhardt:2018ouy}, but also some correlators \cite{Eberhardt:2019ywk, Hikida:2020kil, Dei:2020zui} (including higher genus correlators \cite{Eberhardt:2020akk, Knighton:2020kuh}) have been compared. 

While this model is under very good control, it also has some downsides. The first of these is the non-geometric and non-local nature of the theory. A small tension means that strings are very floppy and generically can wind around asymptotic regions or cycles of the geometry with little cost of energy. The generic state of the theory has lots of winding strings. This is in particular true for the graviton and thus there does not seem to be a local notion of geometry. However, it is somewhat premature to disregard classical geometry entirely. When treating the string perturbatively, we start with a sigma model on a fixed background and of course the string should `feel' the background geometry. Thus, we can still ask questions about sums over geometries. While the concept of a background geometry is not very well-defined in the regime in which we are working, the concept of a worldsheet theory is. For a large background, these are equivalent -- any classical background gives rise to a worldsheet theory. Thus, we will essentially replace the notion of summing over geometries by sums over different worldsheet theories.

While questions such as `which manifolds should we sum over to obtain the boundary partition function?' have a clear answer in this framework, the result is somewhat difficult to interpret from a semiclassical gravity point of view. Indeed, the answer to this question that we shall advocate in this paper is that \emph{the string partition function is independent of the background bulk manifold, large stringy corrections around the given background ensure that all the other semi-classical bulk geometries are automatically taken into account.} This even extends to wormhole geometries. So instead of Figure~\ref{fig:wormhole sum}, the correct answer in the tensionless string is to take either of the two geometries and consider all the stringy excitations on it -- the result will be the same! In this sense, the two contributions in Figure~\ref{fig:wormhole sum} should not be included separately, since we would count the same state multiple times. We already conjectured this to be the case in \cite{Eberhardt:2020bgq}. 

This answer was already anticipated long ago from a purely boundary perspective in free $\mathcal{N}=4$ SYM \cite{Sundborg:1999ue}. There it was seen that the thermal partition function of free $\mathcal{N}=4$ SYM exhibits a phase transition due to an exponential number of light strings. This is the Hawking-Page transition emulated by a large number of stringy corrections. However, a bulk description was missing at the time.

This proposal seems counterintuitive at first glance. The action of a classical background is of order $\mathcal{O}(G_\text{N}^{-1})$, whereas quantum corrections around it should be of order 1 in $G_\text{N}$. This seems to make it impossible for the above statement to be true. This naive argument is circumvented as follows. It is true that the, say, torus partition function of a single string is of order one, but only as long is that string is `short'. The contribution can be enhanced by a factor of $N$ by taking a string that either winds $N$ times around a cycle or asymptotic region of the geometry or $N$ strings that each wind once around it. In the example at hand there can be a very large quantum correction to the classical result due to strings that effectively wind $G_\text{N}^{-1}$ times such cycles. This is essentially the Hagedorn transition, since the very large number of light strings can lead to a macroscopic contribution. These are heavy enough to backreact on the geometry and change it effectively into a different geometry.\footnote{In \cite{Amado:2016pgy, Amado:2017kgr}, the question was examined in vector and matrix models and a similar result was found.}

\begin{figure}
\begin{center}
\begin{minipage}{.49\textwidth}
\includegraphics[width=\textwidth]{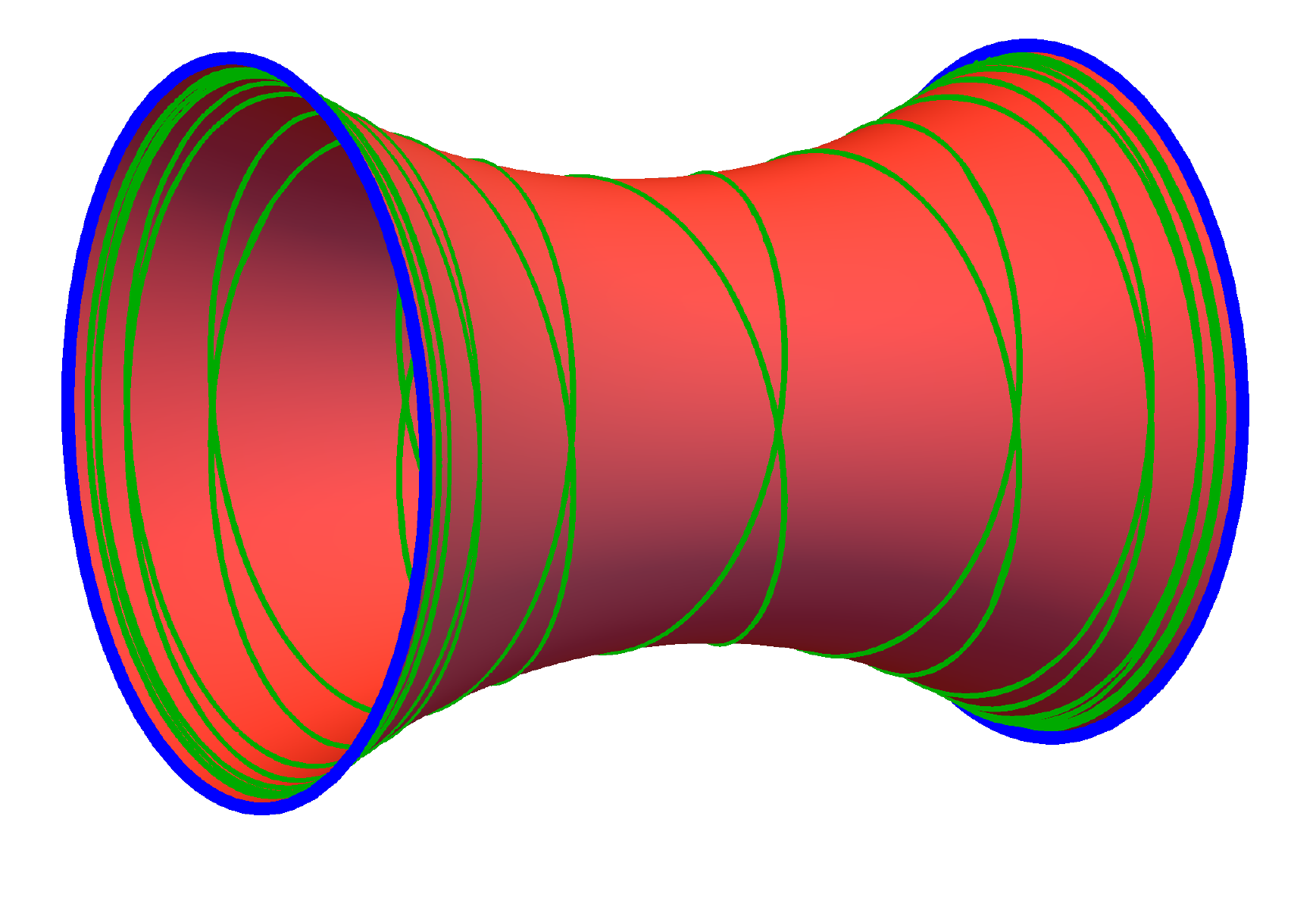}
\end{minipage}
\begin{minipage}{.49\textwidth}
\includegraphics[width=\textwidth]{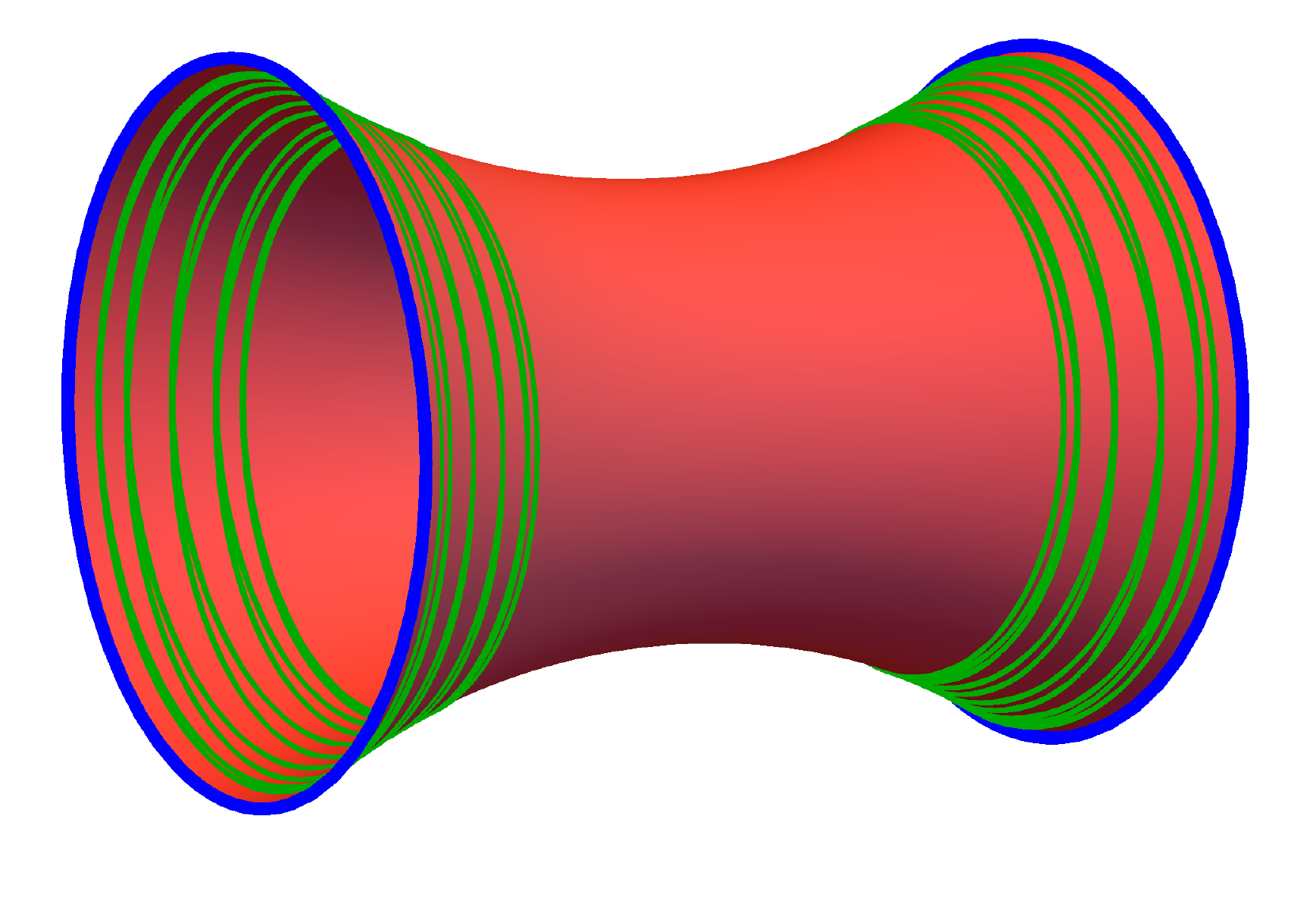}
\end{minipage}
\end{center}
\caption{The wormholes with perturbative string excitations on top of them. These pictures are cartoons, since the actual model that we will consider is three-dimensional and instead of the green worldline curves, we will have worldsheets wrapping the geometry. It turns out that the model realizes only the possibility in the right picture: there are no perturbative string excitations that connect the two asymptotic regions, but only separate string excitations that stay close to the two boundaries.} \label{fig:string on wormhole}
\end{figure}

This still does not explain intuitively why the wormhole partition function factorizes. The answer that we find to this question is perhaps a bit disappointing. Naively, one could have expected that there are string configurations like the first picture of Figure~\ref{fig:string on wormhole} that stretch between the two boundaries and cause a  correlation between the two boundary theories, thus leading to non-factorization. However, such strings actually do not exist in the model. It turns out that all the strings of the model stay close to the boundary of the space as in the second picture of Figure~\ref{fig:string on wormhole}. Since the geometry is asymptotically $\AdS$, they actually stay asymptotically far out and do not explore the bulk. This picture makes it intuitively clear how factorization is achieved. One could even say that there is no bulk, since an observer in such a stringy universe would have no way of detecting it. 

We should mention that the actual mechanism that achieves this `localization' of the string worldsheets to the boundary is quite surprising and works thanks to an actual localization in the moduli space of Riemann surfaces. The string path integral localizes to such Riemann surfaces that cover the boundary holomorphically, which reduces the integral over the moduli space to a sum. This localization principle was discussed and proved in \cite{Eberhardt:2019ywk, Eberhardt:2020akk, Dei:2020zui, Knighton:2020kuh} for correlation functions in global $\AdS_3$. We extend the proof to arbitrary higher genus worldsheets and all (possibly singular) hyperbolic three manifolds, which are the manifolds that can serve as background geometry for the string (modulo some technical issues that will be explained). Even though the main focus of this work lies on partition functions, we explain that our proof also goes through for correlation functions, in which case the worldsheet localizes on certain ramified covers of the boundary.  

\begin{figure}
\begin{center}
\includegraphics[width=.5\textwidth]{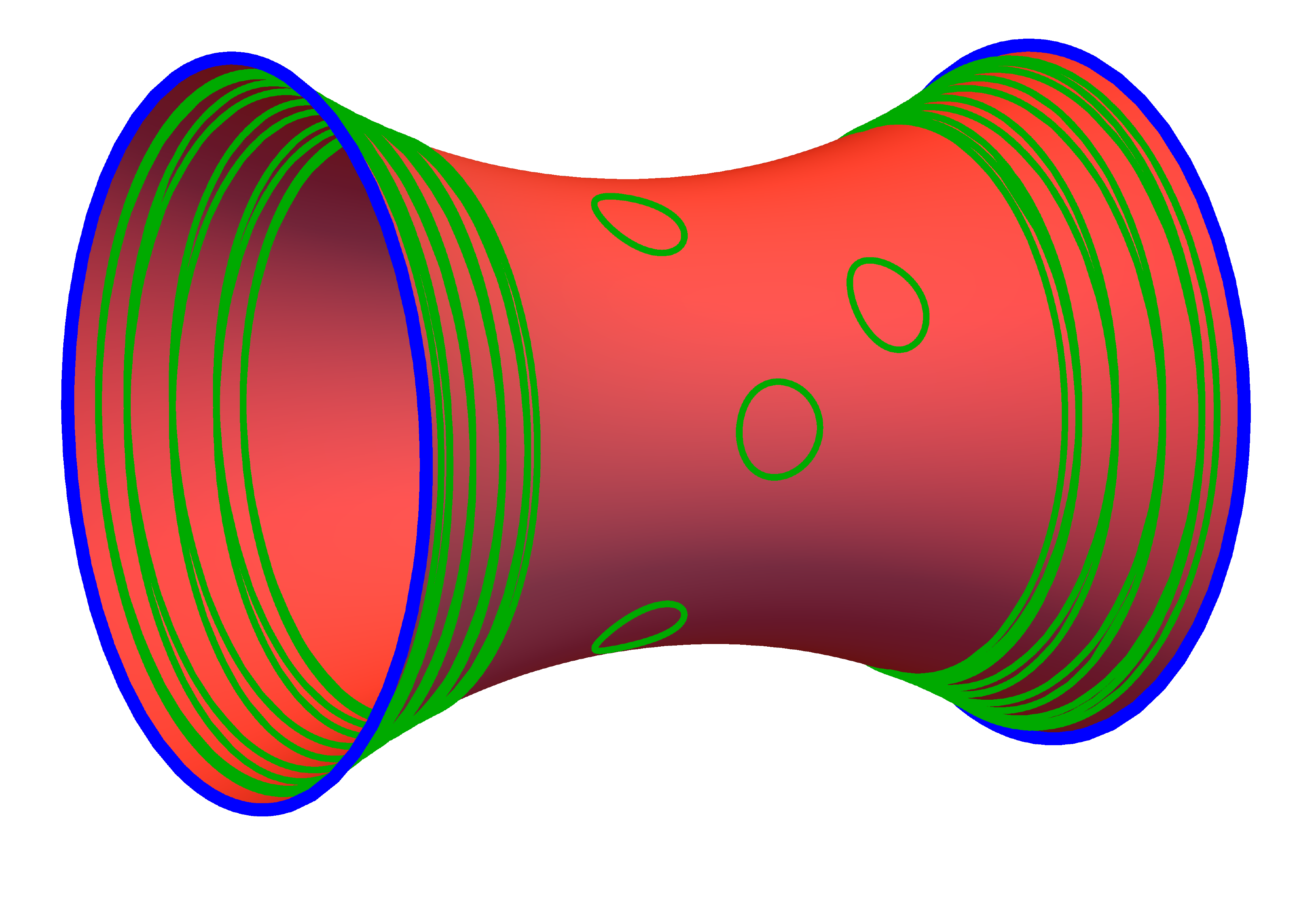}
\end{center}
\caption{A typical string configuration for string theory on the wormhole geometry. As in Figure~\ref{fig:2d bulk}, this picture is drawn in one dimension lower. The small strings in the middle are spheres, whereas the winding strings close to the boundary are genus $g \ge 1$ surfaces.} \label{fig:wormhole with spheres}
\end{figure}

We will not be able to compute the actual value of the string partition function, but explain the general mechanisms behind the independence of the string partition function on the bulk geometry in this model. Even after the localization property has been demonstrated, this is non-trivial because the sphere partition function of the worldsheet theory does not follow the localization principle and is not confined to the boundary of the bulk. Thus, the typical string configuration actually looks roughly like in Figure~\ref{fig:wormhole with spheres}.
It is therefore not true that the quantum corrections around the background know nothing of the background. We argue that their dependence on the bulk geometry is such that they cancel the sphere partition function and the combined string partition function is independent of the chosen bulk manifold. It is also interesting that the sum over all possible covering maps of the boundary can arise in very different manners for different bulk manifolds.

We have just explained that the tensionless string of this model does not `see' the bulk. It is therefore surprising that actually all classical bulk manifolds make an appearance in the boundary theory. For this, it is crucial to work in a grand canonical ensemble, where the number of strings is not fixed, since this is the natural ensemble of string perturbation theory. The dual `CFT' is hence actually $\bigoplus_N \Sym^N(\TT^4)$ with an appropriate chemical potential conjugate to $N$ held fixed. When tuning this chemical potential to special values, the grand canonical partition function is dominated by very large $N$ and classical geometry seems to emerge as a condensate of the winding strings. We confirmed this explicitly only for geometries with a single torus boundary such as thermal $\AdS$, since the simplest wormhole geometry has genus 2 boundaries, which makes explicit computations difficult. The emergence of classical geometries as condensates of `stringy' geometries (i.e.~winding strings) becomes physically clearer once we average the boundary CFT over a suitable set of parameters. As an example, we discuss this for the Narain moduli space of $\TT^4$. Using the results of \cite{Afkhami-Jeddi:2020ezh, Maloney:2020nni}, the averaged string partition function is then expressed as a sum over `micro-geometries', i.e.\ geometries that fill in worldsheets (or connect them by `micro-wormholes'). Such geometries have a very large number of sheets that meet asymptotically at the boundary of the bulk. When these geometries align properly, they can form an emergent macroscopic geometry in the classical sense. 

\medskip

\paragraph{Outline.} The following is an outline of the paper. In the remaining part of the introduction, we first explain our philosophy on how we (approximately) compute string partitions with fixed boundary conditions using string perturbation theory. We then discuss in Section~\ref{subsec:toy model} some of the ideas of this paper with the help of a very simple toy model in two bulk dimensions, where instead of worldsheets, the boundary is covered by worldlines. This model exhibits already some (but not all!) features of the tensionless string and is technically much simpler to treat. The remaining part of the paper is roughly divided into two parts which can be read more or less independently.

Sections~\ref{sec:partition function symmetric orbifold}--\ref{sec:background independence} are more technical and their main goal is to establish the independence of the string partition function on the bulk geometry. Of those, Section~\ref{sec:partition function symmetric orbifold} reviews the computation of the partition function of the symmetric orbifold and Section~\ref{sec:tensionless string on locally AdS3 backgrounds} reviews and develops the formulation of the tensionless string on different bulk manifolds. We use this formalism to demonstrate that the string path integral does indeed localize in the moduli space of Riemann surface. We explore the consequences of this property in Section~\ref{sec:background independence} and explain how to introduce the grand canonical potential from the bulk point of view. These sections make use of some technology from the theory of Riemann surfaces and hyperbolic manifolds. For the benefit of the reader, we collected the relevant material in Appendices~\ref{app:Riemann surfaces}, \ref{app:subgroups}, \ref{app:uniformization}, \ref{app:branched complex projective structures} and \ref{app:hyperbolic 3 manifolds}. Appendix~\ref{app:topologically twisted partition function} contains a discussion of the topologically twisted partition function of the sigma model on $\TT^4$ that is a part of the worldsheet theory.

Section~\ref{sec:stringy and classical geometry} is more physical in nature and discusses the physical interpretation of classical geometries emerging as condensates from `stringy geometries'. We also discuss the effect of introducing an ensemble average in the symmetric orbifold and the `micro-geometries' that we mentioned above. 

We summarize our main findings in Section~\ref{sec:summary discussion} and discuss open problems and future directions.

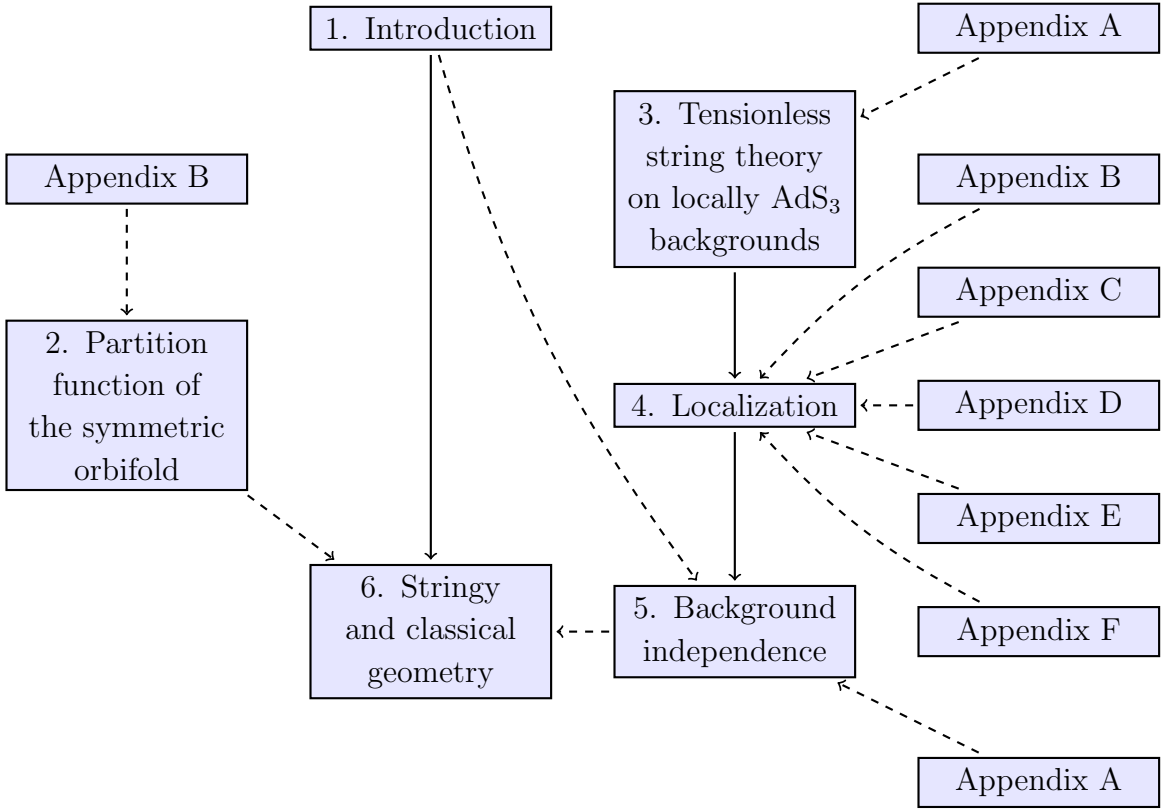
\begin{figure}
\begin{center}
\begin{tikzpicture}[nodes={draw, thick, fill=blue!10}]
\node[rectangle, text width=7em, text centered, outer sep=2pt] (s1) at (0,2) {1. Introduction};
\node[rectangle, text width=7em, text centered, outer sep=2pt] (s2) at (-4,-3) {2. Partition function of the symmetric orbifold};
\node[rectangle, text width=7em, text centered, outer sep=2pt] (s3) at (4,0) {3. Tensionless string theory on locally $\AdS_3$ backgrounds};
\node[rectangle, text width=7em, text centered, outer sep=2pt] (s4)
at (4,-3) {4. Localization};
\node[rectangle, text width=7em, text centered, outer sep=2pt] (s5)
at (4,-6) {5. Background independence};
\node[rectangle, text width=7em, text centered, outer sep=2pt] (s6)
at (0,-6) {6. Stringy and classical geometry};
\node[rectangle, text width=7em, text centered, outer sep=2pt] (A) at (8,2) {Appendix A};
\node[rectangle, text width=7em, text centered, outer sep=2pt] (B) at (8,0) {Appendix B};
\node[rectangle, text width=7em, text centered, outer sep=2pt] (C) at (8,-1.5) {Appendix C};
\node[rectangle, text width=7em, text centered, outer sep=2pt] (D) at (8,-3) {Appendix D};
\node[rectangle, text width=7em, text centered, outer sep=2pt] (E) at (8,-4.5) {Appendix E};
\node[rectangle, text width=7em, text centered, outer sep=2pt] (F) at (8,-6) {Appendix F};
\node[rectangle, text width=7em, text centered, outer sep=2pt] (A2) at (8,-8) {Appendix A};
\node[rectangle, text width=7em, text centered, outer sep=2pt] (B2) at (-4,0) {Appendix B};

\draw[thick,->] (s1) to (s6);
\draw[thick,->,dashed, bend right=10] (s1) to (s5);
\draw[thick,->, dashed] (s2) to (s6);
\draw[thick,->] (s3) to (s4);
\draw[thick,->] (s4) to (s5);
\draw[thick,->, dashed] (s5) to (s6);
\draw[thick,->, dashed] (A) to (s3);
\draw[thick,->, dashed, bend right=10] (B) to (s4);
\draw[thick,->, dashed] (C) to (s4);
\draw[thick,->, dashed] (D) to (s4);
\draw[thick,->, dashed] (E) to (s4);
\draw[thick,->, dashed, bend left=10] (F) to (s4);
\draw[thick,->, dashed] (A2) to (s5);
\draw[thick,->, dashed] (B2) to (s2);
\end{tikzpicture}
\end{center}
\caption{The basic organization of the paper. Arrows indicate strong logical dependencies and we recommend to read the relevant sections first. Dashed lines indicate weaker dependencies.  } \label{fig:organization}
\end{figure}

\paragraph{Suggested reading.} To simplify the reading process, we have depicted the dependencies of the various sections in Figure~\ref{fig:organization}. We suggest that the reader can jump directly after the introduction to section \ref{sec:stringy and classical geometry} and check back on the other sections as needed. Appendices contain background information and are included to make the paper self-contained. They are mostly not necessary to understand the main text.

\subsection{Computing string partition functions}
Here, we make some comments about computing the full non-perturbative string path integral with fixed boundary conditions from the bulk point of view. We discuss this using string perturbation theory. The question should properly be addressed within string field theory. but to make computation feasible we use string perturbation theory as an approximation. We will treat spacetime as Euclidean.

The problem in string perturbation theory is that we treat the background geometry as fixed and consider stringy excitations around the background. 
Let us fix some asymptotic boundary conditions of the bulk spacetime manifolds (such as asymptotically $\AdS$). Then to compute the string partition function with these boundary conditions, we should in principle sum over all bulk manifolds with the appropriate boundaries and include stringy corrections around those bulk geometries:
\be 
Z_\text{string}( \mathcal{N})=\sum_{\text{bulk saddle geometries $\mathcal{M}$ with $\partial \mathcal{M}=\mathcal{N}$}} Z_\text{string}(\mathcal{M})\ . \label{eq:string partition function manifold sum}
\ee
Here, $Z_\text{string}(\mathcal{M})$ is the perturbative string partition function around the background $\mathcal{M}$ and $Z_\text{string}(\mathcal{N})$ is the string partition function with fixed boundary conditions. This is essentially how we would compute compute a semiclassical gravity partition function.
There are several problems with this:
\begin{enumerate}
\item String theory also includes non-perturbative objects (in $g_\text{string}$) and they should in principle also be included in the string partition function on a fixed manifold $\mathcal{M}$. 
\item In general, we expect some backreaction of the string on the geometry. Thus, when including very heavy string excitations in the partition function, they can change the background geometry. Hence we should only include `light' string excitations around a fixed background.
\item String theory contains also other massless fields besides the metric. Thus, the sum over geometries should rather be a sum over supergravity backgrounds.
\item The background geometries $\mathcal{M}$ should be saddles, i.e.~satisfy the supergravity equation of motion. These equations get $\alpha'$ corrected and in principle there could be also be also background values for all the massive string modes. The more correct statement would be to sum instead of over bulk manifolds over different worldsheet sigma-models with the correct asymptotic boundary condition.
\item In the gravitational path integral, we should sum over \emph{all} bulk manifolds, whether they are saddles or not. Of course, saddle geometries lead to a dominant contribution. In string perturbation theory, we do not know even in principle how to include non-saddle geometries, since these do not correspond to consistent worldsheet theories.
\end{enumerate}
These issues make it clear that \eqref{eq:string partition function manifold sum} can at best be an approximation (or at least the definition of $Z_\text{string}(\mathcal{M})$ is not straightforward).

In the example that we consider in this paper, we will argue that \eqref{eq:string partition function manifold sum} fails much more profoundly. For the tensionless string on a manifold $\mathcal{M}_3 \times \text{S}^3 \times \TT^4$ the sum over manifolds is superfluous. It is already fully contained $Z_\text{string}(\mathcal{M})$, as long as we include also heavy string excitations. We are lucky that in this instance the answer we compute turns out to be exact and no effect of backreaction has to be taken into account. So instead of the sum in \eqref{eq:string partition function manifold sum}, we have $Z_\text{string}(\mathcal{N})=Z_\text{string}(\mathcal{M})$ for \emph{any} bulk manifold $\mathcal{M}$ with $\partial \mathcal{M}=\mathcal{N}$. Indications for this in other models were also found in \cite{Furuuchi:2005qm, Amado:2016pgy, Amado:2017kgr}.
\subsection{A simple toy model} \label{subsec:toy model}
Some of the physical intuition for the setup in this paper can understood from a very simple toy model. Let us consider a quantum mechanical model with a one-dimensional Hilbert space. The unique state in this model is taken to have energy $E_0$. We now consider $N$ copies of the model and gauge the obvious $S_N$-symmetry that permutes the copies. This gauging implements physically the intuition that the different copies are indistinguishable.

\begin{figure}
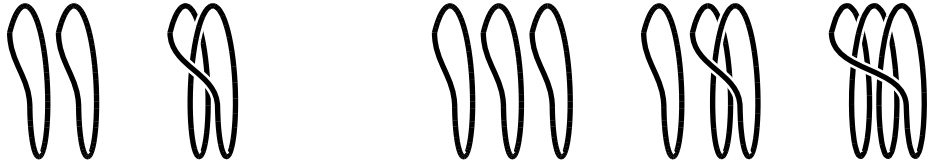

\begin{center}
\circlecovering{1}\circlecovering{1}\qquad \circlecovering{2}\qquad\qquad\qquad \circlecovering{1}\circlecovering{1}\circlecovering{1}\qquad \circlecovering{1}\circlecovering{2}\qquad \circlecovering{3}
\end{center}
\caption{The $S_N$-bundles in the toy model for $N=2$ and $N=3$.} \label{fig:circle covering maps N2 N3}.
\end{figure}

Let us compute the partition function of this theory. The $N$-fold product of the original theory has still a one-dimensional Hilbert space and the unique state is invariant under the $S_N$ symmetry. Thus, the partition function is simply
\be 
Z_N=\tr\left(\mathrm{e}^{-\beta H} \right)=\mathrm{e}^{-\beta N E_0}\ .
\ee
Let us see how this simple result arises from a path integral point of view. The Euclidean quantum mechanics is considered on a thermal circle of length $\beta$. Gauging of $S_N$-symmetry instructs us to sum over all $S_N$-bundles over the thermal circle. The $N$ copies of the model can be viewed as $N$ separate thermal circles. Gauging of $S_N$ sums over all possible joinings of these circles. We displayed the possibilities for $N=2$ and $N=3$ in Figure~\ref{fig:circle covering maps N2 N3}.
In general such a bundle is determined by a partition of $N$ that labels the lengths of the connected components of the bundle.
These $S_N$-bundles have nontrivial symmetry factors that one needs to take into account. For a partition $N=\sum_{m=1}^\infty N_m m$, the symmetry factor is 
\be 
\prod_m \frac{1}{N_m! \, m^{N_m}}\ , \label{eq:symmetry factor}
\ee
which accounts for the fact that we can perform a cyclic relabelling of the covering within each connected components and permute identical components. We then have
\be 
Z_N=\mathrm{e}^{-\beta N E_0}\sum_{\text{partitions } N=\sum_m N_m m} \prod_m \frac{1}{N_m! m^{N_m}}=\mathrm{e}^{-\beta N E_0}\ .
\ee
The last identity follows from the fact that partitions label conjugacy classes of $S_N$ and the size of these conjugacy classes is $N!$ times the symmetry factor \eqref{eq:symmetry factor}.

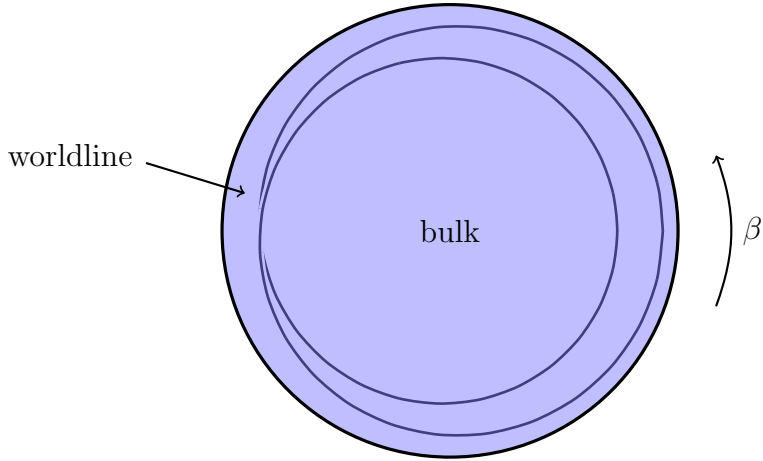
\begin{figure}
\begin{center}
\begin{tikzpicture}
\draw[rubout={line width=1pt,halo=1pt},decorate,red] plot[variable=\x,domain=0:720,samples=55,smooth] ({(2.5+.3*cos(\x/2))*cos(\x)},{(2.5+.3*cos(\x/2))*sin(\x)}); 
\fill[blue!50!white,opacity=.5] (0,0) circle (3);
\draw[very thick] (0,0) circle (3);
\draw[->,bend right=20, thick] (3.5,-1) to node[right] {$\beta$} (3.5,1);
\node at (0,0) {bulk};
\node at (-5,1) {worldline};
\draw[->,thick] (-4,.9) to (-2.7,.5);
\end{tikzpicture}
\end{center}
\caption{We fill the boundary thermal circle with a bulk spacetime. On this bulk spacetime, we let free particles propagate. In the picture, the worldline represents the second covering space of Figure~\ref{fig:circle covering maps N2 N3}.} \label{fig:2d bulk}
 \end{figure}

\paragraph{Grand canonical ensemble.} We now want to interpret this system holographically. For this, the path integral point of view is useful. We view the covering spaces of the boundary circle as particles that propagate near the boundary of the Euclidean bulk spacetime. This is depicted in Figure~\ref{fig:2d bulk}. In this sense, our very simple toy model is dual to free particles in the bulk. Hence the spacetime theory would be a QFT on $\AdS_2$. In a QFT, the number of particles in the background is usually not fixed. Thus, we would like to consider both sides of the `holographic correspondence' in the grand canonical ensemble, where $N$ is not fixed. 
 Instead, we fix a corresponding fugacity variable $p$. The boundary grand canonical partition partition function reads
 \be 
 \mathfrak{Z}=\sum_{N=0}^\infty p^N Z_N=\sum_{N=0}^\infty p^N\mathrm{e}^{-\beta N E_0}=\frac{1}{1-p \mathrm{e}^{-\beta E_0}}\ . \label{eq:2d grand canonical partition function}
 \ee
\paragraph{Worldline action.} We now argue that this model can indeed be taken seriously by constructing a worldline action for the particle. Let us choose polar coordinates to describe the bulk. For definiteness, let us take the bulk to have a flat metric
\be 
\mathrm{d}s^2=\mathrm{d}r^2+r^2 \mathrm{d}\phi^2\ ,
\ee
where $0 \le r \le \frac{\beta}{2\pi}$ and $0 \le \phi \le 2\pi$. Then we take the length of the worldline as its action:
\be 
S_\text{worldline}=E_0\int \mathrm{d}\tau \ \sqrt{G_{\mu\nu}\partial_\tau X^\mu \partial_\tau X^\nu}\ .
\ee
where $X(\tau)=(r(\tau),\phi(\tau))$ are the embedding coordinates in spacetime and $G_{\mu\nu}$ is the spacetime metric. Of course, the equations of motion of this theory allow only for straight worldlines, but one could cure this by adding a very steep Mexican hat-type potential that confines the worldline to the boundary region of the bulk. Then possible worldlines are labelled by their topological winding number near the boundary of spacetime. This can be viewed as an (approximate) type of localization: only worldlines that cover the boundary isometrically are allowed to contribute to the partition function. Hence the length of the worldsheet is $d \beta$ for some integer $d \ge 1$. 
The connected partition function of this model is simply\footnote{We are suppressing some details here that are not important to understand the physics. There are one-loop determinants around the classical solutions that should be taken into account for a complete description. }
\be 
\mathfrak{Z}^\text{conn}_\text{bulk}=\sum_{d=1}^\infty \frac{1}{d} \mathrm{e}^{-\beta d E_0}=-\log \left(1-\mathrm{e}^{-\beta E_0} \right)\ . 
\ee
We can further refine this partition function by counting the winding number. We will do this here in an-hoc manner and simply posit that
\be 
\mathfrak{Z}_\text{bulk}^\text{conn}=\sum_{d=1}^\infty \frac{p^d}{d} \mathrm{e}^{-\beta d E_0}=-\log \left(1-p\mathrm{e}^{-\beta E_0} \right)\quad \Longrightarrow \quad \mathfrak{Z}_\text{bulk}=\frac{1}{1-p \mathrm{e}^{-\beta E_0}}\ ,
\ee
in agreement with what we found in the boundary theory.

\begin{figure}
\begin{center}
\begin{tikzpicture}
\draw[thick, dashed] (0,0) [partial ellipse=0:180:2 and .5];
\draw[thick] (0,0) [partial ellipse=180:360:2 and .5];
\draw[thick] (-2,0) ..controls (-2,3) and ( 2,3).. (2,0);
\draw[red, thick] (0,.3) [partial ellipse=180:360:1.97 and .5];
\draw[red, thick] (0,.6) [partial ellipse=180:360:1.93 and .5];
\draw[red, thick,in=90,out=90, looseness=.4,dashed] (1.97,.3) to (-1.93,.6);
\draw[red, thick,in=90,out=90, looseness=.4,dashed] (1.93,.6) to (-1.97,.3);
\node at (3,1) {=};
\draw[thick, dashed] (6,0) [partial ellipse=0:180:2 and .5];
\draw[thick] (6,0) [partial ellipse=180:360:2 and .5];
\draw[thick] (4,0) ..controls (4,4) and ( 8,4).. (8,0);
\draw[thick,bend right=30] (5,2) to (7,2);
\draw[thick,bend left=30] (5.3,1.85) to (6.7,1.85);
\draw[red, thick] (6,.3) [partial ellipse=180:360:1.97 and .5];
\draw[red, thick] (6,.6) [partial ellipse=180:360:1.93 and .5];
\draw[red, thick,in=90,out=90, looseness=.4,dashed] (7.97,.3) to (4.07,.6);
\draw[red, thick,in=90,out=90, looseness=.4,dashed] (7.93,.6) to (4.03,.3);
\end{tikzpicture}
\end{center}
\caption{The indistinguishability of different bulk manifolds. Since the red worldline stays close to the boundary, different topologies are not distinguishable for it.} \label{fig:2d bulk independence}
\end{figure}
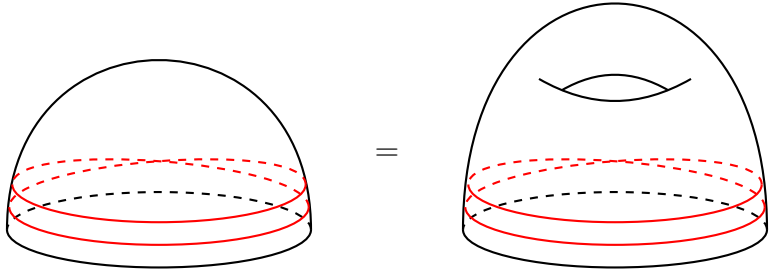

\paragraph{Bulk independence.} We see that the bulk result didn't depend a lot on the details of the bulk. We could have equally well chosen a different bulk manifold, such as a genus 1 surface with a circle removed. We only cared about the properties of the bulk manifold near its boundary (thanks to the Mexican hat potential). This is essentially because of our choice of potential that pushes the worldline out towards the boundary of the bulk. In pictures, this is Figure~\ref{fig:2d bulk independence}.

\paragraph{Condensation.} One might say that the `bulk' theory essentially knows nothing about the bulk, since all the degrees of freedom seem to be localized near the boundary. This is not entirely true. These worldline geometries do reflect some parts of the classical geometry. For this, we notice that the grand canonical partition function has a pole at $p=\mathrm{e}^{\beta E_0}$. We shall argue that the meaning of this pole is the following. When tuning $p$ close to $\mathrm{e}^{\beta E_0}$, worldline geometries with large $N$ dominate the definition of the grand canonical partition function \eqref{eq:2d grand canonical partition function}. This means that the typical bulk picture features an extremely large number of worldlines. On the pole, these worldlines essentially reconstruct the classical bulk geometry, or in a precise sense that we discuss in Section~\ref{subsec:condensation} the bulk geometry is a condensate of the worldlines. In our toy example, we do not know anything about the effective spacetime theory, so we cannot say which classical bulk geometry this should be. In the string theory example that we examine in this paper, we see all bulk geometries emerging that satisfy the classical equations of motion. The location of the pole in $p$ is related to the on-shell spacetime action of the geometry. The residue is essentially the one-loop determinant around that geometry (that is trivial in our toy example). This allows one to reconstruct the grand canonical partition function also alternatively from all the classical partition functions.
\paragraph{Euclidean wormholes.} One can consider our toy model also on wormhole geometries. It is  clear that what we said for the bulk independence continues to hold for wormhole geometries. If several boundaries are present, we can introduce different chemical potentials for the various boundaries and the grand canonical partition function simply factorizes into its constituents. In particular, the multi-boundary grand canonical partition function in out model is simply
\be 
\prod_{i=1}^n \frac{1}{1-p_i \mathrm{e}^{-\beta_i E_0}}\ .
\ee
The poles in this partition function should again correspond to different bulk `geometries'. However, they clearly only account for disconnected geometries. For the string the situation is much more complicated, since (saddle) wormholes are only expected for genus $ \ge 2$ boundaries. We have not been able to determine whether the Euclidean wormholes leave some imprint on the grand canonical wormhole partition function. See the discussion in Section~\ref{subsec:wormhole}.

\paragraph{Averaging.} In this final paragraph, we modify our theory a bit. We want to introduce an average of theories, but with our simple theory with a single state, this is not possible.\footnote{Averaging over $E_0$ would essentially average over the size of the bulk, which is not what we want.}
We replace our simple quantum mechanical theory with another quantum mechanical theory $X$. We will assume that $X$ itself has some parameters over which we can average (for example the SYK model). We then take $N$ copies of $X$ and gauge the permutation symmetry. For a single realization, we can again express the boundary partition function as a sum over all possible covering maps of the boundary circle. But we can now interpret the partition functions on the covering maps themselves using holography. Once we average over the parameters of the theory, they are computed holographically by filling in the covering geometry in all possible ways, thus leading to a sort of `micro-geometry' consisting of $N$ sheets. This includes wormholes that connect different disconnected components of the covering space. The model that is constructed in this way is in some sense stringy, since it has exponentially more geometries (in $N$) then $X$ itself. Moreover, these `micro-geometries' can align themselves to lie on top of each other and form one macroscopic geometry. This leads to a more concrete picture of the condensation process.

\section{Partition function of symmetric orbifolds at higher genus} \label{sec:partition function symmetric orbifold}
Let us review the partition function of the symmetric orbifold $\mathrm{Sym}^N(X)$ for some base theory $X$, following \cite{Bantay:1998fy, Haehl:2014yla}. 

\subsection{Setup}
We consider a Riemann surface $\Sigma_g$ of genus $g$. Because of the conformal anomaly, the partition function does not only depend on the moduli of the surface, but also on an explicit metric. Let us fix the hyperbolic metric, i.e.\ the metric with constant negative curvature $-1$ on $\Sigma_g$ (or a flat metric in the case $g=1$) and write
\be 
\Sigma_g=\mathbb{H}^2/\Gamma_g^\text{F}\ ,
\ee
for some discrete (Fuchsian) subgroup of $\Gamma_g^\text{F} \subset \PSL(2,\mathbb{R})$ and $\HH^2$ the upper half plane.\footnote{The discussion also holds for $g=1$, in which case the universal covering space is $\mathbb{C}$ and the relavant group $\Gamma_1=\mathbb{Z} \times \mathbb{Z}$ acts by translations.} $\Gamma_g^\text{F}$ is determined up to overall conjugation. The upper half plane with the Poincar\'e metric induces the hyperbolic metric on $\Sigma_g$.

\subsection{Grand canonical ensemble} \label{subsec:grand canonical ensemble symm orbifold}
It is convenient to compute partition functions in a grand canonical ensemble. Holographically, the grand canonical ensemble corresponds to the situation where the number of strings in the background is not kept fixed. We will discuss this in Section~\ref{sec:background independence} and \ref{sec:stringy and classical geometry} in detail. From a CFT point of view, we are computing the generating function of the partition function
\be 
\mathfrak{Z}_{\text{Sym}(X)}=\sum_{N=0}^\infty p^N Z_{\mathrm{Sym}^N(X)}\ , \label{eq:grand canonical partition function}
\ee
where we suppressed the dependence on the moduli of the surface. Here $p=\mathrm{e}^{2\pi i \sigma}$ is the corresponding fugacity. We set by convention $Z_{\mathrm{Sym}^0(X)}=1$.

In general, the genus $g$ partition function of a permutation orbifold with group $S_N$ on a genus $g$ surface can be written as follows:
\be 
\mathfrak{Z}_{\text{Sym}(X)}\left(\HH^2/\Gamma_g^\text{F}\right)=\exp\left(\sum_{\begin{subarray}{c}\text{subgroups $\mathrm{H}$ of $\Gamma_g^\text{F}$} \\
\text{up to conjugation} \end{subarray}} \frac{p^{[\Gamma_g^\text{F}:\mathrm{H}]}}{[\Gamma_g^\text{F}:\mathrm{H}]} Z_X\left( \HH^2/\mathrm{H}\right)\right)\ ,
\ee
where $Z_X$ is the partition function of the seed theory $X$. Here, we assumed that the theory $X$ is bosonic, see the discussion below for the case with fermions. 
\paragraph{Relation to covering maps.} More geometrically, the sum over subgroups of $\pi_1(\Sigma_g)$ up to conjugation can be viewed as a sum over all possible unbranched connected covering surfaces of the original theory. $[\Gamma_g^\text{F}:\mathrm{H}]$ is the degree of the corresponding covering map:
\be 
\mathfrak{Z}_{\text{Sym}(X)}(\Sigma_g)=\exp\left(\sum_{N_\text{c}=1}^\infty \frac{p^{N_\text{c}}}{N_\text{c}} \sum_{\begin{subarray}{c}\text{connected covering surfaces} \\ \widetilde{\Sigma}_{N_\text{c}(g-1)+1}\text{ of degree }N_\text{c} \end{subarray}} Z_X \left(\widetilde{\Sigma}_{N_\text{c}(g-1)+1}\right)\right)\ . \label{eq:symmetric orbifold partition function}
\ee
It does not matter whether the sum also extends over infinite coverings, because the denominator ensures that they do not contribute. The exponential generates also disconnected covering surfaces and inserts the correct combinatorial symmetry factors. Here and in the following we use $N_\text{c}$ to refer to the degree of connected covering maps and $N$ to refer to the degree of disconnected covering maps. In this form, the formula even holds true without specifying a constant curvature metric, because there is a natural metric on the covering surface that is the pull back of the metric on the base space along the covering map. Enumerating subgroups of a Fuchsian group $\Gamma_g^\text{F}$ systematically is quite difficult, but counting the number of terms is possible. We have collected some relevant facts in Appendix~\ref{app:subgroups}. There we also explained the role of the symmetric group in this construction.

\paragraph{Fermions.} Finally, let us consider the case where $X$ contains fermions and we want to compute the partition function of the symmetric orbifold on $\Sigma_g$ with a fixed spin structure. Such a spin structure on the base surface corresponds to a spin bundle $S$, whose pullback along the covering map induces a natural spin structure on the covering surface. This is the spin structure wich enters the right hand side of eq.~\eqref{eq:symmetric orbifold partition function}.

\section{Tensionless string theory on locally  \texorpdfstring{$\text{AdS}_3$}{AdS3} backgrounds} \label{sec:tensionless string on locally AdS3 backgrounds}
In this section, we set up the general framework to describe the tensionless string on Euclidean backgrounds of the form $\mathcal{M}_3 \times \mathrm{S}^3 \times \mathbb{T}^4$, where $\mathcal{M}_3$ is a hyperbolic 3-manifold, i.e.\ a space that is locally Euclidean $\AdS_3$. This section is partially a review and is based on \cite{Eberhardt:2018ouy, Dei:2020zui, Knighton:2020kuh}. We employ the hybrid formalism \cite{Berkovits:1999im} that continues to be well-defined in the tensionless limit. We start with global $\text{AdS}_3$, in which case the space has $\PSL(2,\mathbb{R})\times \PSL(2,\mathbb{R})$ symmetry (or in the Euclidean case $\PSL(2,\CC)$ symmetry). 
\subsection{The hybrid formalism}
The hybrid formalism starts with the following worldsheet theory:
\be 
\PSU(1,1|2)_k \oplus \text{top. twisted } \mathbb{T}^4 \oplus \text{ghosts}\ .
\ee
Here, $\PSU(1,1|2)_k$ is the WZW model on the supergroup $\PSU(1,1|2)$. The $\mathbb{T}^4$ sigma model has $\mathcal{N}=4$ supersymmetry and is topologically twisted and hence contributes $c=0$ towards the central charge. Finally, the ghost sector consists of the usual $bc$ ghosts of the bosonic sector and an additional $\rho$-ghost that replaces the $\beta\gamma$-ghost. It is a timelike free boson with screening charge and contributes $c=28$ towards the central charge. 

The parameter $k$ corresponds to the amount of NS-NS flux in the background and the tensionless string is obtained for $k=1$. We will make this choice in the remainder of the paper.

\paragraph{BRST cohomology.} The BRST operator is relatively complicated to write down in these variables, but we shall not have need of its explicit form. The hybrid string is formulated as an $\mathcal{N}=4$ topological string. Thus, there is a topologically twisted $\mathcal{N}=4$ algebra on the worldsheet whose supercharges we denote by
\begin{align}
\text{conformal weight 1:}&\qquad G^+\ , \quad \tilde{G}^+\ , \\
\text{conformal weight 2:}&\qquad G^-\ , \quad \tilde{G}^-\ ,
\end{align}
since as usual due to the topological twist, the conformal weights of the supercharges are shifted. This theory actually has \emph{two} BRST operators given by $G^+_0$ and $\tilde{G}^+_0$. The physical state subspace of the Hilbert space is identified with the double cohomology (similarly for the right-movers):
\be 
G^+_0 \ket{\Phi}=0 \ , \qquad \tilde{G}^+_0 \ket{\Phi}=0 \ , \qquad \ket{\Phi} \sim \ket{\Phi}+G^+_0\tilde{G}^+_0\ket{\Psi}\ .
\ee
Physical states are moreover the top components of spin $\frac{1}{2}$ multiplets of the R-symmetry of the algebra. This is necessary for the integrated vertex operator
\be 
\int G^{-} \bar{G}^{-} \Phi
\ee
to be an $\mathfrak{su}(2)$ singlet.
The second cohomology achieves the restriction of physical states to the small Hilbert space, which in the RNS formalism is often imposed by hand. 

\paragraph{Correlation functions.} Correlation functions are defined as in the $\mathcal{N}=4$ topological string \cite{Berkovits:1994vy, Berkovits:1999im}
\be 
\int_{\mathcal{M}_g} \left \langle \prod_{i=1}^{3g-3} |G^{-}(\mu_i)|^2 \left[ \int \tilde{G}^+ \bar{\tilde{G}}^+ \right]^{g-1} \int J\bar{J} \prod_{i=1}^n \int G^{-} \bar{G}^{-} \Phi_i \right \rangle \ .
\ee
Here $J$ is the Cartan-element of the R-symmetry $\SU(2)$ (that has conformal weight 1). $G^-(\mu)$ is the usual pairing between Beltrami differentials and conformal fields of weight 2.
For the $\mathcal{N}=4$ algebra, there are many inequivalent ways of doing the topological twist. They are related by rotating the supercharges. Effectively, one just replaces some of the $G^-$ by $\tilde{G}^-$ or equivalently, some of the $\tilde{G}^+$ by $G^+$.
It was explained in \cite{Dei:2020zui} that (for $k=1$) there is a unique choice up to equivalence that is non-vanishing. The amplitudes are defined by
\be 
\int_{\mathcal{M}_g} \left \langle \prod_{i=1}^{g-1} |G^{-}(\mu_i)|^2\prod_{i=g}^{3g-3} |\tilde{G}^{-}(\mu_i)|^2  \left[ \int \tilde{G}^+ \bar{\tilde{G}}^+ \right]^{g-1} \int J\bar{J} \prod_{i=1}^n \int \tilde{G}^{-} \bar{\tilde{G}}^{-} \Phi_i \right \rangle \ . \label{eq:N4 correlation function}
\ee
Compared to the previous formula, some supercharges $G^-$ have been replaced with $\tilde{G}^-$ (and similarly for the right-movers). The result is independent on the precise choice of these switches (up to normalization).

\subsection{Free field realization}\label{subsec:free field realization}
The formalism simplifies considerably for $k=1$, since there is a free field realization of $\PSU(1,1|2)_1$ in terms of symplectic bosons and free fermions \cite{Goddard:1987td, Eberhardt:2018ouy}. We follow the conventions of \cite{Dei:2020zui}. The free fields have the following defining (anti-)commutation relations:
\begin{align}
[\xi^\alpha_r,\eta^\beta_s]=\varepsilon^{\alpha\beta} \delta_{r+s,0}\ , \qquad \{\psi^\alpha_r,\chi^\beta_s\}=\varepsilon^{\alpha\beta} \delta_{r+s,0}\ .
\end{align}
Here, $\xi^\alpha$ and $\eta^\beta$ are spin-$\frac{1}{2}$ symplectic bosons and $\psi^\alpha$ and $\chi^\beta$ are usual spin-$\frac{1}{2}$ fermions. $\alpha$, $\beta$ take values in $\{+,-\}$ and are $\SU(2)$ spinor indices. These free fields generate the superalgebra $\mathfrak{u}(1,1|2)_1$. The generators of this superalgebra are identified with the bilinears
\begin{subequations}
\begin{align}
J^3_m&=-\tfrac{1}{2} (\eta^+\xi^-)_m-\tfrac{1}{2} (\eta^-\xi^+)_m\ , & K^3_m&=-\tfrac{1}{2} (\chi^+\psi^-)_m-\tfrac{1}{2} (\chi^-\psi^+)_m\ , \\
J^\pm_m&=(\eta^\pm\xi^\pm)_m\ , & K^\pm_m&=\pm(\chi^\pm\psi^\pm)_m\ , \\
S_m^{\alpha\beta+}&=(\chi^\beta \xi^\alpha)_m\ , & S_m^{\alpha\beta-}&=-(\eta^\alpha\psi^\beta)_m \ ,  \\
U_m&=-\tfrac{1}{2} (\eta^+\xi^-)_m+\tfrac{1}{2} (\eta^-\xi^+)_m\ , & V_m&=-\tfrac{1}{2} (\chi^+\psi^-)_m+\tfrac{1}{2} (\chi^-\psi^+)_m\ .
\end{align}\label{eq:u112-algebra}
\end{subequations}
It is also convenient to define the combinations $Z=U+V$ and $Y=U-V$. 
The superalgebra $\mathfrak{u}(1,1|2)_1$ contains the algebra $\mathfrak{su}(1,1|2)_1$ as a subalgebra that consists of all the currents except for $Y$. $\mathfrak{su}(1,1|2)_1$ in turn is obtained as a central extension of $\mathfrak{psu}(1,1|2)_1$, the central generator being $Z$.
Thus, we essentially need to set $Z=0$ in order to recover $\mathfrak{psu}(1,1|2)_1$ from $\mathfrak{u}(1,1|2)_1$. This is done by introducing the following BRST operator:
\be 
\mathcal{Q}=c Z\ , \label{eq:U1 BRST operator}
\ee
where we introduce a new $bc$ ghost system with $h(b)=1$ and $h(c)=0$. Note that $[\mathcal{Q}, Y(z)]=\partial c(z)$. Thus $\mathcal{Q}$ has the effect of removing both $Y$ and $Z$ from the free field realization. Fortunately the $Z$--$Z$ OPE does not have a central term and hence we have indeed $\mathcal{Q}^2=0$. 

\paragraph{Representations.} Representations of $\mathfrak{psu}(1,1|2)_1$ are straightforward to describe in the free field representation. We will be brief here, for more details see \cite{Eberhardt:2018ouy, Dei:2020zui}. Unflowed representations are are identified with the R-sector representation of the free fields.\footnote{The spin structures of the symplectic bosons and the fermions are coupled since the supercharges $S^{\alpha\beta\gamma}$ have to be single-valued.} Positive modes of the free fields annihilate the primary state $\ket{m_1,m_2}$. Zero modes of the symplectic bosons act as
\begin{align} 
\xi_0^+ \ket{m_1,m_2}&=\ket{m_1,m_2+\tfrac{1}{2}}\ , & \eta_0^+ \ket{m_1,m_2}&=2 m_1 \ket{m_1+\tfrac{1}{2},m_2}\ , \\
\xi_0^- \ket{m_1,m_2}&=-\ket{m_1-\tfrac{1}{2},m_2}\ , & \eta_0^- \ket{m_1,m_2}&=-2 m_2 \ket{m_1,m_2-\tfrac{1}{2}}\ .
\end{align}
The $\mathfrak{sl}(2,\mathbb{R})$ spin of the representation is given by $j=m_1-m_2$.
On top of this, the fermions span a $2^2=4$-dimensional Clifford module. We define
\be 
\chi_0^+\ket{m_1,m_2}=\psi_0^+ \ket{m_1,m_2}=0
\ee
so that the four states are obtained by action of $\chi^-_0$ and $\psi_0^-$. The four states consist of two R-symmetry $\mathfrak{su}(2)$ singlets and an $\mathfrak{su}(2)$ doublet.
Evaluating the zero mode $Z_0$ on the states leads to
\be 
Z_0=\begin{cases}
m_1-m_2-\tfrac{1}{2}\ , &\text{$\mathfrak{su}(2)$ doublet}\ , \\
m_1-m_2-\frac{1}{2}\pm \tfrac{1}{2} \ , &\text{$\mathfrak{su}(2)$ singlets}\ .
\end{cases}
\ee
Thus, the only possible BRST invariant representation of $\mathfrak{psu}(1,1|2)_1$ has the bosonic subrepresentations $(j=0,\mathbf{1}) \oplus (j=1,\mathbf{1}) \oplus (j=\tfrac{1}{2},\mathbf{2})$ \cite{Eberhardt:2018ouy}.

\paragraph{Spectral flow.} More representations are obtained from acting with spectral flow on the symplectic bosons. There are two independent spectral flow symmetries that we denote by $\sigma^{(+)}$ and $\sigma^{(-)}$.
\begin{subequations}
\begin{align}
\sigma^{(\pm)}(\xi^\mp_r)&=\xi^\mp_{r \pm \frac{1}{2}}\ , & \sigma^{(\pm)}(\eta^\pm_r)&=\eta^\pm_{r \mp \frac{1}{2}}\ , \\
\sigma^{(\pm)}(\psi^\mp_r)&=\psi^\mp_{r \mp \frac{1}{2}}\ , & \sigma^{(\pm)}(\chi^\pm_r)&=\chi^\pm_{r \pm \frac{1}{2}}\ .
\end{align}
\end{subequations}
Unspecified actions are trivial. The composition $\sigma =\sigma^{(+)}\circ \sigma^{(-)}$ is the usual spectral flow action on $\mathfrak{psu}(1,1|2)_1$; it leaves the generators $Y$ and $Z$ invariant. There is also the opposite composition $\hat{\sigma}=\sigma^{(+)} \circ (\sigma^{(-)})^{-1}$. The existence of these spectral flow automorphism allows one to define spectrally flowed representations that are obtained by composing the unflowed representations with the spectral flow automorphism. Focussing on the the spectral flow automorphism $\sigma$, this allows one to define the corresponding vertex operators\footnote{We suppress here again the right-movers, the vertex operator would be properly denoted by $V^w_{j,h,\bar{h}}(x,\bar{x},z,\bar{z})$.} \footnote{These vertex operators are affine primary (in a spectrally flowed sense) with respect to the $\sl(2,\RR)_1$ subalgebra. It will not be important for us what the precise properties of the vertex operators are with respect to the $\su(2)_1$ subalgebra and we may take them to be affine primary (in the unflowed sense). } 
\be 
V^w_{j,h}(x,z)\ .
\ee
Here, $w \in \mathbb{Z}_{\ge 1}$ is the amount of spectral flow.\footnote{We can restrict to $w \ge 1$, since only these fields correspond to fields in spacetime.} We traded the labels $m_1$ and $m_2$ for the more physical labels of $\mathfrak{sl}(2,\mathbb{R})$ spin $j$ and spacetime conformal weight $h$.
The additional label $x$ corresponds to the position of the vertex operator in spacetime. We take $x$ to be complex (and the corresponding variable for the right-movers is the complex conjugate), which specifies the reality of the theory and implies that we are working in Euclidean $\text{AdS}_3$.

Due to the existence of the second spectral flow operator $\hat{\sigma}$, the same $\mathfrak{psu}(1,1|2)_1$ representation actually appears multiple times in the free field representation. Indeed, by spectrally flowing with respect to $\hat{\sigma}$, one replicates the same representation of $\mathfrak{psu}(1,1|2)_1$. In particular, we can look at the spectrally flowed images of the vacuum representation. This leads to copies of the vacuum representation of $\mathfrak{psu}(1,1|2)_1$, but they are \emph{not} in the vacuum w.r.t.~the free fields. In particular, we shall make use of the field $W(u)$ that is obtained by twice flowing the vacuum representation w.r.t.\ $\hat{\sigma}$ \cite{Dei:2020zui}. Focussing on the bosons, it has defining OPEs
\be 
\xi^\alpha(z) W(u) \sim \mathcal{O}(z-u)\ , \qquad \eta^\alpha(z) W(u) \sim \mathcal{O}((z-u)^{-1})\ .
\ee
 \paragraph{String correlators.}
Let us now discuss how the prescription for correlators in the hybrid formalism \eqref{eq:N4 correlation function} combines with the free field realization. It was discovered in \cite{Dei:2020zui}, that the replacement $G^- \to \tilde{G}^-$  effectively shifts the $\mathfrak{sl}(2,\mathbb{R})$ spin $j$ of the vertex operators down by 1 unit. It does not matter how we distribute these shifts on the vertex operators. This potentially violates the conservation of the $Y$-current of the free field realization. In order to get a non-zero result for the correlator one has to insert $n+2g-2$ additional $W$-fields. Thus, one is lead to study the correlators
\be 
\left \langle \prod_{\alpha=1}^{2g-2} W(u_\alpha) \prod_{i=1}^n V_{j_i,h_i}^{w_i} (x_i,z_i) \right \rangle
\ee
in the $\text{PSU}(1,1|2)_1$ WZW model. 
Conservation of the $\U(1)$ current $Y$ imposes the condition
\be 
\sum_i j_i=2-2g-\frac{n}{2}\ . \label{eq:j constraint}
\ee
These correlators are the main object of study in the hybrid formalism. Even though we are mainly interested in studying the partition functions, we see that we have to insert at least one vertex operator satisfy this constraint (except for $g=1$). Thus, we will keep most parts of this paper general and they also apply to correlation functions.

Until now, we have not properly implemented the gauging of the $\U(1)$-current. Besides introducing a pair of $bc$-ghosts with BRST operator \eqref{eq:U1 BRST operator}, we also have to integrate over the moduli space of flat $\U(1)$-bundles over the Riemann surface, as indicated by the presence of $g$ zero modes of $b$. The moduli space of line bundles on the Riemann surface is the Jacobian, defined by
\be 
\Jac(\Sigma_g)=\mathbb{C}^g/(\mathbb{Z}^g \oplus \boldsymbol{\Omega} \mathbb{Z}^g)\ ,
\ee
where $\boldsymbol{\Omega}$ is the period matrix. We have collected some background on Riemann surfaces in Appendix~\ref{app:Riemann surfaces}.
 The isomorphism between the moduli space of line bundles and the Jacobian is given by the Abel Jacobi map. 
 
 The prescription of the correlation functions for the hybrid string \eqref{eq:N4 correlation function} already incorporates $g$ additional integrals besides the moduli space integrals. We can interpret these integrals as an integral over the Jacobian. 
Of these $g$ integrals, $g-1$ integrals are charged under the $\U(1)$-current of the compactification manifold. This is exactly as it has to be: because the $\mathbb{T}^4$ sigma-model is topologically twisted, its charge conservation is anomalous. Thus to get a non-trivial partition function we have to insert charged operators in the correlator. This is precisely achieved by the description \eqref{eq:N4 correlation function}.
\subsection{Orbifolds} \label{subsec:orbifolds}
We now want to reduce this theory from global Euclidean $\text{AdS}_3$ to a quotient space $\text{AdS}_3 /\Gamma$ for a discrete group $\Gamma \subset \PSL(2,\CC)$. We do not necessarily require $\Gamma$ to act properly discontinuously, which means that the quotient space can have orbifold singularities. Consistency of the theory seems to require the deficit angles to have the form $2\pi(1-M^{-1})$ for $M \in \ZZ_{\ge 2}$.
Groups $\Gamma$ that act properly discontinuously are known as Kleinian groups and we collected some background material in Appendix~\ref{app:hyperbolic 3 manifolds}. 

\paragraph{Spin structure.} There is a small subtlety here since we are considering a supersymmetric theory. We also want to specify a spin structure on the target manifold. This is essentially achieved by lifting $\Gamma \subset \PSL(2,\CC)$ to a subgroup of $\SL(2,\CC)$. The different liftings correspond to the different spin structures. In the following, we consider the spin structure part of the spacetime geometry and hence always work with the lift $\Gamma \subset \SL(2,\CC)$. This lift is discussed further in the context of uniformizations in Appendix~\ref{app:uniformization}.

\paragraph{Action on worldsheet fields.} Being a subgroup of $\SL(2,\mathbb{C})$, the action of $\Gamma$ on the various fields in the worldsheet theory is described by the action of the generators $J_0^a$. Thus, we are essentially taking a usual orbifold in CFT. In particular, $\Gamma$ does not act on the fields $W(u_\alpha)$. $\Gamma$ does however act on the vertex operators $V_{j_i,h_i}^{w_i} (x_i,z_i)$ and correspondingly, one can define twisted vertex operators. They are not needed to discuss partition functions. We will only consider untwisted vertex operators in this paper. 
The partition function is obtained in the special case when we chose $w_i=1$ and $h_i=0$ for all vertex operators. In this case the vertex operators correpond to the vacuum of the dual CFT and are hence actually independent of $x$. Note that it is not possible to choose $n=0$, since it would be impossible to satisfy the constraint \eqref{eq:j constraint} (except for $g=1$).

\paragraph{Twisted sectors.} When computing the orbifold correlators, we are effectively performing a gauging by a discrete subgroup $\Gamma \subset \SL(2,\mathbb{C})$. 
This gauging is achieved by summing over all principal $\Gamma$-bundles over the Riemann surface $\Sigma$. We can write
\be 
\left \langle \cdots \right \rangle=\frac{1}{|\Gamma|}\sum_{\text{$\Gamma$-bundles } \rho_\Gamma} \left \langle \cdots \right \rangle_{\rho_\Gamma}\ ,
\ee 
where in $\left \langle \cdots \right \rangle_{\rho_\Gamma}$ all fields have twisted boundary conditions. $\Gamma$-bundles are specified by a homomorphism
\be 
\rho_\Gamma: \pi_1(\Sigma_g) \longrightarrow \Gamma\ ,
\ee
that tells us how the fields are twisted when we move around the cycles of $\Sigma_g$ and hence we simply identify $\rho_\Gamma$ with the bundle.
$\pi_1(\Sigma_g)$ depends of course on a choice of base point. Changing the base point results in an overall conjugation of the homomorphism and specifies the same bundle. Thus $\Gamma$-bundles are in 1-to-1 correspondence with the set $\Hom(\pi_1(\Sigma_g,z_0),\Gamma)/\Gamma$, which we will just write by $\Hom(\pi_1(\Sigma_g),\Gamma)$.  In the following all homomorphisms from $\pi_1(\Sigma_g)$ are understood up to overall conjugation. This is the analogue of the more well-known situation at genus 1, where one has to sum over all twisted boundary conditions. Since $\pi_1(\mathbb{T}^2)=\mathbb{Z}^2$, we have
\be 
\Hom(\pi_1(\mathbb{T}^2),\Gamma)\cong \langle a,b \in \Gamma\,|\, ab=ba \,, \ \ (a,b) \sim (g a g^{-1},g b g^{-1})\rangle\ ,
\ee
where the group elements $a$ and $b$ are the images under the homomorphism $\rho_\Gamma$ of the $\alpha$- and $\beta$-cycle of the torus. 

Recall that we also perform a $\U(1)$-gauging via the BRST operator $\mathcal{Q}$ \eqref{eq:U1 BRST operator} to reduce the free theory to $\PSU(1,1|2)_1$ and correspondingly integrate over all flat $\U(1)$-bundles that can also be specified by a homomorphism $\rho_{\U(1)}:\pi_1(\Sigma_g) \longrightarrow \U(1)$. Actually, we do not need any reality and consider the homomorphism into the complexified group $\rho_{\CC^\times}:\pi_1(\Sigma_g) \longrightarrow \CC^\times$ consisting of all complex numbers excluding zero.\footnote{There is more freedom in the abelian case and several homomorphisms can correspond to the same bundle. See Appendix~\ref{subapp:classification of line bundles}.} It is convenient to combine these two homomorphisms as follows
\be 
\rho\equiv\rho_\Gamma \otimes \rho_{\CC^\times}:\pi_1(\Sigma_g) \longrightarrow \Gamma \times \CC^\times \subset \SL(2,\CC) \times \CC^\times\ .
\ee
\paragraph{Normalization.} We should also mention that the orbifold has another effect -- it multiplies the partition function by $|\Gamma|^{-1}$. Since the orbifold group is infinite, this factor is naively zero. However, we expect that upon regularization, we can get a non-zero value. Since we will in any case not be able to determine the precise value of the partition function (or correlation functions), we will not discuss this factor further. For simple geometries it can be calculated explicitly \cite{Eberhardt:2020bgq}.

\subsection{Summary}
Let us summarize the most important features of the formalism. The form of the string partition function/correlation function \eqref{eq:N4 correlation function} is basically dictated by demanding charge conservation. The $g$ additional integrals can be interpreted as integrals over the Jacobian and implement precisely the gauging that reduces the free fields to $\mathfrak{psu}(1,1|2)_1$.

The natural prescription for the string correlation function in the free-field realization takes the following schematic form:
\be 
\sum_{\text{$\Gamma$-bundles }\rho_\Gamma}\int_{\mathcal{M}_{g,n}} \int_{\Jac(\Sigma_g)} \left \langle  \prod_{\alpha=1}^{2g-2+n} W(u_\alpha) \partial H(z) \prod_{\beta=1}^{g-1} \mathrm{e}^{-i H}(v_\beta) \prod_{i=1}^n V_{j_i,h_i}^{w_i}(x_,z_i)\right \rangle_{\rho_\Gamma\otimes \rho_{\CC^\times}}\ . \label{eq:string correlator prescription}
\ee
Here, we have suppressed insertions of the ghosts $\sigma$ and $\rho$ (but they are needed to obtain a non-vanishing correlator). We have also suppressed right-movers. The insertion of the fields $W(u_\alpha)$ is necessary in the free-field realization. $\partial H$ is the R-symmetry $\U(1)$ current of the internal CFT on $\TT^4$. Since the internal CFT is topologically twisted, the $\U(1)$ charge conservation is anomalous which necessitates the inclusion of these terms. We also have inserted the $\partial H$-current, which comes from the $J$-current in \eqref{eq:N4 correlation function} and is needed for a non-vanishing correlator, see Appendix~\ref{app:topologically twisted partition function}. We have not been very precise about these terms. Since we will not be able to compute them fully, we only need the qualitative structure. The summation over $\Gamma$-bundles reduces the theory from global $\text{AdS}_3$ to $\AdS_3/\Gamma$.
The integrand does depend on the locations of $u_\alpha$ and $v_\beta$, but does so in a trivial free way. The spins $j_i$ satisfy the following constraint:
\be 
\sum_i j_i=2-2g-\frac{n}{2}\ .
\ee
In the following, we demonstrate that the integrand localizes in the total space $\mathcal{M}_{g,n} \times \Jac(\Sigma_g)$.\footnote{Of course, this space is not a direct product, but we continue to denote it like this for notational simplicity.} Thus, the correlation functions reduce to a discrete sum instead of integrals.

\section{Localization}\label{sec:localization}
We now show that the worldsheet partition function localizes on covering surfaces of the boundary. Our argument generalizes the argument of \cite{Eberhardt:2019ywk, Eberhardt:2020akk, Dei:2020zui, Knighton:2020kuh} and proceeds in several steps. Readers only interested in the result may skip to Section~\ref{subsec:worldsheet partition function}.
\subsection{The argument} \label{subsec:localization argument}
\paragraph{Insertion of $\xi^\pm$.} 
The strategy is to consider the expressions
\be 
\lambda^\pm(z)=\left\langle \xi^\pm(z) \prod_{\alpha=1}^{2g-2+n} W(u_\alpha) \prod_{i=1}^n V_{j_i,h_i}^{w_i}(x_i,z_i) \right \rangle_\rho\ .
\ee
The subscript $\rho$ means that we compute the correlation function with twisted boundary conditions that are specified by the homomorphism
\be 
\rho=\rho_\Gamma \otimes \rho_{\CC^\times}: \pi_1(\Sigma_g) \longrightarrow \Gamma \times \CC^\times \subset \SL(2,\mathbb{C}) \times \CC^\times \ ,
\ee
as described in Section~\ref{subsec:orbifolds}.
We stress that $\SL(2,\mathbb{C}) \times \CC^\times$ (and not $\GL(2,\mathbb{C})$) is the correct group once we remember that there are also the free fermions $\psi^\pm$ (and $\chi^\pm$) that have twisted boundary conditions in the path integral, see the free field realization \eqref{eq:u112-algebra}. Those twisted boundary conditions are specified by the homomorphism $\rho_{\CC^\times}$. 

It is sometimes useful to combine $\boldsymbol{\lambda}=(-\lambda^-,\lambda^+)$, which takes values in the two-dimensional holomorphic vectorbundle $S \otimes E_\rho$, where $S$ is a fixed spin structure and $E_\rho$ is the flat bundle determined by $\rho$.
For a fixed spin structure $S$ on the worldsheet, any other spin structure can be obtained by tensoring with a $\mathbb{Z}_2$-bundle. 
 Viewing $\mathbb{Z}_2 \subset \CC^\times$ as a subgroup, we can combine the necessary sum over spin structures with the sum (or integral) over non-trivial $\U(1)$-bundles. We will see below that there is a natural spin structure $S$, but for now we keep it arbitrary.
To summarize, let us collect the properties of $\lambda^\pm(z)$:
\begin{enumerate}
\item $\boldsymbol{\lambda}(z)$ is a section of the holomorphic vectorbundle $S \otimes E_\rho$.
\item Both components of $\boldsymbol{\lambda}(z)$ have single zeros at $z=u_\alpha$.
\item $\boldsymbol{\lambda}(z)$ has only poles near $z_i$. More precisely, the behaviour is
\begin{subequations}
\begin{align} 
\lambda^+(z)&=\mathcal{O}\left((z-z_i)^{-\frac{w_i+1}{2}}\right)\ , \\
(\lambda^-(z)+x_i \lambda^+(z))&=\mathcal{O}\left((z-z_i)^{\frac{w_i+1}{2}}\right)\ .
\end{align}
\end{subequations}
This follows from translating the representations described in Section~\ref{subsec:free field realization} into OPEs. This is done explicitly in \cite{Dei:2020zui}.
\end{enumerate}

The existence of such a holomorphic section is extremely constraining. Of course, we could simply have identically $\boldsymbol{\lambda}\equiv 0$, but this also implies the vanishing of the full partition function. This follows by considering the OPE limit $z \to z_i$. The leading term in the OPE is
\be 
\xi^+(z) V_{j_i,h_i}^{w_i}(x_,z_i)=(z-z_i)^{-\frac{w_i+1}{2}} V_{j_i-\frac{1}{2},h_i+\frac{1}{2}}^{w_i}(x_,z_i)
\ee
and thus the leading term in the singularity of $\lambda^+(z)$ captures the correlation function of the primaries itself. Vanishing of $\boldsymbol{\lambda}$ would hence imply vanishing of the full correlator.
So let us assume that $\boldsymbol{\lambda} \not\equiv 0$ and see what it implies. 

\paragraph{Construction of a meromorphic 1-form.}
Given this data, we can construct a meromorphic 1-form $\omega(z)$ with twisted $\U(1)$ boundary conditions as follows. More precisely, $\omega$ is a meromorphic section of $K \otimes L_{\rho_{\CC^\times}}$, where $K$ is the canonical bundle and $L_{\rho_{\CC^\times}}$ the flat line bundle determined by the restriction of $\rho$ to the $\CC^\times$ subgroup. The poles of $\omega(z)$ are precisely given by the insertion points $z=z_i$ (and are single poles) and $\omega(z)$ has single zeros at all the $z=u_\alpha$ (and no other zeros).
We set
\be 
\omega(z)=\sqrt{\lambda^-(z)\partial \lambda^+(z)-\lambda^+(z) \partial \lambda^-(z)}\ .
\ee
To show that this is indeed a well-defined 1-form, we check the following properties:
\begin{enumerate}
\item $\lambda^-(z)\partial \lambda^+(z)-\lambda^+(z) \partial \lambda^-(z)$ is a meromorphic section of the line bundle $K^2 \otimes L_{\rho_{\CC^\times}}^2$. For this, one simply has to check that the $\SL(2,\mathbb{C})$-part of the homomorphism $\rho$ cancels out in this combination. Moreover, even though we have not used a covariant derivative, this expression transforms covariantly.
\item $\lambda^-(z)\partial \lambda^+(z)-\lambda^+(z) \partial \lambda^-(z)$ has (at most) double poles at $z=z_i$ and no other poles. 
\item $\lambda^-(z)\partial \lambda^+(z)-\lambda^+(z) \partial \lambda^-(z)$ has (at least) double zeros at $z=u_\alpha$.
\item $\lambda^-(z)\partial \lambda^+(z)-\lambda^+(z) \partial \lambda^-(z)$ has precisely these zeros and poles with all double multiplicity. The line bundle $K^2 \otimes L_{\rho_{\CC^\times}}^2$ has degree $2\deg(K)=4g-4$ ($L_{\rho_{\CC^\times}}$ is flat and hence does not contribute to the degree). The number of zeros minus the number of poles of any meromophic section of this bundle is hence $4g-4$ (counted with multiplicity). We found a maximal list of poles and a minimal list of zeros for $\lambda^-(z)\partial \lambda^+(z)-\lambda^+(z) \partial \lambda^-(z)$ and this argument shows that this list is complete.
\item $\lambda^-(z)\partial \lambda^+(z)-\lambda^+(z) \partial \lambda^-(z)$ possesses a well-defined square root. This follows from the fact that all its zeros and poles are second order and thus taking the square root is well-defined up to an overall sign.
\end{enumerate}
This shows all the desired properties. There is a small caveat: the square root is only guaranteed to be a section of $K \otimes L_{\rho_{\CC^\times}} \otimes L_{\ZZ_2}$, where $L_{\ZZ_2}$ is a $\ZZ_2$-bundle that squares to the trivial line bundle. We will resolve this problem below and show that $L_{\ZZ_2}$ is absent when choosing a suitable spin structure.

$\omega(z)$ is a simple quantity, since we can apply the technology of line bundles and divisors to it. Note first that the existence of such a meromorphic section $\omega(z)$ is very constraining. It is a meromorphic differential with $n$ poles, but $2g-2+n$ prescribed zeros. By the Riemann-Roch theorem, this is generically impossible and thus non-vanishing of $\omega(z)$ imposes a non-trivial constraint.
We can quantify this very precisely. Fix any meromorphic 1-form $\tilde{\omega} \in K \otimes L_{\rho_{\CC^\times}} (\otimes L_{\ZZ_2})$. We suppress the $\ZZ_2$-factor in the following. Then the ratio $\omega(z)/\tilde{\omega}(z)$ is a meromorphic function on the Riemann surface. Correspondingly, it's divisor is principal. 
 The divisor of $\omega(z)$ is
\be 
D=-\sum_{i=1}^n z_i+\sum_{\alpha=1}^{2g-2+n} u_\alpha\ .
\ee
Thus, we need that the divisor $D-K-D(L_{\rho_{\CC^\times}})$ to be principal. By the Abel-Jacobi theorem, this is equivalent to the statement that the image under the Abel-Jacobi map vanishes. This specifies the line bundle $L_{\rho_{\CC^\times}}$ uniquely and shows that there is \emph{exactly one} line bundle for which $\omega(z)$ can be non-vanishing. The Abel-Jacobi map can be evaluated more explicitly for  $L_{\rho_{\CC^\times}}$, which is explained in Appendix~\ref{subapp:classification of line bundles}.
\paragraph{Constructing $\gamma(z)$.}
As a next step, we construct a map $\gamma(z)$ that will turn out to be a branched covering map from the worldsheet to the boundary. We define it as
\be 
\gamma(z)\equiv-\frac{\lambda^-(z)}{\lambda^+(z)}\ .
\ee
$\gamma(z)$ has again a number of properties that are straightforward to check:
\begin{enumerate}
\item $\gamma(z)$ is a (multi-valued) function on the Riemann surface $\Sigma_g$. $\pi_1(\Sigma_g)$ acts on it by M\"obius transformations.
\item $\partial \gamma(z)\ne 0$ and $\partial (\gamma(z)^{-1})\ne 0$ for all $z\ne z_i$. This follows from 
\be 
\partial \gamma(z)=\frac{\lambda^+(z)\partial \lambda^-(z)-\lambda^-(z) \partial \lambda^+(z)}{\lambda^+(z)^2}=\frac{\omega(z)^2}{\lambda^+(z)^2}\ .
\ee
Zeros of $\partial \gamma(z)$ originate either from zeros of $\omega(z)$ or from poles of $\lambda^+(z)$. In both cases, they cancel out by our previous analysis. The argument for $\gamma(z)^{-1}$ is analogous.
\item $\gamma(z)=x_i+\mathcal{O}((z-z_i)^{w_i})$ for $i=1,\dots,n$.
\end{enumerate}
The second property means that $\gamma(z)$ maps into $\mathbb{CP}^1$ and is branched over the $z_i$.
To state the first point more clearly, let us uniformize the worldsheet Riemann surface using a Fuchsian uniformization $\Sigma_g=\HH^2/\Gamma_g^\text{F}$.\footnote{For worldsheet genus 1, the uniformization is $\mathbb{C}/(\mathbb{Z} \times \mathbb{Z})$, but all the following arguments are unchanged.} \footnote{In order to be consistent with our notation, we denote by $\Gamma_g^\text{F} \subset \PSL(2,\RR)$ the genus $g$ Fuchsian uniformization group and by $\Gamma \subset \SL(2,\CC)$ the orbifold group.} We have then $\pi_1(\Sigma_g)\cong \Gamma_g^\text{F}$. $\gamma(z)$  can be viewed as a single-valued map from the upper half plane $\HH^2$ to $\CP^1$. It is an equivariant map in the following sense:
\be 
\gamma(g(z))=\rho_\Gamma(g)(\gamma(z))
\ee
for $z \in \HH^2$ and $g \in \pi_1(\Sigma_g)$. Here $\rho_\Gamma(g)\in \SL(2,\CC)$ acts on $\gamma(z)$ via M\"obius transformations. Such a map is known as a (branched) \emph{developing map} on the Riemann surface.\footnote{Other names are \emph{deformation} or \emph{geometric realization}.} It defines a branched complex projective structure on the surface $\Sigma_g$. We have collected some facts about (branched) complex projective structures in Appendix~\ref{app:branched complex projective structures}.
\paragraph{Reconstructing $\lambda^\pm(z)$.}
Using $(\lambda^+(z),\lambda^-(z))$ we have constructed the two quantities $\omega(z)$ and $\gamma(z)$. The two are actually equivalent to the original data, since we can recover 
\be 
\boldsymbol{\lambda}(z)=\frac{\omega(z)}{\sqrt{\partial \gamma(z)}} \begin{pmatrix}
\gamma(z) \\ 1
\end{pmatrix}\ . \label{eq:lambda reconstructed}
\ee
The square root of $\partial \gamma(z)$ is well-defined, because $\partial \gamma(z)$ has no zeros and all poles are double poles away from $z=z_i$. The quantities $\omega(z)$ and $\gamma(z)$ are far more convenient, since they separate the $\CC^\times$-part and the $\SL(2,\mathbb{C})$-part of the problem. Thus, we will continue to work with them without losing any information. 

\paragraph{Spin structure.}
We have been slightly cavalier with the square root, they could in principle introduce signs in both $\omega(z)$ and $\sqrt{\partial \gamma(z)}$ around the cycles of the Riemann surface. This does not happen, provided that we choose the correct spin structure on the worldsheet. Let us see this in more detail. $\partial \gamma(z)$ satisfies
\be 
\partial g(z)(\partial\gamma)(g(z))=(\partial\rho(g))(\gamma(z))\partial\gamma(z)\ ,
\ee
for $g \in \Gamma$, where we view group elements sometimes as M\"obius transformations.
We are trying to define a square root of this transformation behaviour. We know how to take the square root of $(\partial\rho(g))(\gamma(z))$ -- this is dictated by the lift of $\Gamma$ from $\PSL(2,\CC)$ to $\SL(2,\CC)$. For
\be 
\rho(g)(z)=\frac{a z+b}{c z+d}\ , \qquad (\partial \rho(g))(z)=\frac{1}{(cz+d)^2}\ ,
\ee
we can \emph{define} the square root to be 
\be 
 \sqrt{(\partial \rho(g))(z)}\equiv \frac{1}{cz+d}\ .
\ee
To make sense of $\sqrt{\partial \gamma(z)}$, we also need to define a square root $\sqrt{\partial g(z)}$. This is analogous to the above situation: $\sqrt{\partial g(z)}$ is not well-defined for $\Gamma_g^\text{F} \subset \PSL(2,\RR)$, but only once we lift it to $\widetilde{\Gamma}_g^\text{F} \subset \SL(2,\RR)$, which defines a spin structure on the worldsheet. This is explained also in Appendix~\ref{subapp:Fuchsian uniformization}. We thus conclude that every such map $\gamma(z)$ naturally induces a spin structure on the worldsheet. With these definitions,
\be 
\frac{1}{\sqrt{\partial \gamma(z)}} \begin{pmatrix}
\gamma(z) \\ 1
\end{pmatrix}
\ee
is by construction a section of $S^{-1} \otimes E_\rho$ and hence $\omega$ is a section of $K \otimes L_{\rho_{\CC^\times}}$, thus eleminating the possibility of an additional $\ZZ_2$ bundle that could appear in the square root.
In the beginning of this Section, we fixed a spin structure $S$. We now see that this is not arbitrary and we should identify $S$ with the induced spin structure so that the formula \eqref{eq:lambda reconstructed} becomes correct.
\paragraph{Localization in moduli space of Riemann surfaces.}
We again note that the existence of such a map $\gamma(z)$ is extremely constraining. Viewing $\gamma(z)$ as a map on the upper half-plane, we can look at its Schwarzian
\be 
S(\gamma)(z)=\frac{\partial^3 \gamma(z)}{\partial \gamma(z)}-\frac{3 (\partial^2 \gamma(z))^2}{2(\partial \gamma(z))^2}\ ,
\ee
which defines a meromorphic quadratic differential on the Riemann surface. It is holomorphic away from $z=z_i$ because $\partial \gamma(z)\ne 0$. Near $z=z_i$, it has a double pole 
\be 
S(\gamma)(z)=-\frac{w^2_i-1}{2(z-z_i)^2}+\mathcal{O}((z-z_i)^{-1})\ . \label{eq:Schwarzian differential OPE}
\ee
 It is also periodic around the cycles of the Riemann surface because the Schwarzian derivative is invariant under M\"obius transformations. Thus $S(\gamma)(z)$ is indeed a meromorphic section of $K^2$.

From here, we already see that there will be a further localization in the parameters of the problem. Let us first discuss the case  $w_i=1$ for all vertex operators, where the Schwarzian is a holomorphic quadratic differential. The homomorphism $\rho_{\Gamma}$ depends on $6g-3$ complex parameters (for $g \ge 2$), corresponding to the choice of the matrices $\rho(\alpha_1),\dots,\rho(\alpha_g)$, $\rho(\beta_1),\dots,\rho(\beta_g)$. The $-3$ comes from the fact
that the matrices have to obey a single constraint:
\be 
\prod_{I=1}^g [\rho(\alpha_I),\rho(\beta_I)]=1\ .
\ee
Another 3 parameters are redundant, because they correspond to overall conjugation. Thus, the representation variety $\Hom(\pi_1(\Sigma_g),\SL(2,\mathbb{C}))$ (up to overall conjugation) has complex dimension $6g-6$. Most  $\rho_\Gamma$'s will not be associated to a map $\gamma(z)$. Such maps are in 1-1 correspondence with quadratic differentials $S(\Gamma)(z)$ (again up to overall composition with a M\"obius transformation). The (complex) dimension of the space of quadratic differentials is only $3g-3$. Thus, we conclude that there are $3g-3$ constraints that have to be obeyed in order for such a map to exist. This shows that when $\rho_\Gamma$ is fixed, there are only discrete points in the moduli space of Riemann surfaces for which $\lambda^\pm(z)$ can be non-zero. If we also want to integrate over the positions of the vertex operators, then there are additional constraints because we also want to require $\gamma(z_i)=x_i$. This leads to $n$ further constraints and thus the string integrand localizes in $\mathcal{M}_{g,n}$.

In the case with poles, one has to be careful, because the correspondence between meromorphic quadratic differentials that satisfy \eqref{eq:Schwarzian differential OPE} and developing maps $\gamma$ is no longer 1--1. Instead, the quadratic differentials have to satisfy an extra condition, that is called integrability \cite{Hehjal}. This condition ensures that $\gamma$ has trivial monodromy around the insertion points.\footnote{If we would compute correlation functions with twisted vertex operators, then we would specify non-trivial monodromy around the insertion points.}
This imposes $n$ conditions on the quadratic differential. Another $n$ conditions are imposed by requiring that the solution $\gamma(z)$ satisfies $\gamma(z_i)=x_i$.\footnote{Of course these are actually $n-3$ constraints because M\"obius transformations of the $x_i$ are invisible for the Schwarzian. We already took M\"obius transformations into account and in order to have a uniform presentation we count them as $n$ constraints.} Thus the space of quadratic differentials with these properties is $3g-3-n$. Comparing again with the the dimension of $\Hom(\pi_1(\Sigma_g),\SL(2,\mathbb{C}))$, we see that $3g-3+n$ conditions have to be obeyed, which shows that generically $\lambda^\pm(z)$ can only exist on isolated points in $\mathcal{M}_{g,n}$.
Due to the non-abelian nature of the problem, this locus is much harder to quantify than for the case of line bundles.
\paragraph{Relation to covering maps.}
Ideally, we would like to conclude that the map $\gamma(z)$ is a covering map to the boundary surface. This is almost true, but unfortunately our input is not enough to decide this question.

The missing part in establishing this is to show that $\gamma(z)$ only maps to the region of discontinuity $\Omega$ of the boundary. See Appendix~\ref{app:hyperbolic 3 manifolds} for an explanation of the region of discontinuity. For global $\AdS_3$, this problem is non-existent, since $\Omega=\CP^1$. If this were the case, then we could compose with the canonical covering map $\pi:\Omega \longrightarrow \Sigma_G$ in order to construct a covering map (here $\Sigma_G$ is the boundary Riemann surface of genus $G$). In the unbranched case (i.e.~when $w_i=1$ for every $i$), 
it is known that the following conditions are equivalent \cite{Gunning}:
\begin{enumerate}
\item $\gamma(z) \ne \mathbb{CP}^1$
\item $\gamma$ is a covering map on its image.
\item $\rho_{\Gamma}(\pi_1(\Sigma_g))$ acts discontinuously on the image (i.e. $\gamma(z)$ maps into $\Omega$).
\end{enumerate}
Unfortunately, there are situations where neither of these conditions is satisfied. See \cite{Hehjal} for an explicit counterexample.

Thus, while this argument shows that a non-trivial $\boldsymbol{\lambda}(z)$ can only exist when both the $\U(1)$-bundle localizes and the complex structure of the Riemann surface localizes, we cannot exactly predict the localization locus. However, when the worldsheet is a covering surface of the boundary, we can explicitly construct such a section $\boldsymbol{\lambda}(z)$, see e.q.~\eqref{eq:lambda reconstructed} (this holds for arbitrary correlators).

Physically, there is reason to believe that only the covering maps should appear in the localization locus. If a map does not satisfy these conditions, then there is a point $z_*$ on the worldsheet such that $\gamma(z_*) \not \in \Omega$. In a small neighborhood of $z_*$, we can still view $\Gamma(z)$ as a map from the worldsheet to the boundary surface, but the map is undefined at $z=z_*$. In fact, the behaviour of $\gamma(z)$ is very similar to an essential singularity: in an arbitrarily small neighborhood of $z_*$, the map takes every possible value in $\CP^1$ infinitely many times (with the possible exception of up to two points according to Picard's theorem). A good analogy is the ``covering map'' $\CP^1 \longrightarrow \CP^1$ given by $\gamma(z)=\mathrm{e}^{\frac{1}{z}}$. While $\gamma(z)$ satisfies $\partial \gamma(z) \ne 0$ (and $\partial (\gamma(z)^{-1})\ne 0$) everywhere, it is undefined at $z=0$ and has an essential singularity there. Of course, we know that CFT correlators on the sphere cannot have essential singularities which is why this issue does not arise. For non-trivial boundaries, this is not automatic, but we still find it reasonable that CFT correlators are free from essential singularities in this sense.
We could view such a map $\gamma(z)$ also as a covering map of infinite degree. Covering spaces are suppressed by a factor of the degree in the partition function of the symmetric orbifold, see eq.~\eqref{eq:symmetric orbifold partition function}. Thus, also from this point of view, it is natural that these non-covering maps do not contribute. We shall assume this in the following.

\subsection{Special cases}\label{subsec:special cases}
The discussion simplifies considerably when either the worldsheet or the boundary surface has genus 1. We discuss these cases here separately. We restrict to the unbranched case.
\paragraph{Boundary genus 1.} In this case, we can conjugate $\rho_{\Gamma}$ and let it map into the affine group. Thus, $\gamma(z)$ defines in this case a complex affine structure (following the terminology of \cite{Gunning}.) It is a known result that such an affine structure only exists when also the worldsheet has genus 1. This implies that in spaces with a boundary torus, all higher genus corrections have to vanish. This was already conjectured to be the case in \cite{Eberhardt:2020bgq}. 

This result is simple to prove, so let us repeat the proof here. If $\rho_{\Gamma}$ maps in the group of affine transformations, $\partial \gamma(z)$ only transforms muliplicatively around the cycles of the Riemann surface. Hence $\partial \gamma(z)$ is an element of a line bundle $K \otimes L$. The line bundle $L$ captures the multiplicative factors by which $\partial \gamma(z)$ transforms. Since $L$ is again specified by a homomorphism (the projection of $\rho_{\Gamma}$ to the rotational subgroup of the affine group), $L$ is a flat line bundle. Thus, $\deg(K \otimes L)=2g-2$. As such $\partial \gamma(z)$ has $2g-2$ more zeros than poles. But we have seen that $\partial \gamma(z)$ has no zeros and hence $g \le 1$.

In the case where both worldsheet and boundary have genus 1, localization on covering maps is easy to prove explicitly. We uniformize the worldsheet torus as usual as $\mathbb{C}/(\mathbb{Z}\oplus \mathbb{Z}\tau)$. There are three different cases:
\begin{enumerate}
\item The boundary torus is obtained by Schottky uniformization: $\mathbb{T}^2=(\mathbb{CP}^1\setminus \{0,\infty\})/\mathbb{Z}$. The group action in the boundary identifies the boundary coordinate $x \in \CP^1$ as
\be 
x \sim \mathrm{e}^{2\pi i t}x\ ,
\ee
where $t$ is the boundary modular parameter. Thus, we search for a map $\gamma(z)$ (viewed as a meromorphic map on $\mathbb{C}$) that satisfies $\partial \gamma(z) \ne 0$, $\partial (\gamma(z)^{-1}) \ne 0$ and
\be 
\gamma(z+1)=\mathrm{e}^{2\pi i c t} \gamma(z)\ , \qquad \gamma(z+\tau)=\mathrm{e}^{-2\pi i d t} \gamma(z)\ ,
\ee
for two integers $c$ and $d$. Consider $f(z)=\frac{\gamma(z)}{\partial \gamma(z)}$. $f(z)$ is clearly periodic and is thus an elliptic function. $f(z)$ has no poles, since the only possible poles are located at the poles of $\gamma(z)$ and cancel out. This implies that $f(z)$is bounded and hence constant. Thus $\gamma(z)=B \exp\left(2\pi i A z\right)$. 
We can now solve for the periodicity conditions. The first implies that $A=ct-a$ for some integer $a$. The second periodicity condition implies the relation
\be 
(c\tau+d)t=a\tau+b
\ee
for integers $a,b,c,d$. Notice that the exponential map automatically maps in $\Omega=\mathbb{CP}^1\setminus\{0,\infty\}$ and thus we can construct a well-defined map from the worldsheet to the boundary torus.  
Thus the localization locus indeed corresponds to all covering surfaces of the boundary torus.
\item The boundary torus is obtained by the following quotient:
\be 
x \sim \mathrm{e}^{\frac{2\pi itm}{M}+\frac{2\pi i n}{M}}x
\ee
for $m \in \mathbb{Z}$ and $n \in \mathbb{Z}_M$. The corresponding bulk geometries are conical defects. The maps with these periodicity conditions are again exponential maps and one finds the same localization locus as before.
\item The boundary torus is obtained from the standard uniformization
\be 
x\sim x+1\sim x+t\ .
\ee
Thus, we search now for a map satisfying $\partial \gamma(z) \ne 0$, $\partial (\gamma(z)^{-1}) \ne 0$ and
\be 
\gamma(z+1)=ct-a+\gamma(z)\ , \qquad \gamma(z+\tau)=-dt+b+ \gamma(z)
\ee
for four integers $a,b,c,d$. $(\partial \gamma(z))^{-1}$ is an elliptic function without poles and hence constant. Thus $\gamma(z)=A z$ for some $A$. The first periodicity condition implies $A=ct-a$ and the second leads to the same condition
\be 
(c\tau+d)t=a\tau+b
\ee
as above.
\end{enumerate}
\paragraph{Worldsheet genus 1.} This case is also similar to the previous one. Let us assume that the boundary has genus $G \ge 2$, since we already analyzed the genus 1 case. Since $\pi_1(\Sigma_g)$ is abelian, also the image $\rho_{\Gamma}(\pi_1(\Sigma_g))$ is abelian. We may hence use an overall conjugation and conjugate the image $\rho_{\Gamma}(\pi_1(\Sigma_g))$ into the affine group. Abelian (Kleinian) subgroups of $\SL(2,\mathbb{C})$ are precisely given by the three cases that we discussed above.\footnote{There are also a couple of other possibilities, that are excluded by inspection such as the order 2 abelian subgroup generated by inversion $x \mapsto -\frac{1}{x}$.} We thus learn again that $\gamma(z)$ takes one of the simple forms that we described above. In any case, the image of $\gamma(z)$ is $\mathbb{CP}^1$ with either one or two points removed. However, for a boundary with genus $G \ge 2$, the limit set has necessarily infinitely points. Thus, we conclude that the image of $\gamma$ does not lie in the region of discontinuity $\Omega$ of the boundary. So while the worldsheet partition function indeed localizes, the localizing surface here does not correspond to a covering map. This exemplifies the problem that we encountered in the last subsection. 

\paragraph{Worldsheet genus 0.} Finally, let us also comment on the genus 0 case. In this case, our argument completely trivializes. Since $\pi_1(\Sigma_0)$ is the trivial group, there are no non-trivial $\Gamma$-bundles over which we could sum and no non-trivial moduli in which the partition function could localize. In fact, one can see that the existence of $\lambda^\pm$ does not impose any constraints.\footnote{We consider a two-point function with $w_1=w_2=1$, since this the lowest number of fields where the formalism makes sense.} Thus, we expect the sphere partition function to be generically non-vanishing.
\subsection{Worldsheet partition functions} \label{subsec:worldsheet partition function}
We can now say something about the full worldsheet correlation function (with the additional operator insertions that appear in \eqref{eq:string correlator prescription}). 
It should take the following qualitative form:
\begin{multline} 
\left \langle  \prod_{\alpha=1}^{2g-2+n} W(u_\alpha) \partial H(z)\prod_{\beta=1}^{g-1} \mathrm{e}^{-i H}(v_\beta) \prod_{i=1}^n V_{j_i,h_i}^{w_i}(x_,z_i)\right \rangle_{(\Sigma_{g,n},L_g)}
\\
=\sum_{(\Sigma',L')} \delta^{(6g-6+2n)}(\Sigma_{g,n}-\Sigma')\delta^{(2g)}(L_g-L') \\
\times\left \langle  \prod_{\alpha=1}^{2g-2+n} W(u_\alpha) \partial H(z)\prod_{\beta=1}^{g-1} \mathrm{e}^{-i H}(v_\beta) \prod_{i=1}^n V_{j_i,h_i}^{w_i}(x_,z_i)\right \rangle_{(\Sigma',L')}'\ . \label{eq:correlator localization}
\end{multline}
Here, the notation emphasizes the dependence of the correlator on the Riemann surface $\Sigma_g$ (or the punctured Riemann surface $\Sigma_{g,n}$) and the line bundle $L_g$ that specifies the periodicity properties of the free fields. The prime on the correlator on the RHS indicates that the delta function is factored out. Let us be summarize what we have exactly shown:
\begin{enumerate}
\item We have shown that the correlation function vanishes, except when $(\Sigma_{g,n},L_g)$ coincides with special (punctured) Riemann surfaces with corresponding line bundles.
\item From counting dimensions, we have shown that these special Riemann surfaces are isolated points in the $\mathcal{M}_{g,n} \times \Jac(\Sigma_g)$. Thus, the sum appearing on the RHS is indeed a discrete sum.
\item Whenever $\Sigma'$ is a ramified cover of the boundary surface with ramification indices specified by spectral flow, there exists a pair $(\Sigma',L')$ that satisfies all the constraints and that should appear in the sum.
\item We have \emph{not} shown that all the surfaces that satisfy the constraints we have analyzed are ramified covers and in general this is not true. But we have argued that nonetheless only such surfaces should appear in the partition function and assume this in the following to be the case.
\item We have only shown that the correlation function vanishes generically. As such it is a distribution with point-like support in $\Sigma_{g,n} \times \Jac(\Sigma_{g,n})$. Such distributions are a finite sum of $\delta$-functions and derivatives \cite[Theorem 6.25]{Rudin}. It is natural to assume that no derivatives appear. This is the case for genus 0 correlators for global $\AdS$, which under much better analytic control \cite{Eberhardt:2019ywk, Dei:2020zui}, but we do not know an argument for this in our more general setting. We shall assume this in the following.
\end{enumerate}
Let us in the following normalize the delta function such that
\begin{align} 
\int_{\mathcal{M}_{g,n}} 
\delta^{(6g-6+2n)} (\Sigma_{g,n}-\Sigma')f(\Sigma_{g,n})&=f(\Sigma')\ , \\
\int_{\Jac(\Sigma_{g})} \delta^{(2g)}(L-L')f(L)&=f(L')\ .
\end{align}
for any functions $f$. We view the delta-functions as top-forms on the respective spaces so that no measure is necessary.
The primed correlator appearing in \eqref{eq:correlator localization} should in particular be understood as a function on moduli space.

\paragraph{Remaining correlator.} Let us now discuss the remaining correlator. As we already stressed, we do not know how to compute it, but we know a few qualitative properties that it should satisfy. The first is a simple counting argument for the degrees of freedom. It was already explained in \cite{Gaberdiel:2018rqv, Eberhardt:2018ouy} that the tensionless string has only four transverse oscillators and these correspond roughly to the oscillators of $\mathbb{T}^4$. In other words, the remaining correlation function should essentially only capture the topologically twisted partition function of $\mathbb{T}^4$. This topologically twisted partition function has a direct relation with the untwisted partition function (evaluated on a specific metric), provided that the insertion points $v_\beta$ are chosen judiciously. This is explained in Appendix~\ref{app:topologically twisted partition function}.

\paragraph{Spin structure.} The question remains with  which spin structure the $\TT^4$ partition function should be evaluated. The natural guess (that is confirmed in the specific example of a torus in \cite{Eberhardt:2020bgq}) is to take the induced spin structure on the worldsheet that we discussed in Section~\ref{subsec:localization argument}, since there is no other distinguished spin structure on the worldsheet. This spin bundle in fact coincides with the pull-back bundle $\gamma^*S$ of the spin bundle on the boundaries that is defined by lifting $\Gamma \subset \PSL(2,\CC)$ to $\SL(2,\CC)$ as discussed in Section~\ref{subsec:orbifolds}. 

\paragraph{Full integrand.}
The upshot of this is that the worldsheet integrand should take the following form:
\be 
\mathcal{I}=\sum_{(\gamma,L')} \delta^{(6g-6+2n)}(\Sigma_{g,n}-\Sigma_\gamma)\delta^{(2g)}(L_g-L') Z_\text{classical}(\Sigma_\gamma) Z^{\TT^4}(\Sigma_\gamma,\gamma^*S)\ , 
\ee
where $S$ is the spin structure of the boundary component to which $\gamma$ maps and $\Sigma_\gamma$ the covering surface.
\paragraph{Classical part.} We have included a factor $Z_\text{classical}(\Sigma_\gamma)$ in the ansatz for the integrand \eqref{eq:full integrand}. Such a factor necessarily is present because of the conformal anomaly. The central charge of the (not topologically twisted) sigma model on $\TT^4$ is $c=6$ and consequently $Z^{\TT^4}(\Sigma_\gamma,\gamma^*S)$ depends on the metric on the worldsheet. This metric dependence is cancelled by $Z_\text{classical}(\Sigma_\gamma)$. However, this property does not determine $Z_\text{classical}(\Sigma_\gamma)$ completely. Comparing thermal $\text{AdS}_3$ with the BTZ black hole and the conical defect, it was seen in \cite{Eberhardt:2020bgq} explicitly that $Z_\text{classical}$ \emph{does} depend on the precise background. However, we should expect that $Z_\text{classical}$ does actually not depend on the moduli of the worldsheet and we have
\be 
\mathcal{I}=Z_\text{classical}\sum_{(\gamma,L')} \delta^{(6g-6+2n)}(\Sigma_{g,n}-\Sigma_\gamma)\delta^{(2g)}(L_g-L')  Z^{\TT^4}(\Sigma_\gamma,\gamma^*S)\ . \label{eq:full integrand}
\ee
One argument for this is modular invariance. This expression is modular invariant. If the classical part would also depend on the moduli, then it would need to do so in a way that preserves modular invariance, which is typically impossible for a `classical' partition function. This independence is also observed in simple examples \cite{Eberhardt:2020bgq}.\footnote{These cases are a bit special because the worldsheet torus can cover the boundary torus with an arbitrarily high degree and the classical part depends on the degree of the covering map. This issue does not arise for the higher genus situations.} 
\section{Background independence} \label{sec:background independence}
After having established that the worldsheet partition function of the tensionless string localizes in the moduli space of Riemann surfaces, we want to go a step further and argue that the \emph{string} partition function is actually independent of the precise geometry in which we place the string, but depends only on the boundary geometry.
\subsection{Some topology} \label{subsec:topology fundamental group}
Before we start, let us review some concepts of the fundamental group. Let us fix a geometry $\mathrm{AdS}_3/\Gamma$ an orbifold group $\Gamma$. Let us denote the boundary surface by $\Sigma_G$, which we assume for now to be connected. 
The boundary is obtained as $\Omega/\Gamma$, where $\Omega \subset \mathbb{CP}^1$ is the region of discontinuity of the action of $\Gamma$ on $\CP^1$. $\Omega$ is an intermediate covering space of $\Sigma_\Gamma$. Thus, we have the hierarchy of coverings
\be 
\widetilde{\Sigma}_G \longrightarrow \Omega \longrightarrow \Sigma_G\ ,
\ee
where $\widetilde{\Sigma}_G=\HH^2$ for $ G \ge 2$. The covering map $\Omega \longrightarrow \Sigma_G$ is given by identifying points in the orbits of the group action. Such a covering space is called \emph{regular} (or \emph{normal}). $\Gamma$ is the group of deck transformations of this covering map, i.e. the group of automorphisms of $\Omega$ which leave the covering map unchanged. Regularity implies that there is a short exact sequence of groups \cite[Proposition 1.40]{Hatcher}\footnote{We are not precise about the base points here. All statements should be understood  with a fixed base point.} 
\be 
1 \longrightarrow \pi_1(\Omega)  \longrightarrow \pi_1(\Sigma_G) \longrightarrow \Gamma \longrightarrow 1\ . \label{eq:fundamental groups short exact sequence}
\ee
In particular, there is a canonical projection map $p:\pi_1(\Sigma_G)\longrightarrow \Gamma$. 

For disconnected boundaries, the relevant groups are strictly speaking no longer groups because we can define a fundamental group for every connected component on the boundary. In this case there is a projection $p:\pi_1(\Sigma_{G_i}^{(i)}) \longrightarrow \Gamma$ for every component $\Sigma_{G_i}^{(i)}$ of the boundary.

\subsection{Sum over covering maps} \label{subsec:sum over covering maps}
In this subsection, we explain how string theory on different backgrounds can be equivalent. 
Twisted sectors of the orbifold partition function with orbifold group $\Gamma$ are labelled by (not necessarily injective) homomorphisms
\be 
\rho:\pi_1(\Sigma_g) \longrightarrow \Gamma\ ,
\ee
up to overall conjugation. We used to denote this homomorphism by $\rho_\Gamma$ in the last section, but since $\rho_{\CC^\times}$ will not appear again, we simply denote it by $\rho$.

We also argued that the worldsheet partition function localizes on covering surfaces.
Covering surfaces in a given twisted sector are specified by injective homomorphisms $\iota:\pi_1(\Sigma_g) \longrightarrow \pi_1(\Sigma_G)$ such that $p \circ \iota=\rho$. In case the boundary is disconnected, $\iota$ could map into the fundamental group of any component.
To understand this, suppose that we have given an injective homomorphism $\iota:\pi_1(\Sigma_g) \longrightarrow \pi_1(\Sigma_G)$ with the property $p \circ \iota=\rho$. The image of $\iota$ defines a subgroup of $\pi_1(\Sigma_G)$ and hence a covering space. However, such a homomorphism actually specifies a \emph{marked} covering space, i.e.~a covering space together with canonical generators for the fundamental group. These canonical generators are given by the images under $\iota$ of the generators of $\pi_1(\Sigma_g)$. Thus, we will actually count identical covering spaces, but with different markings several times. This is appropriate if we want the partition function to be a function on Teichm\"uller space and not on moduli space. Let us also recall that we understand all homomorphisms $\iota:\pi_1(\Sigma_g) \longrightarrow \pi_1(\Sigma_G)$ up to inner automorphisms, i.e.~overall conjugation.

In a commutative diagram, we can summarize the situation as follows:
\be 
\begin{tikzpicture}[baseline={([yshift=-.5ex]current bounding box.center)}]
\node (A) at (0,2) {$\pi_1(\Sigma_g)$};
\node (B) at (-4,0) {1};
\node (C) at (-2,0) {$\pi_1(\Omega)$};
\node (D) at (0,0) {$\pi_1(\Sigma_G)$};
\node (E) at (2,0) {$ \Gamma$};
\node (F) at (4,0) {1};
\draw[thick,->] (B) to (C);
\draw[thick,->] (C) to (D);
\draw[thick,->] (D) to node[above] {$p$} (E);
\draw[thick,->] (E) to (F);
\draw[thick,->] (A) to node[left] {$\iota$} (D);
\draw[thick,->] (A) to node[right] {$\,\rho$} (E);
\end{tikzpicture}\ ,
\ee
where the horizontal maps are exact.
Thus, the worldsheet partition function naturally is of the following form for a connected boundary
\be 
Z_\text{worldsheet}=\sum_{\rho: \pi_1(\Sigma_g) \longrightarrow \Gamma} \sum_{\begin{subarray}{c} \iota:\pi_1(\Sigma_g) \longrightarrow \pi_1(\Sigma_G) \\ \text{injective}, \, p \circ \iota=\rho\end{subarray}} Z_{\iota}\ . \label{eq:worldsheet partition function organization}
\ee
For a disconnected boundary, the change is very minimal. Since the worldsheet can cover either boundary, we get an additional sum over boundary components:
\be 
Z_\text{worldsheet}=\sum_i\sum_{\rho: \pi_1(\Sigma_g) \longrightarrow \Gamma} \sum_{\begin{subarray}{c} \iota:\pi_1(\Sigma_g) \longrightarrow \pi_1(\Sigma_G^{(i)}) \\ \text{injective}, \, p \circ \iota=\rho\end{subarray}}  Z_{\iota}\ . \label{eq:worldsheet partition function organization disconnected}
\ee

\subsection{Examples} \label{subsec:examples}
Let us exemplify this structure with some examples.
\paragraph{Cusp geometry.} This is a particularly simple case that actually was not discussed in \cite{Eberhardt:2020bgq}. In this case, the boundary of the space is a single torus and the orbifold group is $\mathbb{Z} \oplus \mathbb{Z}$, acting on the boundary by identification of $x \sim x+1 \sim x+t$. In this case $\Omega=\mathbb{C}$ and hence $\pi_1(\Omega)=\{0\}$ is the trivial group. So, $p$ is simply the identity map, which implies that $\rho=\iota$. Thus, out of the two sums that are present in \eqref{eq:worldsheet partition function organization}, actually only the sum over twisted sectors is present. In every twisted sector, the partition function localizes on exactly one configuration. This configuration already appeared in our discussion in Section~\ref{subsec:special cases}. In this case, $\rho=\iota$ is a homomorphism from $\mathbb{Z}^2$ to $\mathbb{Z}^2$ and hence naturally identified with an integer $2 \times 2$ matrix. 

\paragraph{Thermal $\text{AdS}_3$.} In this case $\Omega=\CP^1\setminus \{0,\infty\}$ and the short exact sequence of fundamental groups reads
\be 
1 \longrightarrow \pi_1(\Omega)\cong \langle \alpha \rangle \longrightarrow \pi_1(\Sigma_G) \cong \langle \alpha \rangle \oplus \langle \beta\rangle \longrightarrow \mathbb{Z} \cong \langle \beta \rangle \longrightarrow 1\ .
\ee
Here, we wrote $\langle \alpha \rangle$ for the abelian group $\{n\alpha\,|\, n \in \mathbb{Z}\}$.  The map from $\pi_1(\Omega)$ to $\pi_1(\Sigma_G)$ is then the obvious inclusion map and the second map is the projection map that sends $\alpha \mapsto 0$ and $\beta \mapsto \beta$. Homomorphisms $\iota$ are still naturally identified with $2 \times 2$ integer matrices.
Under the identification $\mathbb{Z}^2 \cong \langle \alpha \rangle \oplus \langle \beta \rangle $ both in the boundary and the bulk, we can write
\be 
\iota=\begin{pmatrix}
a & b \\ c & d 
\end{pmatrix}\ , \qquad p \circ \iota=\begin{pmatrix}
c & d
\end{pmatrix}\ .
\ee
Thus, twisted sectors are naturally labelled by the two integers $(c,d)$ and the localization in each twisted sector is labelled by two integers $(a,b)$. This is precisely the structure that was observed in \cite{Eberhardt:2020bgq}. For the Euclidean BTZ black hole black hole, the situation is analogous, except that the role of $\alpha$ and $\beta$ is interchanged and hence $p \circ \iota= \begin{pmatrix}
a & b
\end{pmatrix}$.

\paragraph{Conical defect in thermal $\AdS_3$.} This case is very similar. The short exact of fundamental groups is now
\be 
1 \longrightarrow \pi_1(\Omega)\cong \langle \alpha \rangle \longrightarrow \pi_1(\Sigma_G) \cong \langle \alpha \rangle \oplus \langle \beta\rangle \longrightarrow \mathbb{Z} \oplus \mathbb{Z}_M \cong  \langle \beta \rangle\oplus \langle \theta \rangle  \longrightarrow 1\ .
\ee
The first map is now the embedding $\alpha \mapsto (M\alpha,0)$. $\theta$ is the additional generator of order $M$. Under the same identification of $\iota$ with a $2 \times 2$ matrix as before, we have
\be 
p \circ \iota=\begin{pmatrix}
a \bmod M & b \bmod M \\ c  & d
\end{pmatrix}\ .
\ee

\paragraph{Handlebodies.} 
Moving on to a single higher-genus boundary, our description becomes less explicit. 
The orbifold group is a Schottky group $\Gamma_G^\text{S}$ of genus $G$ and as such is a free group with generators $B_1$, \dots, $B_G$.
The fundamental group of the boundary surface is generated by $\alpha_1$, \dots, $\alpha_G$, $\beta_1$, \dots, $\beta_G$ with the usual relation $\prod_I [\alpha_I,\beta_I]=1$. According to the short exact sequence \eqref{eq:fundamental groups short exact sequence}, the fundamental group of the domain $\Omega$ is the smallest normal subgroup of $\pi_1(\Sigma_G)$ that contains $\alpha_1$, \dots, $\alpha_G$. The surjection $\pi_1(\Sigma_G)\longrightarrow \Gamma_G^\text{S}$ simply sends $\alpha_I \longmapsto 1$ and $\beta_I \longmapsto B_I$. This defines the handlebody that belongs to this marking. For any different marking of the boundary surface, we obtain a different handlebody. Unfortunately, describing homomorphisms $\rho:\pi_1(\Sigma_g) \longrightarrow \Gamma_G^\text{S}$ is quite complicated and we will not be more explicit than this.

\subsection{Grand canonical ensemble} \label{subsec:grand canonical ensemble string}
Until now, we have not discussed how the chemical potential of the grand canonical ensemble enters the worldsheet theory. In the boundary theory, we have discussed it in Section~\ref{subsec:grand canonical ensemble symm orbifold}. In general, the boundary of the hyperbolic 3-manifold in question might have several components. In this case, from a boundary point of view, we can actually introduce several chemical potentials. From a worldsheet point of view, these chemical potentials enter in a slightly assymmetric way. We first discuss the `diagonal' chemical potential (by this we mean the average of all boundary chemical potentials) and postpone the others to Section~\ref{subsec:discrete torsion and more chemical potentials}.
There are two ways of introducing it, which seem to be equivalent. Basically, the chemical potential can be introduced by adding the spacetime identity vertex operator on the worldsheet. This corresponds to the zero mode of the dilaton in spacetime and hence the chemical potential can also be identified with the string coupling constant. We begin with the latter formulation.

\paragraph{Identification with string coupling.}
 The fugacity $p=\mathrm{e}^{2\pi i \sigma}$ is closely related to the string coupling constant. Consider adding the topological term
\be 
\frac{\lambda}{4\pi}\int \sqrt{\gamma} \mathcal{R}
\ee
to the worldsheet action. Here $\gamma$ is the worldsheet metric and $\mathcal{R}$ the corresponding Ricci scalar. This can be evaluated using the Gauss-Bonnet theorem and naively equals $\lambda(2-2g)$. However, the Riemann surface is punctured and one has to be a bit careful about the insertions of vertex operators, where the metric becomes singular. We want to exclude the insertion points from the integration.
Near these insertions, the metric is determined by the condition that the worldsheet is a holomorphic cover of the boundary theory. 

To analyze what happens close to the insertion, let us choose local coordinates such that a vertex operator is inserted close to $z_i=0$ and the corresponding insertion in the boundary is also $x_i=0$. The covering map behaves locally like
\be 
\gamma(z)=a_i z^{w_i}+\mathcal{O}(z^{w_i+1})\ .
\ee
Thus, the metric behaves as
\be 
\mathrm{d}s^2 \propto |\partial \gamma(z)|^2 \mathrm{d}z \, \mathrm{d}\bar{z} =A |z|^{2w_i-2} \mathrm{d}z \, \mathrm{d}\bar{z}+\mathcal{O}(|z|^{2w_i-1}) \ .
\ee
For a metric of the form $\mathrm{d}s^2=\mathrm{e}^{\phi(z,\bar{z})} \mathrm{d}z\, \mathrm{d}\bar{z}$, we have
\be 
\sqrt{\gamma} \mathcal{R}=-\Delta \phi(z,\bar{z})\ .
\ee
In our case, $\phi(z,\bar{z})=(w_i-1)\log|z|^2+\text{regular terms}$, thus leading to
\be 
\sqrt{\gamma} \mathcal{R}=-4\pi (w_i-1) \delta^{(2)} (z)+\text{finite} .
\ee
Integrating over the full Riemann surface leads by the Gauss-Bonnet theorem to the Euler characteristic. Thus, after excising the singular points (or alternatively integrating only over the regular part of the curvature), we obtain
\be 
\frac{\lambda}{4\pi}\int_{\Sigma_g \setminus\{z_1,\dots,z_n\}} \sqrt{\gamma} \mathcal{R}=\lambda \left(2-2g+\sum_{i=1}^n (w_i-1) \right)\ .
\ee
The Riemann-Hurwitz formula connects the degree $N_\text{c}=\deg(\gamma)$, the worldsheet and boundary genus and the ramification indices as follows:
\be 
2-2g=N_\text{c}(2-2G)-\sum_{i=1}^n (w_i-1)\ . \label{eq:Riemann Hurwitz}
\ee
Thus, we can express the parentheses in terms of the degree $N_\text{c}$ of the relevant covering map and the boundary genus $G$ as follows:
\be 
\frac{\lambda}{4\pi}\int_{\Sigma_g \setminus\{z_1,\dots,z_n\}} \sqrt{\gamma} \mathcal{R}=\lambda N_\text{c} (2-2G)\ .
\ee
This is exactly what we want: this term in the action weighs different covering maps according to their degree, which is analogous to the situation in the symmetric orbifold \eqref{eq:symmetric orbifold partition function}.
For this to work literally, we have to identify
\be 
\lambda(2-2G)=2\pi i \sigma\ .
\ee
In terms of the string coupling constant, we have
\be 
p=g_\text{string}^{2G-2}\ ,
\ee
where $p=\mathrm{e}^{2\pi i \sigma}$ and $g_\text{string}=\mathrm{e}^{-\lambda}$. Here we assume that there is only one boundary surface or that all boundary surfaces have the same genus. If this is not the case, the situation can be corrected by including off-diagonal chemical potentials, see Section~\ref{subsec:discrete torsion and more chemical potentials}.

\paragraph{The spacetime identity operator.} Alternatively, the grand canonical partition function is obtained by adding the following term to the worldsheet action \cite{Porrati:2015eha}:
\be 
\delta S=2\pi i\sigma \int_\Sigma I(z,\bar{z})\ . \label{eq:action deformation}
\ee
Here, $I(z,\bar{z})$ is the worldsheet operator that corresponds to the spacetime identity. Note that $\delta S$ is a an $\SL(2,\mathbb{C})$ singlet and as such is a marginal operator that we can add to the worldsheet theory. $I(z,\bar{z})$ is actually quite easy to construct, its unintegrated version corresponds to the vertex operator
\be 
I^{(0)}(z,\bar{z})=V^{w=1}_{j,h=0}(x,\bar{x},z,\bar{z})
\ee
Since $h=0$, the RHS does not depend on $x$ and gives hence a well-defined operator $I^{(0}(z,\bar{z})$. The worldsheet conformal weight also vanishes.
Thus, $I^{(0)}(z,\bar{z})$ corresponds to an unintegrated vertex operator on the worldsheet. To obtain the integrated version, we apply the descent formalism. For the $\mathcal{N}=4$ topological string, one applies the twisted supercharges $\tilde{G}^-_{-1}\bar{\tilde{G}}^-_{-1}$ to $I^{(0)}(z,\bar{z})$. One needs to pick $\tilde{G}^-_{-1}$ and not $G^-_{-1}$ to preserve conservation of the various ghost currents of the full worldsheet theory. 
Since we want the vertex operator to have also vanishing $Z$-charge (see Section~\ref{subsec:free field realization}), we need to pick also $j=\frac{1}{2}$. Applying $\tilde{G}^-_{-1}$ lowers the value of the $\mathfrak{sl}(2,\mathbb{C})$ spin by one unit and hence the integrated vertex operator has $j=-\frac{1}{2}$. This leaves the constraint \eqref{eq:j constraint} unchanged. 

The integrated vertex operator $\mathcal{I}=\int \tilde{G}^-_{-1}\bar{\tilde{G}}^-_{-1}I^{(0)}(z,\bar{z})$ plays then the role of the spacetime identity. Up to normalization, it survives the orbifold, since it is an $\mathfrak{sl}(2,\mathbb{C})$ singlet. As such, it is a central term of the operator algebra. Since all it's OPEs are trivial, it simply produces an eigenvalue when inserted in a correlator,
\be 
\left \langle \mathcal{I} \prod_{\alpha=1}^{2g-2+n} W(u_\alpha)\prod_{i=1}^n V_{j_i,h_i}^{w_i} (x_i,z_i) \right \rangle=\lambda\left \langle  \prod_{\alpha=1}^{2g-2+n} W(u_\alpha)\prod_{i=1}^n V_{j_i,h_i}^{w_i} (x_i,z_i) \right \rangle\ .
\ee
To determine this eigenvalue, one can invoke various consistency conditions. 
First of all, the eigenvalue $\lambda$ cannot depend on either $j_i$, $h_i$, $x_i$ or $z_i$, since these quantities can be changed by acting with an element in the chiral operator algebra on the vertex operators and $\mathcal{I}$ should be central in this operator algebra. Similarly, we could consider any descendant vertex operator that is obtained by the action of the chiral operator algebra. 

From general properties of OPEs, it follows that the eigenvalue $\lambda$ should be additive in the involved quantum numbers.
Adding another unintegrated identity vertex operator should not change the result (since we can pull out the vertex operators one after another). This has the effect of adding both one more $W$ and a $V$. When the boundary theory is a sphere, every vertex operator $V^{w_i}_{j_i,h_i}(x_i,z_i)$ contributes $\tfrac{1}{2}w_i$ to the eigenvalue \cite{Giveon:2001up, Porrati:2015eha}. Since $\mathcal{I}$ is a $\SL(2,\CC)$ singlet under the orbifold, this should not change in the topologically non-trivial situation. These arguments together show that $V^{w_i}_{j_i,h_i}(x_i,z_i)$ contributes $\frac{1}{2}w_i$ to the eigenvalue of $\mathcal{I}$ and each $W$ contributes $-\frac{1}{2}$, thus giving in total
\be 
\lambda=\frac{A}{2} \left(\sum_{i=1}^n w_i-(2g-2+n) \right)\ ,
\ee
where we allowed for an arbitrary normalization $A$ of $\mathcal{I}$.
We set $A=(1-G)^{-1}$, where $G$ is the boundary genus so that\footnote{We assume that $G\ne 1$. For $G=1$, both the parenthesis and the prefactor vanish so that the ratio can still be non-trivial, but is more subtle to define.} 
\be 
\lambda=\frac{1}{2-2G}\left(2-2g+\sum_{i=1}^n (w_i-1) \right)\ .
\ee
This fits with the sphere result of \cite{Porrati:2015eha}. Of course, this constant only makes sense when the corresponding correlator is non-vanishing, otherwise the statement is void. For this a covering map from the worldsheet to the boundary has to exist. 

From the Riemann Hurwitz formula \eqref{eq:Riemann Hurwitz}, we conclude that $\lambda=N_\text{c}=\deg(\gamma)$, i.e.~the eigenvalue of the spacetime identity operator is again the degree of the associated covering map.\footnote{This statement makes again sense for a boundary torus.} Thus we see that adding the identity operator is equivalent to changing the string coupling. This makes sense because both these operators can be identified with the dilaton zero mode. Since the string coupling is easier to introduce, we will continue to work with the string coupling.

\paragraph{Sphere partition function.} We saw that the chemical potential $p$ of the grand canonical ensemble is essentially mapped to the string coupling constant on the worldsheet according to $p=g_\text{string}^{2G-2}$. It is important to keep in mind that the whole discussion only applies for worldsheet genera $g \ge 1$. There is also a non-vanishing sphere partition function that we need to treat separately. We do not know how the sphere partition function is computed in general, because the volume of the residual M\"obius symmetry is difficult to regularize. However, we expect it to capture the `classical part' of the string partition function. We thus expect that
\be 
Z_0 \propto I_\text{bulk}(\HH^3/\Gamma) \propto \vol(\HH^3/\Gamma)
\ee
is proportional to the volume of spacetime. Here $I_\text{bulk}(\mathcal{M})$ is the spacetime on-shell action evaluated on hyperbolic manifold $\mathcal{M}$.\footnote{Such a local spacetime action does not exist in this regime. When we talk about the spacetime effective action, we mean the supergravity action which applies in the regime $k \gg 1$ and continue it to $k=1$. This seems to give the correct results, but is not a very satisfying procedure.}
This volume is anomalous and has to be regularized. The regularization depends on the boundary metric (and not only on its conformal class), which gives rise to the conformal anomaly. This computation is standard, see \cite{Krasnov:2000zq,Takhtajan:2002cc}. We normalize $\HH^3$ by setting the Ricci scalar to $-6$. 
The constant of proportionality is then essentially given by the central charge of the boundary theory
\be 
Z_0=\frac{c}{6\pi}\vol(\HH^3/\Gamma)\equiv -c \, I_\text{bulk}(\HH^3/\Gamma)\ , \label{eq:Ibulk}
\ee
so that $\mathrm{e}^{Z_0}$ captures the conformal anomaly of a central charge $c$ CFT. We normalize $I_\text{bulk}(\HH^3/\Gamma)$ such that it precisely accounts for the conformal anomaly of a $c=1$ CFT.
As we shall see below, we should take $c=6$, the central charge of a single $\TT^4$, since the effect of the grand canonical potential is already incorporated in the string coupling $g_\text{string}^{-2}$ that multiplies the sphere partition function.
\subsection{Discrete torsion and more chemical potentials} \label{subsec:discrete torsion and more chemical potentials}
We constructed the `diagonal' chemical potential above. In general, we can however have different chemical potentials at the boundaries of the bulk. From a worldsheet perspective, these additional chemical potentials can be introduced by considering non-trivial discrete torsion. We first recall the concept of discrete torsion and then apply it to our situation.
\paragraph{Discrete torsion.} For an orbifold CFT, the genus $g$ partition function takes in general the following form
\be 
\frac{1}{|\Gamma|}\sum_{\rho: \pi_1(\Sigma_g) \rightarrow \Gamma} \varepsilon(\rho) Z_\rho\ ,
\ee
where $\Gamma$ is again the orbifold group and homomorphisms $\rho$ up to conjugation specify $\Gamma$-bundles.
$\varepsilon(\rho)$ is a phase depending on the $\Gamma$-bundle over which we sum. It is constrained through various consistency conditions and in general allowed values are classified by the Schur multiplier of the orbifold group -- the cohomology group $\mathrm{H}^2(\Gamma,\U(1))$. Given a 2-cocycle $\phi \in \mathrm{H}^2(\Gamma,\U(1))$, one can define \cite{Vafa:1986wx, Vafa:1994rv, Sharpe:2000ki}
\be 
\varepsilon(\rho)=\prod_{I=1}^g \frac{\phi(\rho(\alpha_I),\rho(\beta_I))}{\phi(\rho(\beta_I),\rho(\alpha_I))}\ ,
\ee
where $\alpha_1,\dots,\alpha_g$, $\beta_1,\dots,\beta_g$ are a canonical homology basis of the worldsheet, see \eqref{eq:canonical homology basis}.

\paragraph{Schur multiplier for Kleinian groups.} Let us discuss the Schur multiplier for Kleinian groups, i.e.~the orbifold groups $\Gamma$ that reduce $\HH^3$ to $\HH^3/\Gamma$.  Let us assume that $\mathcal{M} \cong \HH^3/\Gamma$ is a non-singular manifold.\footnote{Otherwise we have to employ orbifold cohomology in the following discussion.} The group cohomology of $\Gamma$ can be computed by noticing that $\HH^3/\Gamma$ is an Eilenberg-MacLane space $K(\Gamma,1)$, meaning that $\pi_1(\HH^3/\Gamma)=\Gamma$ and $\pi_n(\HH^3/\Gamma)=0$ for $n>1$. It is known that the group cohomology coincides with the usual singular cohomology of its corresponding Eilenberg-MacLane space. Hence the Schur multiplier can be interpreted more geometrically as the singular cohomology group $\mathrm{H}^2(\HH^3/\Gamma,\U(1))$.

From a worldsheet action point of view, an element $B \in\mathrm{H}^2(\HH^3/\Gamma,\U(1))$ is a two-form with $\mathrm{d}B=0$ (up to exact forms). Hence we can add the following topological term to the worldsheet action:
\be 
2\pi i \int_{\Sigma} g^*B\ , 
\ee
where $g:\Sigma \longrightarrow \HH^3$ is the embedding coordinate of the string in $\HH^3/\Gamma$. If $B \in \mathrm{H}^2(\HH^3/\Gamma,\ZZ)$, then this term is in $2\pi i \ZZ$ and hence has no effect in the path integral. Thus, inequivalent topological terms are indeed classified by the cohomology group $\mathrm{H}^2(\HH^3/\Gamma,\U(1))$. 
We explain in Appendix~\ref{app:hyperbolic 3 manifolds} that $\mathrm{H}^2(\HH^3/\Gamma,\U(1))\cong \U(1)^R$ is torsion free. It is naturally related to the torsion free homology group $\mathrm{H}_2(\HH^3/\Gamma,\ZZ)\cong \ZZ^R$. $R$ can be computed in terms of $\Gamma$ and the boundary genera of the manifold $\mathcal{M}$, see eq.~\eqref{eq:rank second homology group}.

In the following, we will discuss some very natural generators of $\mathrm{H}_2(\HH^3/\Gamma,\ZZ)$, namely the boundary components of $\HH^3/\Gamma$. Let $\partial\HH^3/\Gamma= \Sigma^{(1)} \sqcup \cdots\sqcup \Sigma^{(n)}$ be the boundary components of the space. Then each $\Sigma^{(i)}$ defines a cocycle in $[\Sigma^{(i)}] \in \mathrm{H}_2(\HH^3/\Gamma,\ZZ)$. Of course, $\sum_{i=1}^n [\Sigma^{(i)}] $ is the boundary of $\HH^3/\Gamma$ and is thus null-homologous. This is the only relation in homology and thus we obtain $n-1$ generators of homology (and hence cohomology), exactly the number of missing chemical potentials. In the examples we consider, these generators fully generate the second (co)homology group, but we discuss in Appendix~\ref{app:hyperbolic 3 manifolds} an example with an additional generator for $\mathrm{H}_2(\HH^3/\Gamma,\ZZ)$.\footnote{We do not understand the meaning of additional generators of $\mathrm{H}_2(\HH^3/\Gamma,\ZZ)$ in general. However, for the conical defect, $\mathrm{H}^2(\Gamma,\U(1))$ has an additional torsion generator and its effect can be absorbed in the definition of the boundary chemical potential \cite{Eberhardt:2020bgq}. Thus, it has no physical effect in this case.}
\paragraph{Identification with off-diagonal chemical potentials.} We are now arguing that additional chemical potentials that can be introduced for multiple boundaries can indeed be identified with the discrete torsion parameters. Let us pick as generators of $\mathrm{H}_2(\HH^3/\Gamma,\ZZ)$ the fundamental classes $[\Sigma^{(1)}]$, \dots, $[\Sigma^{(n-1)}]$. We are omitting the last boundary to keep the generators independent (we have $[\Sigma^{(n)}]=-[\Sigma^{(1)}]-\cdots-[\Sigma^{(n-1)}]$). As we mentioned, there could be more generators that do not originate from the boundaries which we shall ignore. Let now $B^{(1)}$, \dots, $B^{(n-1)}$ the corresponding generators of $\mathrm{H}^2(\HH^3/\Gamma,\ZZ)$ such that $\int_{\Sigma^{(i)}} B^{(j)}=\delta_{ij}$ for $i=1,\dots,n-1$. Then we can consider the topological terms
\be 
S_\text{top}=2\pi i \sum_{i=1}^{n-1} \sigma_i \int_{\Sigma} g^*B^{(i)}\ .
\ee
We can evaluate this term on the configurations that appear in the path integral. Let us assume that the worldsheet $\Sigma$ covers the boundary $\Sigma^{(i)}$ holomorphically (for $i=1,\dots,n-1$), then the term evaluates to
\be 
S_\text{top}=2\pi i \deg(\gamma)\sum_{j=1}^{n-1} \sigma_j \int_{\Sigma^{(i)}} B^{(j)}=2\pi i \deg(\gamma) \sigma_i\ .
\ee
If it covers instead $\Sigma^{(n)}$, we obtain
\be 
S_\text{top}=2\pi i \deg(\gamma)\sum_{i=1}^{n-1} \sigma_i \int_{\Sigma^{(n)}} B^{(i)}
=-2\pi i \deg(\gamma) \sum_{i=1}^{n-1} \sigma_i\ .
\ee
This confirms that these terms indeed correspond to the off-diagonal chemical potentials.

\paragraph{The quasi-Fuchsian wormhole.} For the two-sided wormhole that is obtained as $\HH^3/\Gamma_G^\text{QF}$ for $\Gamma_G^\text{QF}$ a genus $G$ quasi-Fuchsian group (see Appendix~\ref{subapp:Kleinian groups examples} for the relevant definitions), we can be quite explicit. There is a single generator of $\mathrm{H}^2(\HH^3/\Gamma_G^\text{QF},\ZZ)$, which we can take to be the fundamental class of either boundary component. Thus, the topological term weighs contributions that cover the left boundary with opposite phases than those that cover the right boundary. Eq.~\eqref{eq:worldsheet partition function organization disconnected} takes the form
\begin{multline} 
Z_\text{worldsheet}=\left(p_1^{\frac{1}{2}}p_2^{-\frac{1}{2}}\right)^{\frac{2g-2}{2G-2}}\sum_{\rho: \pi_1(\Sigma_g) \longrightarrow \Gamma_G^\text{QF}} \sum_{\begin{subarray}{c} \iota:\pi_1(\Sigma_g) \longrightarrow \pi_1(\Sigma_{G}^{(1)}), \\\text{injective}, \, p \circ \iota=\rho \end{subarray}} Z_{\iota}\\
+\left(p_1^{-\frac{1}{2}}p_2^{\frac{1}{2}}\right)^{\frac{2g-2}{2G-2}}\sum_{\rho: \pi_1(\Sigma_g) \longrightarrow \Gamma_G^\text{QF}} \sum_{\begin{subarray}{c} \iota:\pi_1(\Sigma_g) \longrightarrow \pi_1(\Sigma_{G}^{(2)}), \\ \text{injective}, \, p \circ \iota=\rho\end{subarray}} Z_{\iota}\ . \label{eq:Fuchsian worldsheet partition function}
\end{multline}
Upon setting the string coupling constant to $g_\text{string}^{2G-2}=p_1^{\frac{1}{2}} p_2^{\frac{1}{2}}$, we can arrange it that covering maps mapping to the left (right) boundary are weighted by the fugacities $p_1$ ($p_2$).

\subsection{Full string partition function}
We shall now restrict the discussion to partition functions and tie various observations together.
Our ultimate goal is to compute the string partition function in the background $\mathrm{AdS}_3/\Gamma \times \S^3 \times \TT^4$ in the grand canonical ensemble. Let us first restrict to backgrounds with a single boundary so that there is no discrete torsion. The grand canonical string partition function takes the general form
\be 
\tilde{\mathfrak{Z}}=\exp\left(\sum_{g=1}^\infty g_\text{string}^{2g-2} Z_g \right)=\exp\left(\sum_{g=1}^\infty p^{\frac{2g-2}{2G-2}} Z_g \right)\ ,
\ee
where $Z_g$ is the genus $g$ worldsheet partition function. The localization property implies that that the worldsheet genus takes the form $g=N_\text{c}(G-1)+1$ (with the exception of genus $g=0$). To take into account the sphere partition function with $g=0$, we proceed as follows. The sphere partition function in the canonical ensemble should provide a contribution $-6\deg(\gamma) I_\text{bulk}$, where the factor 6 comes from the fact that the seed theory of the boundary has central charge $c=6$, see eq.~\eqref{eq:Ibulk}. Thus, in the grand canonical ensemble, the effect of the sphere partition function is a renormalization
\be 
p \longmapsto p \,\mathrm{e}^{-6 I_\text{bulk}}
\ee 
or $\sigma \longmapsto \sigma+\frac{3i}{\pi} I_\text{bulk}$.

Thus, the string partition function becomes
\be 
\mathfrak{Z}=\exp \left(\sum_{N_\text{c}=1}^\infty \left(p \mathrm{e}^{-6 I_\text{bulk}}\right)^n Z_{N_\text{c}(G-1)+1}\right)\ .
\ee
The partition function $Z_{N_\text{c}(G-1)+1}$ is in turn obtained by integrating \eqref{eq:full integrand}. We refine \eqref{eq:full integrand} by splitting the sum over covering maps as discussed in Section~\ref{subsec:sum over covering maps}, see eq.~\eqref{eq:worldsheet partition function organization}:
\be 
Z_g=Z_{\text{classical},\,g}\int_{\mathcal{M}_g} \sum_{\rho: \pi_1(\Sigma_g) \longrightarrow \mathrm{G}} \sum_{\begin{subarray}{c}\iota:\pi_1(\Sigma_g) \longrightarrow \pi_1(\Sigma_G), \\ \text{injective},\, p \circ \iota=\rho \end{subarray}} \delta^{(6g-6)}(\Sigma_g,\Sigma_{\iota}) Z^{\TT^4}(\Sigma_{\iota},S_{\iota})\ .
\ee
Here, we changed the notation slightly and denote by $\Sigma_{\iota}$ the marked covering surface that is determined by the homomorphism $\iota$. We integrate over $\Sigma_g$ in the expression. Finally, $S_{\iota}$ is the induced spin structure on $\Sigma_{\iota}$.
\paragraph{Markings and integration.} We should note that the surface $\Sigma_{\iota}$ comes naturally with a marking, i.e.~a set of generators $\alpha_1$, \dots, $\alpha_g$, $\beta_1$, \dots, $\beta_g$ of $\pi_1(\Sigma_g)$ up to \emph{inner} automorphisms. The ambiguity up to an inner automorphism comes from the fact that we haven't chosen a basepoint for $\pi_1(\Sigma_g)$ and neither for the boundary. Thus, the same surface appears several times in the sum of the integrand, since we are also summing over markings. Recall that the moduli space of Riemann surfaces is related to Teichm\"uller space as follows:
\be 
\mathcal{M}_g=\mathcal{T}_g/\MCG(\Sigma_g)\ ,
\ee
where $\MCG(\Sigma_g)=\Out(\pi_1(\Sigma_g))$ is the mapping class group. We have collected a few relevant facts in Appendix~\ref{subapp:Teichmuller space}. $\mathcal{T}_g$ is the moduli space of marked Riemann surfaces. Hence, we can denote the result of the integral schematically as
\be 
Z_g=Z_{\text{classical},\,g}\sum_{\rho: \pi_1(\Sigma_g) \longrightarrow \Gamma} \sum_{\begin{subarray}{c} \iota :\pi_1(\Sigma_g) \longrightarrow \pi_1(\Sigma_G),\\
 \text{injective}, \,  p \circ \iota=\rho \end{subarray}} Z^{\TT^4}(\Sigma_{\iota},S_{\iota}) \bigg/\Out(\pi_1(\Sigma_g))\ ,
\ee
which means that we pick only one arbitrary marking for each covering surface. If we combine the two sums into a single sum of arbitrary homomorphisms of $\pi_1(\Sigma_g)$ to $\pi_1(\Sigma_G)$, then this `gauging' by $\Out(\pi_1(\Sigma_g))$ is easy to implement. An injective homomorphism $\iota:\pi_1(\Sigma_g) \longrightarrow \pi_1(\Sigma_G)$ up to conjugation and outer automorphism is fully characterized by its image $\iota(\pi_1(\Sigma_g)) \subset \pi_1(\Sigma_G)$ up to conjugation. Thus the sum becomes a sum over subgroups of $\pi_1(\Sigma_G)$ up to conjugation of finite index $d$. Putting the pieces together, we obtain for the full string partition function
\be 
\mathfrak{Z}=\exp \left(\sum_{N_\text{c}=1}^\infty \left(p \mathrm{e}^{-6 I_\text{bulk}}\right)^{N_\text{c}} Z_{\text{classical},\,N_\text{c}(G-1)+1} \sum_{\begin{subarray}{c}\mathrm{H}\subset\pi_1(\Sigma_G)\text{ up to conjugation}, \\ [\pi_1(\Sigma_G):\mathrm{H}]=N_\text{c} \end{subarray} } Z^{\TT^4}(\Sigma_{\mathrm{H}},S_{\mathrm{H}})\right)\ .
\ee
Here we changed notation again slightly to account for the fact that we are labelling (unmarked) covering surfaces and their spin structures by subgroups of $\pi_1(\Sigma_G)$ (up to conjugation). 
\paragraph{The classical part.} We see that this formula is very close to the partition function of the symmetric orbifold \eqref{eq:symmetric orbifold partition function}. It becomes equal provided that
\be 
Z_{\text{classical},\, N_\text{c}(G-1)+1}=\frac{\mathrm{e}^{6N_\text{c}I_\text{bulk}}}{N_\text{c}}\ . \label{eq:Z classical}
\ee
This was observed to be true in cases with a torus boundary in \cite{Eberhardt:2020bgq}, but we do not know a general argument for this formula. 
\paragraph{Background independence.}
In particular, once we make this identification, the background dependence of the string partition function completely cancels out. $I_\text{bulk}$ depends on the precise orbifold group that we use to engineer the bulk manifold, whereas the sum over subgroups only depends on the boundary surface, but not on the  bulk three-manifold. While we are not able to give a general proof of this phenomenon, we hopefully elucidated the mechanism behind it.

\paragraph{Disconnected boundary.} Let us now consider the generalization to a disconnected boundary. We consider for simplicity the case of a wormhole obtained by a quasi-Fuchsian group, with boundary components two genus $G$ surfaces, that we denote by $\Sigma_G^{(1)}$ and $\Sigma_G^{(2)}$. The discussion is entirely analogous once we identify $g_\text{string}^{2G-2}=p_1^{\frac{1}{2}}p_2^{\frac{1}{2}}$ with the diagonal chemical potential, see eq.~\eqref{eq:Fuchsian worldsheet partition function}. The sphere partition function again renormalizes the chemical potentials. Since in the canonical ensemble, the sphere partition function should again lead to the contribution $-6 \deg(\gamma)I_\text{bulk}$, it does not distinguish covering maps that cover the left- or the right boundary. This means that only the diagonal chemical potential is renormalized as follows:
\be 
p_1^{\frac{1}{2}}p_2^{\frac{1}{2}} \longmapsto p_1^{\frac{1}{2}}p_2^{\frac{1}{2}}  \,\mathrm{e}^{6 I_\text{bulk}}\ ,
\ee
where $p_1$ and $p_2$ are the fugacities of the left and right boundaries. In order to reproduce the symmetry orbifold partition function, both terms in \eqref{eq:worldsheet partition function organization disconnected} have the classical contribution  $Z_{\text{classical},d(G-1)+1}^{(i)}=d^{-1}\mathrm{e}^{6d I_\text{bulk}}$, as in eq.~\eqref{eq:Z classical}. Thus, the string partition function on the wormhole manifestly factorizes,
\be 
\mathfrak{Z}_\text{wormhole}(p_1,p_2)=\mathfrak{Z}_{\text{Sym}}(p_1)\mathfrak{Z}_{\text{Sym}}(p_2)\ .
\ee
\section{Stringy and classical geometry} \label{sec:stringy and classical geometry}
In this section, we would like interpret our results more geometrically. In the previous sections, we provided evidence that the worldsheet partition function is independent of the bulk geometry and actually only depends on the boundary geometry.

\subsection{Stringy geometry}
We have seen that the symmetric orbifold partition function is naturally expressed in terms of (possibly disconnected) covering spaces and these are interpreted holographically as the worldsheet. Thus, there is apparently no such concept as bulk geometry, there is only the geometry of the covering space, i.e.~the worldsheet.

Even though our results defy the intuition of semiclassical gravity, they are still very geometric -- as long as one replaces the concept of a bulk manifold by the collection of worldsheets, which is a notion of stringy geometry in this concept. The worldsheets themselves can in good approximation be treated semiclassically.
\subsection{Condensation}\label{subsec:condensation}
Sometimes, a collection of these stringy geometries can be interpreted as classical geometries. For this to be meaningful, we want to talk about large $N$ in the symmetric orbifold. Since we have argued that string theory describes the grand canonical ensemble, $N$ gets replaced by the fugacity $p=\mathrm{e}^{2\pi i \sigma}$. Initially, the grand canonical ensemble is only well-defined if we take $\Im \sigma$ big enough, since otherwise the definition of the grand canonical partition function \eqref{eq:grand canonical partition function} does not converge. We define the function $(\Im \sigma)_\text{min}$ as the minimal chemical potential for which the grand canonical potential is still well-defined. This is an interesting function on the moduli space of the boundary surface(s).
For $\Im \sigma$ close to $(\Im \sigma)_\text{min}$, the grand canonical partition function is dominated by contributions from large $N$ and we can expect semiclassical bulk geometry to emerge.

Let us consider a bulk background geometry with a torus boundary such as thermal $\AdS_3$ or the Euclidean BTZ black hole. In this case, we can be very explicit and $(\Im \sigma)_\text{min}$ is only a function of the boundary modular parameter $\tau_\text{bdry}$. The behaviour of the symmetric orbifold partition function in the canonical ensemble is known explicitly in a large $N$ limit \cite{Keller:2011xi}:
\be 
\log Z_{\text{Sym}^N(\TT^4)}^\text{NS}(\tau_{\text{bdry}})\sim N\pi  \max_{(c,d)=1,\ c+d\text{ odd}} \frac{\Im \tau_{\text{bdry}}}{|c\tau_{\text{bdry}}+d|^2}\ .
\ee
The maximum is taken over all coprime pairs of integers whose sum is odd.
We considered the NS spin structure. The expression is manifestly invariant under the relevant modular group
\be 
\Gamma_\text{NS}=\left\{\left. \begin{pmatrix}
a & b \\ c & d
\end{pmatrix}\, \right|\, a+d\text{ even},\ b+c\text{ even},\  c+d\text{ odd}\right\}\ .
\ee
 This translates into
\be 
(\Im \sigma)_\text{min}=\frac{1}{2}\max_{(c,d)=1,\ c+d\text{ odd}} \frac{\Im \tau_{\text{bdry}}}{|c\tau_{\text{bdry}}+d|^2}\ .
\ee
This function has phase transitions. Conventionally, these terms are interpreted as classical bulk geometries that dominate the partition function in different regimes. For purely imaginary $\tau_\text{bdry}$, there is a phase transition at $\tau=i$ corresponding to the transition between thermal $\text{AdS}_3$ ($(c,d)=(0,1)$) and the non-rotating Euclidean BTZ black hole ($(c,d)=(1,0)$).

This function arises from the symmetric orbifold as follows. The torus partition function takes the form
\be 
\mathfrak{Z}_{\Sym(\TT^4)}^{\text{NS}}=\exp \left( \sum_{m,w=1}^\infty \sum_{r \in \ZZ/\mathbb{Z}w}\frac{p^{m w}}{mw} Z^{\TT^4} \begin{bmatrix}
\frac{r+m}{2} \\ \frac{w}{2} 
\end{bmatrix}\left(\frac{m \tau_\text{bdry}+r}{2} \right) \right)\ .\label{eq:torus symmetric orbifold partition function}
\ee
This is a way of writing a sum over all covering tori, compare to eq.~\eqref{eq:symmetric orbifold partition function} for the general case. The degree of the relevant covering map takes the form $N_\text{c}=mw$. The parenthesis with entries $\frac{r+m}{2}\in \frac{1}{2}\ZZ/\ZZ$ and $\frac{w}{2}\in \frac{1}{2}\ZZ/\ZZ$ indicates the spin structure of the $\TT^4$ partition function. Intuitively, he parameter $w$ is the winding number of the worldsheet around the spatial cycle of the boundary torus and the parameter $m$ is the winding number around the temporal cycle.
Close to $(\Im \sigma)_\text{max}$, some covering maps are dominating the partition function. In the example of a boundary torus, these are simple to describe. For the term $(c,d)=(0,1)$, the relevant connected covering surface is the torus with modular parameter $\tau_\text{covering}=N_\text{conn} \tau_\text{bdry}$, where $N_\text{conn}$ is the degree of the (connected) covering map. This corresponds to the terms $w=1$, $r=1$ and $m=N_\text{c}$ in the eq.~\eqref{eq:torus symmetric orbifold partition function}. Taking the exponential into account, we get a collection of (possibly disconnected) dominating worldsheets for every partition $N=N_1+N_2+\dots$ of $N$. This is simply the untwisted sector of the symmetric orbifold, which describes $N$ fundamental strings that wind each once around thermal $\text{AdS}_3$, appropriately symmetrized to account for their statistics. For the other terms, different covers dominate. For instance, in the case of $(c,d)=(1,0)$, the relevant covering surface has modular parameter $\tau_\text{covering}=\frac{\tau_\text{bdry}}{N_\text{conn}}$, which is dominant in the limit $\tau_\text{bdry} \to 0$. In general, for every choice of coprime $(c,d)$ with $c+d$ odd, there is exactly one dominating connected covering surface of every degree.\footnote{The relation between $(c,d)$ and $(m,w,r)$ is in general cumbersome to describe and we omit it. But there is a very clear geometrical interpretation of the relation which is given by the rules \ref{item:rule bulk geometry 1}--\ref{item:rule bulk geometry 3} below.}

We remark that for each degree there are only finitely many covering surfaces, and thus for every finite degree (i.e.~any finite $N$), there are some classical geometries that the stringy geometries fail to distinguish. In the grand canonical ensemble, all the different geometries are resolved eventually. This qualitative feature should also hold for more complicated bulk/boundary geometries.

\paragraph{Poles from bulk geometries.} What we just observed experimentally in the symmetric orbifold is expected to hold true generally. Assume that the canonical boundary partition function is dominated by a semiclassical single bulk geometry. From the general ideas of holography, we expected that
\be 
Z_{\text{Sym}^N(\TT^4)} \sim \mathrm{e}^{-6NI_\text{bulk}}\ ,
\ee
up to an order 1 quantum correction. Here, we normalized $I_\text{bulk}$ as in Section~\ref{subsec:grand canonical ensemble string}. The explicit factor of $N$ in the exponent comes from the fact that the Newton-constant satisfies $G_\text{N} \propto N^{-1}$ as follows from the Brown-Hennaux formula \cite{Brown:1986nw}. For the grand canonical partition function, this implies that
\be 
\mathfrak{Z}_{\text{Sym}(\TT^4)} \sim \frac{1}{1-p \mathrm{e}^{-6I_\text{bulk}}}\ .
\ee
Thus, we expect the appearance of a pole at $p=\mathrm{e}^{6 I_\text{bulk}}$ (or at $\sigma=-\frac{3i}{\pi} I_\text{bulk}$).

\paragraph{Symmetric orbifold of the monster.} This behaviour was also observed in a toy model -- the symmetric product orbifold of the (holomorphic) Monster CFT \cite{deLange:2018mri}. Holomorphicity gives much stronger control in this case and in fact the grand canonical partition function is determined by Borcherds formula:
\be 
\mathfrak{Z}_{\text{Sym}(\mathbb{M})}=\frac{\mathrm{e}^{-2\pi i \sigma}}{j(\sigma)-j(\tau_\text{bdry})}\ , \label{eq:Monster partiton function}
\ee
where $j$ is the Klein $j$-invariant. In this case, condensation to thermal $\text{AdS}_3$ happens near $\sigma=\tau_\text{bdry}$. In fact, it was argued in \cite{deLange:2018mri} that this condensation is a Bose-Einstein condensation -- near $\sigma=\tau_\text{bdry}$, a finite fraction of CFTs is in the ground state. The formula has also poles whenever $\sigma$ is related to $\tau_\text{bdry}$ by a modular transformation, which can be interpreted as condensation of black hole geometries. This qualitative behaviour carries over to the non-holomorphic symmetric orbifold of $\TT^4$.

\paragraph{Analytic continuation and pole structure.} The toy model makes it clear that it is useful to consider the grand canonical partition function as a meromorphic function in $\sigma$. While the definition of the symmetric orbifold partition function \eqref{eq:symmetric orbifold partition function} only converges for $\Im \sigma>(\Im \sigma)_\text{min}$, it is easy to analytically continue in $\sigma$. This is most conveniently done using the product formula of the symmetric orbifold partition function \cite{Dijkgraaf:1996xw, Maldacena:1999bp}. Through analytic continuation, one can actually see that the grand canonical partition function has additional poles whenever
\be 
\sigma=\frac{i \Im \tau_\text{bdry}}{2M^2}\ ,
\ee
where $M$ is an odd positive integer. In fact, the grand canonical partition function has (at least) the following poles: 
\begin{align}
\sigma=\frac{i }{2M^2}  \frac{\Im \tau_{\text{bdry}}}{|c\tau_{\text{bdry}}+d|^2} \label{eq:torus partition function poles}
\end{align}
for $c+d$ is odd and $M$ odd.\footnote{These additional poles did not show up in the Monster toy example because the orbifold projection acts assymmetrically and projects them out.} 

\paragraph{Interpretation as bulk geometries.} We can interpret \emph{all} these poles as bulk geometries. Integers $M >1$ correspond to conical defect bulk geometries with deficit angle $2\pi(1-M^{-1})$. The constraint on $c+d$ and $M$ odd comes from the spin structure, since only in this case there is a bulk spin structure compatible with the given boundary spin structure \cite{Maloney:2007ud}. Thus, we argued that whenever the chemical potential approaches a pole, the tensionless strings form a condensate and the classical bulk geometry emerges.

We should note that there can be even more poles for $\sigma$ in the upper half plane.\footnote{Formally, there are even more poles in the lower half-plane. However, one can see quite easily that the poles accumulate on the rationals (this is obvious in the Monster example eq.~\eqref{eq:Monster partiton function}). Thus, it is not clear to us whether analytic continuation to the lower half-plane is sensible.} There is a pole in the grand canonical partition function for every scalar state in the CFT $\TT^4$ that satisfies $\Delta<\frac{1}{2}$. Such states may or may not be present depending on the precise shape of the torus. We will see a similar problem appearing for higher genus partition functions.
For these values of the chemical potential, there is a singularity appearing from a large number of excited tensionless strings. We do not know how to interpret these singularities from a semiclassical bulk perspective.

\begin{figure}
\begin{center}
\begin{tikzpicture}
\draw[thick,dashed,blue] (0,0) -- (4,2);
\draw[thick,blue] (4,2) -- (10,5);
\draw[domain=1.5:4, smooth, variable=\x, thick, blue] plot ({\x}, {8/\x});
\draw[domain=4:10, smooth, variable=\x, thick, dashed,blue] plot ({\x}, {8/\x});
\foreach \x in {4,4.2,...,6} {
	\draw[<-] ({\x},{.5*\x+.1}) to ({\x},{.5*\x+.3});
}
\foreach \x in {6.2,6.4,...,10} {
	\draw[<-] ({\x},{.5*\x+.1}) to ({\x},{.5*\x+.7});
}
\foreach \x in {1.6,1.8,...,3.8} {
	\draw[<-] ({\x},{8/\x+.1}) to ({\x},{8/\x+.7});
}
\node[rotate=26.5] at (5,3.05) {condensation};
\node[blue] at (11,5) {$(\Im \sigma)_\text{min}$};
\node[align=center] at (5,5) {grand canonical partition \\ function converges};
\draw[domain=0:10, smooth, variable=\x, thick, dashed,blue] plot ({\x}, {8*\x/(1^2*\x*\x+2^2*16)});
\draw[domain=0:10, smooth, variable=\x, thick, dashed,blue] plot ({\x}, {8*\x/(1^2*\x*\x+4^2*16)});
\draw[domain=0:10, smooth, variable=\x, thick, dashed,blue] plot ({\x}, {8*\x/(2^2*\x*\x+1^2*16)});
\draw[domain=0:10, smooth, variable=\x, thick, dashed,blue] plot ({\x}, {8*\x/(2^2*\x*\x+3^2*16)});
\draw[domain=0:10, smooth, variable=\x, thick, dashed,blue] plot ({\x}, {8*\x/(2^2*\x*\x+5^2*16)});
\draw[domain=0:10, smooth, variable=\x, thick, dashed,blue] plot ({\x}, {8*\x/(3^2*\x*\x+2^2*16)});
\draw[domain=0:10, smooth, variable=\x, thick, dashed,blue] plot ({\x}, {8*\x/(3^2*\x*\x+4^2*16)});
\draw[domain=0:10, smooth, variable=\x, thick, dashed,blue] plot ({\x}, {8*\x/(4^2*\x*\x+1^2*16)});
\draw[domain=0:10, smooth, variable=\x, thick, dashed,blue] plot ({\x}, {8*\x/(4^2*\x*\x+3^2*16)});
\draw[domain=0:10, smooth, variable=\x, thick, dashed,blue] plot ({\x}, {8*\x/(4^2*\x*\x+5^2*16)});
\draw[domain=0:10, smooth, variable=\x, thick, dashed,blue] plot ({\x}, {8*\x/(5^2*\x*\x+2^2*16)});
\draw[domain=0:10, smooth, variable=\x, thick, dashed,blue] plot ({\x}, {8*\x/(5^2*\x*\x+4^2*16)});

\draw[domain=.15:10, smooth, variable=\x, thick, dashed,red,samples=50] plot ({\x}, {8*\x/(1^2*\x*\x)/9});
\draw[domain=0:10, smooth, variable=\x, thick, dashed,red] plot ({\x}, {8*\x/(0^2*\x*\x+1^2*16)/9});
\draw[domain=0:10, smooth, variable=\x, thick, dashed,red] plot ({\x}, {8*\x/(1^2*\x*\x+2^2*16)/9});
\draw[domain=0:10, smooth, variable=\x, thick, dashed,red] plot ({\x}, {8*\x/(1^2*\x*\x+4^2*16)/9});
\draw[domain=0:10, smooth, variable=\x, thick, dashed,red] plot ({\x}, {8*\x/(2^2*\x*\x+1^2*16)/9});
\draw[domain=0:10, smooth, variable=\x, thick, dashed,red] plot ({\x}, {8*\x/(2^2*\x*\x+3^2*16)/9});
\draw[domain=0:10, smooth, variable=\x, thick, dashed,red] plot ({\x}, {8*\x/(2^2*\x*\x+5^2*16)/9});
\draw[domain=0:10, smooth, variable=\x, thick, dashed,red] plot ({\x}, {8*\x/(3^2*\x*\x+2^2*16)/9});
\draw[domain=0:10, smooth, variable=\x, thick, dashed,red] plot ({\x}, {8*\x/(3^2*\x*\x+4^2*16)/9});
\draw[domain=0:10, smooth, variable=\x, thick, dashed,red] plot ({\x}, {8*\x/(4^2*\x*\x+1^2*16)/9});
\draw[domain=0:10, smooth, variable=\x, thick, dashed,red] plot ({\x}, {8*\x/(4^2*\x*\x+3^2*16)/9});
\draw[domain=0:10, smooth, variable=\x, thick, dashed,red] plot ({\x}, {8*\x/(4^2*\x*\x+5^2*16)/9});
\draw[domain=0:10, smooth, variable=\x, thick, dashed,red] plot ({\x}, {8*\x/(5^2*\x*\x+2^2*16)/9});
\draw[domain=0:10, smooth, variable=\x, thick, dashed,red] plot ({\x}, {8*\x/(5^2*\x*\x+4^2*16)/9});
\draw[thick,->] (0,0) to (11,0) node[above] {$\Im \tau_{\text{bdry}}$};
\draw[thick,->] (0,0) to (0,6) node[above] {$\Im \sigma$};
\end{tikzpicture}
\end{center}
\caption{The behaviour of the grand canonical torus partition function for purely imaginary $\tau_\text{bdry}$. Blue lines correspond to poles with $M=1$ and red lines to poles with $M=3$.} \label{fig:torus phase diagram}
\end{figure}
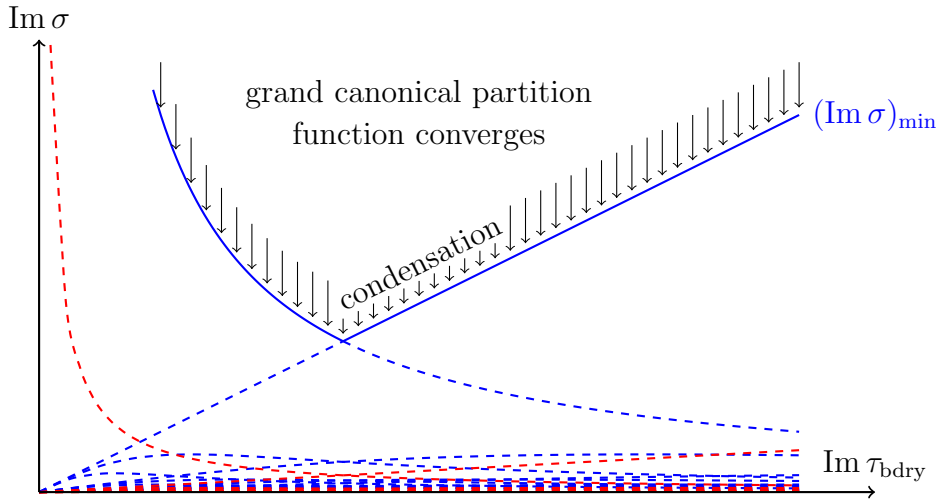

The phase diagram of the torus partition function for purely imaginary $\tau_\text{bdry}$ takes the form as depicted in Figure~\ref{fig:torus phase diagram}. Near the edge, stringy geometry condenses into classical geometry. One can analytically continue below the solid blue line that represents $(\Im \sigma)_\text{min}$ to observe also the other condensing geometries.

\paragraph{Residue.} The residue of the grand canonical potential can be interpreted as the infinite $N$ stringy one-loop determinant around the respective background. Let us focus on the thermal $\text{AdS}_3$ pole at $\sigma=\frac{i}{2} \Im \tau_\text{bdry}$. Writing
\be 
Z^{\TT^4}(\tau_\text{bdry})=\mathrm{e}^{-\pi \Im \tau_\text{bdry}} Z_\text{qu}^{\TT^4}(\tau_\text{bdry})\ ,
\ee
we separate out the classical part of the $\TT^4$ partition function. The residue takes the form
\begin{multline} 
\Res_{\sigma=\frac{i}{2} \Im \tau_\text{bdry}} \mathfrak{Z}_{\text{Sym}(\TT^4)}^{\text{NS}} \\
= \exp \left(\sum_{m,\, w=1}^\infty\!\!{}^{'} \sum_{r\in \ZZ/\ZZ w}\frac{\mathrm{e}^{-\pi m\left(w-\frac{1}{w}\right)\Im \tau_\text{bdry}}}{mw} Z^{\TT^4}_\text{qu}\begin{bmatrix}
\frac{r+m}{2} \\ \frac{w}{2}
\end{bmatrix} \left(\frac{m\tau_\text{bdry}+r}{w}\right)\right)\ .
\end{multline}
The prime signifies that the summation over $m$ for the divergent term is omitted -- this is the term with $w=1$ and the leading term in $Z_\text{qu}^{\TT^4}$ coming from the vacuum. This is the one-loop determinant of the symmetric orbifold $\text{Sym}^\infty(\TT^4)$. It was computed in this form from string theory in \cite{Eberhardt:2018ouy}. Aside from possible convergence issues that we have not analyzed, we can reconstruct the full grand canonical partition function from the infinite $N$ one-loop determinants:
\be 
\mathfrak{Z}^{\text{NS}}_{\text{Sym}(\TT^4)}=\sum_{\text{bulk geometries }\mathcal{M}_3} \frac{Z_{\text{qu}}(\mathcal{M}_3)}{\sigma+\frac{3i}{\pi} I_\text{bulk}(\mathcal{M}_3)}+\text{poles from light scalars in $\TT^4$}\ , \label{eq:farey tail sum}
\ee
where $Z_{\text{qu}}(\mathcal{M}_3)$ is the suitable one-loop determinant for the respective bulk geometry. While this is under good control for a torus boundary, we know much less about the analytic behaviour for higher genus boundaries. There could be branch cuts, higher order poles etc. 
By taking suitable contour integrals, one may also reconstruct the finite $N$ canonical partition function.\footnote{For the symmetric orbifold of the Monster CFT, this was already discussed in \cite{deLange:2018mri}.} We should note that this `Farey tail' relies on holomorphicity in $\sigma$ and not on holomorphicity in the modular parameter and thus does not need to assume holomorphicity of the partition function or the elliptic genus \cite{Dijkgraaf:2000fq, Manschot:2007ha}. 

\paragraph{Rules for the bulk geometry.} Given that the grand canonical ensemble torus partition function can be interpreted as a sum over bulk geometry, we may try to come up with a general rule on how to reconstruct the bulk geometry/topology from the respective family of covering maps. One can analyze the geometry of the covering maps that leads to the poles \eqref{eq:torus partition function poles}.
The relation between condensing worldsheets and bulk geometries is in this case the following:
\begin{enumerate}
\item If every condensing worldsheet wraps a boundary cycle only once, then this boundary cycle is contractible in the emerging classical geometry. \label{item:rule bulk geometry 1}
\item If every condensing worldsheet wraps a boundary cycle exactly $M$ times, then this boundary cycle is bounded by a disk in the bulk with a conical defect with deficit angle $2\pi(1-M^{-1})$. \label{item:rule bulk geometry 2}
\item For a non-contractible cycle in the emerging bulk geometry, there is no bound in the condensate for how many times the cycle is wrapped. \label{item:rule bulk geometry 3}
\end{enumerate}
The second rule can be motivated from the first, since a string that winds $M$ times around a conical deficit with deficit angle $2\pi(1-M^{-1})$ can be viewed by unwrapping the geometry as a string that winds once around the same geometry without deficit angle. 

These rules make it geometrically clear that there is exactly one connected covering map of every degree contributing for the non-singular geometries and one covering map whose degree is a multiple of $M$ for the conical defect geometries. There is a last geometry that we have not discussed: the cusp geometry which we mentioned in Section~\ref{subsec:examples}. It has a somewhat singular role, since it can be understood as the $M \to \infty$ limit of the conical defects. This means that the potential pole it causes in the grand canonical partition function is $\sigma=0$, which is the accumulation point of poles of the conical defects. Thus, it is a delicate issue whether it should be included in the sum over bulk geometries. Since it is obtained as the limit $M \to \infty$ of the conical defects, our rules seem to suggest that is not associated to any stringy geometry of covering maps.

\paragraph{Higher genus boundary.} We expect the behaviour of the higher genus partition functions to be similar, but there are some important qualitative differences. In the higher genus case, the number of connected covering spaces is (super)exponentially large in the degree $N_\text{c}$, compared to roughly $\mathcal{O}(N_\text{c})$ connected covering spaces in the case of a single torus boundary.\footnote{The number of connected covering spaces of degree $N_\text{c}$ is actually $\sigma_1(N_\text{c})$, the divisor function, for a torus boundary.} It is explained in Appendix~\ref{app:subgroups} that there are $\mathcal{O}(2N_\text{c}(N_\text{c}!)^{2g-2})$ connected covering surfaces of degree $N_\text{c}$ if $g\ge 2$ and $N_\text{c}\gg 1$. This enormous qualitative difference is also reflected in the classical geometries: for genus $g \ge 2$, there are many more possible bulk manifolds -- not only handlebodies, but also non-handlebody geometries. 

Correspondingly, the expected pole structure of the grand canonical partition function and the function $(\Im \sigma)_\text{min}$ is much more complicated. Besides the conformal structure, it also depends on the metric on the boundary surface due to the conformal anomaly. Under a Weyl transformation, $(\Im \sigma)_\text{min}$ transforms as
\be 
(\Im \sigma)_\text{min} \longmapsto (\Im \sigma)_\text{min}-12\pi S[\phi]\ ,
\ee
where $S[\phi]$ is the Liouville action capturing the conformal anomaly of the Weyl transformation. This ambiguity does not affect the behaviour of the poles and for definiteness, we might hence fix the hyperbolic metric on the boundary Riemann surface. 

The natural guess for the function $(\Im \sigma)_\text{min}$ is now
\be 
(\Im \sigma)_\text{min} = \frac{3i}{\pi}\max_{\begin{subarray}{c} \text{(singular) hyperbolic 3-manifolds compatible} \\ \text{with boundary spin structure }\mathcal{M}_3 \end{subarray}} I_\text{bulk}(\mathcal{M}_3)\ , \label{eq:maximum bulk manifolds}
\ee
where the maximum is taken over all possibly singular hyperbolic three-manifolds with the correct conformal boundary. It was claimed that this is not true for higher genus partition functions if there is a light enough scalar in the spectrum, which happens e.g.~for a very large or small $\TT^4$ \cite{Keller:2014xba}. Such a condensing bulk geometry would no longer have the form $\text{AdS}_3/\Gamma \times \mathrm{S}^3 \times \TT^4$, but would also involve the internal factors non-trivially. This is similar to what we observed for the pole structure of the torus partition function, except that the `excited geometries' originating from the $\TT^4$ excitations can conceivably even be dominant in the higher genus case.

It is plausible that all the handlebodies themselves can be realized as particular stringy geometries. In fact, there seems to be a natural class of covering maps associated to every handlebody. Recall that a handlebody is specified by a choice of Lagrangian sublattice of $\mathrm{H}_1(\Sigma_G,\ZZ)$ in the boundary, corresponding to the homology cycles that become contractible in the bulk. Let us consider the Lagrangian sublattice generated by the standard generators $\langle \beta_1,\dots,\beta_G \rangle$, which would be the higher genus analogue of thermal $\text{AdS}_3$. In this case, there is a natural class of covering surfaces that are candidates to form a bulk condensate. In general, degree $N$ covering surfaces are specified by homomorphisms $\phi:\pi_1(\Sigma_G) \longrightarrow S_N$, see the discussion in Appendix~\ref{subapp:homomorphisms SN}. We propose to consider homomorphisms for which $\phi(\alpha_I)=\id$, the identity permutation. The resulting covering maps hence have only their $\beta_I$-cycles unwrapped. This is motivated by the rules \eqref{item:rule bulk geometry 1}--\eqref{item:rule bulk geometry 3} above that we observed for a torus boundary. We do not know whether $\phi(\beta_I)$ should satisfy further constraints in order to contribute to the semiclassical geometry. This assignment is also motivated by the fact that the $\beta_I$ cycles are the analogue of the thermal cycle in the torus case.
If we instead want to create a conical defect for the $\alpha_1$ cycle, say, then we would require that $\phi(\alpha_1)$ is some fixed cyclic permutation of length $M$.

It is then plausible that also other non-handlebody solutions geometries lead to poles in the grand canonical partition function. There are however several issues that are unclear to us:
\begin{enumerate}
\item We do not know whether there exists a well-defined analytic continuation. There could be more exotic phenomena for higher genus boundaries such as branch cuts etc. 
\item In the case of a genus $G \ne 1$ boundary, we can add a counter term $\frac{1}{4\pi}\int \mathcal{R} \sqrt{g}=2-2G$ to the boundary action, where $g$ is the boundary metric. This has the effect of renormalizing the partition function. At least at the face of it, we can choose these counterterms independently for all $N$ in the canonical ensemble. Changing these counterterms can change all the locations of the poles in the grand canonical partition function simultaneously and it can also change their order. 
\item We saw in the case of the torus partition function that the cusp geometry is more singular than conical defects, since it would be located at accumulation points of the chemical potential. Such accumulation points should also exist for the higher genus boundaries. At least some of these correspond to analogues of cusp geometries (that we define as geometries with codimension 3 singularities).
\item We do not know the fate of codimension 1 singular bulk geometries. A particularly important codimension 1 singular bulk geometry is obtained by orbifolding the Fuchsian wormhole by the reflection symmetry. The geometry is described in more detail in Appendix~\ref{subapp:Kleinian groups examples}. If such a geometry were to emerge from stringy covering maps, it would need to do so in a way that treats all cycles symmetrically.\footnote{This particular bulk geometry is non-orientable in the sense that the reflection symmetry that we use in the orbifold reverses orientation. Thus, the geometry might not appear for this reason.}
\item Some geometries only exist for specific choices of the boundary moduli. For example, if the boundary surface has an involution without fixed points, then we can consider the quotient of the Fuchsian wormhole by the combined reflection and involution. 
If the correspondence between bulk geometries and poles in the grand canonical partition function holds, then it seems that there must be new `exceptional' poles for these special values of the moduli. 
\end{enumerate}

\subsection{The wormhole} \label{subsec:wormhole}
We have seen that at least in some cases, one can convincingly identify geometries with a single boundary as a family of covering maps. It is an important problem to extend this to geometries with disconnected boundaries. In this case, there are two chemical potentials $\sigma_1$ and $\sigma_2$. 
Of course, the purpose of our discussion of the tensionless string was to establish that the grand canonical partition function factorizes, both from a bulk and a boundary perspective and the disconnected grand canonical partition function simply takes the form $\mathfrak{Z}^{(1)}(\sigma_1) \mathfrak{Z}^{(2)}(\sigma_2)$, where $\mathfrak{Z}^{(i)}$ are the grand canonical partition functions of the two boundaries.
In view of the above discussion the poles of this product grand canonical potential in $(\sigma_1,\sigma_2)$ are obviously just a reflection of all the disconnected geometries. 

We now speculate that even though the grand canonical partition function factorizes, it might still contain the information about all the classical wormholes. To see wormholes, it is much more natural to change the basis of the chemical potentials and define
\be 
\sigma_1=\frac{\sigma}{2}+\theta\ , \qquad \sigma_2=\frac{\sigma}{2}-\theta\ ,
\ee
so that $\sigma$ is the `diagonal' chemical potential and $\theta$ is the `off-diagonal' chemical potential.
We now want to `entangle' the left and right boundaries, since this should create a correlation between the two boundaries. We view this as a version of a stringy ER=EPR \cite{Maldacena:2013xja}.
 In the present instance, this is naturally done by integrating over $\theta$. From the definition of the grand canonical ensemble, this has the following effect:
\begin{align}
\widetilde{\mathfrak{Z}}_{\text{Sym}(\TT^4)}(\sigma)&=\int_{0}^{1} \mathrm{d}\theta\ \mathfrak{Z}^{(1)}_{\text{Sym}(\TT^4)}\left(\frac{\sigma}{2}+\theta\right)\mathfrak{Z}^{(2)}_{\text{Sym}(\TT^4)}\left(\frac{\sigma}{2}-\theta\right)\\
&=\sum_{N_1=0,\, N_2=0}^\infty \int_{0}^{1}\mathrm{d}\theta\  Z^{(1)}_{\text{Sym}^{N_1}(\TT^4)}Z^{(2)}_{\text{Sym}^{N_2}(\TT^4)} \mathrm{e}^{\pi i \sigma(N_1+N_2)} \mathrm{e}^{2\pi i \theta(N_1-N_2)} \\
&=\sum_{N=0}^\infty  Z^{(1)}_{\text{Sym}^{N}(\TT^4)}Z^{(2)}_{\text{Sym}^{N}(\TT^4)} \mathrm{e}^{2\pi i \sigma N} \ .
\end{align}
For this to make sense, we want that 
\be 
\Im \sigma>2(\Im \sigma)_\text{min}^{(1)}\, , \ 2(\Im \sigma)_\text{min}^{(2)}\ ,
\ee
so that the series converges uniformly and we are allowed to interchange the sum and the integral. Here, the superscripts ${}^{(1)}$ and ${}^{(2)}$ refer as usual to the left and right boundary. 
The resulting expression still has poles for every disconnected geometry that lead to a pole earlier. The location of the pole in $\sigma$ is $\sigma^{(1)}_\text{p}+\sigma^{(2)}_\text{p}$, where $\sigma^{(1)}_\text{p}$ and $\sigma^{(2)}_\text{p}$ are two poles of $\mathfrak{Z}^{(1)}_{\text{Sym}(\TT^4)}$ and $\mathfrak{Z}^{(2)}_{\text{Sym}(\TT^4)}$, respectively. Thus, these geometries are naturally still interpreted as the disconnected geometries with bulk action 
\be 
I_\text{bulk}(\mathcal{M}^{(1)}_3 \sqcup \mathcal{M}^{(2)}_3)=I_\text{bulk}(\mathcal{M}^{(1)}_3)+I_\text{bulk}(\mathcal{M}^{(2)}_3)=\sigma^{(1)}_\text{p}+\sigma^{(2)}_\text{p}\ .
\ee
However, it could be possible that $\widetilde{\mathfrak{Z}}_{\text{Sym}(\TT^4)}(\sigma)$ has poles that are not associated to any of the disconnected geometries. For example, if $\mathfrak{Z}_{\text{Sym}(\TT^4)}^{(1,2)}$ has branch cuts, then the integral over $\theta$ leads to an integral around these branch cuts, which is not associated with any pole and hence no disconnected geometry. This does \emph{not} seem to happen for the torus partition function, because it only has simple poles. But this is expected, since there is no connected bulk geometry with two torus boundaries.\footnote{More precisely, there is no hyperbolic 3-manifold that has two tori as conformal boundaries. This is quite simple to see from a realization of the hyperbolic manifold from a Kleinian group, see e.g.~\cite{Eberhardt:2020bgq} for an explanation.} It seems a very difficult problem to explicitly compute $\mathfrak{Z}_{\text{Sym}(\TT^4)}^{(1)}$ and $\mathfrak{Z}_{\text{Sym}(\TT^4)}^{(2)}$ for genus 2 boundaries, do the analytic continuation and understand the pole structure. Hence we have not been able to determine whether wormhole geometries can actually appear in the grand canonical partition function.

\subsection{An ensemble average}\label{subsec:ensemble average}
One could introduce an ensemble average in  the symmetric orbifold.\footnote{I thank Alex Belin for discussion about this.} The presumably easiest way would be by averaging over the Narain moduli space of $\TT^4$, as was done in \cite{Maloney:2020nni, Afkhami-Jeddi:2020ezh} for a bosonic sigma-model on $\TT^4$. This is problematic to do for the partition function of the symmetric orbifold itself, because the average does not converge.\footnote{This is because averages $\langle Z_{\TT^4}^{G_1}\cdots Z_{\TT^4}^{G_n}\rangle$ for a single $\TT^4$ sigma model only converge for $\sum_i G_i \le 2$, where $G_i$ denotes as usual the genus of the surface.}
One can get a convergent result for the logarithm of the torus partition function. This is essentially the \emph{quenched} free energy \cite{Engelhardt:2020qpv} (in contrast to the \emph{annealed} free energy, where one first averages and then takes a logarithm). For a single torus boundary, this quantity takes the form
\be 
\overline{  \log \mathfrak{Z}_{\text{Sym}(\TT^4)}} =\sum_{m,\, w=1}^\infty\sum_{r \in \ZZ/\ZZ w} \frac{p^{mw}}{mw} \ \overline{ Z_{\TT^4}\begin{bmatrix}
\frac{r+m}{2} \\ \frac{w}{2}
\end{bmatrix}\left(\frac{m \tau_\text{bdry}+r}{w} \right)}\ ,
\ee
where overline denotes the ensemble average. The averaged $\TT^4$ partition function can be interpreted as a sum over bulk geometries. These bulk geometries are geometries that fill in the covering space and not the original boundary of $\AdS_3$. We have learned that the covering space is to be interpreted as the worldsheet. Thus, averaged connected partition function can be interpreted as a sum over all fillings of all connected worldsheets. Disregarding convergence issues, the same is true for $\mathfrak{Z}_{\text{Sym}(\TT^4)}$ itself: it can be interpreted as a sum over all possible geometries (in the sense of $\U(1)$-gravity \cite{Maloney:2020nni}) that fills in the possibly disconnected worldsheet.

Such geometries are topological spaces, where one is allowed to make additional identifications on the boundary to obtain the correct boundary manifold. For example, the totally disconnected contribution in thermal $\AdS_3$ would be the space
\be 
\bigsqcup_{i=1}^N \text{ thermal $\AdS_3^{(i)}$} \Big/ \sim\ , \label{eq:thermal AdS3 boundary identification}
\ee
where $\sim$ identifies all the boundaries of the thermal $\AdS_3$ spaces.\footnote{The reader might feel uneasy about this identification. We perform this identification mainly for the physical interpretation's sake, but it doesn't change any physical quantity that we are computing.}   The number of copies is fixed to $N$ in the canonical ensemble, but is again arbitrary in the grand canonical ensemble. We emphasize however that there are a lot of possible geometries of this sort. We could replace some of the thermal \eqref{eq:thermal AdS3 boundary identification} by black holes or could replace pairs (or higher tuples) of $\AdS_3$ spaces by wormholes, etc. We could also replace a pair of thermal $\AdS_3$'s by one thermal $\AdS_3$ whose length of the thermal cycle is twice as large, but which has an additional identification on its boundary torus.

\begin{figure}
\begin{minipage}{.49\textwidth}
\begin{center}
\includegraphics[width=\textwidth]{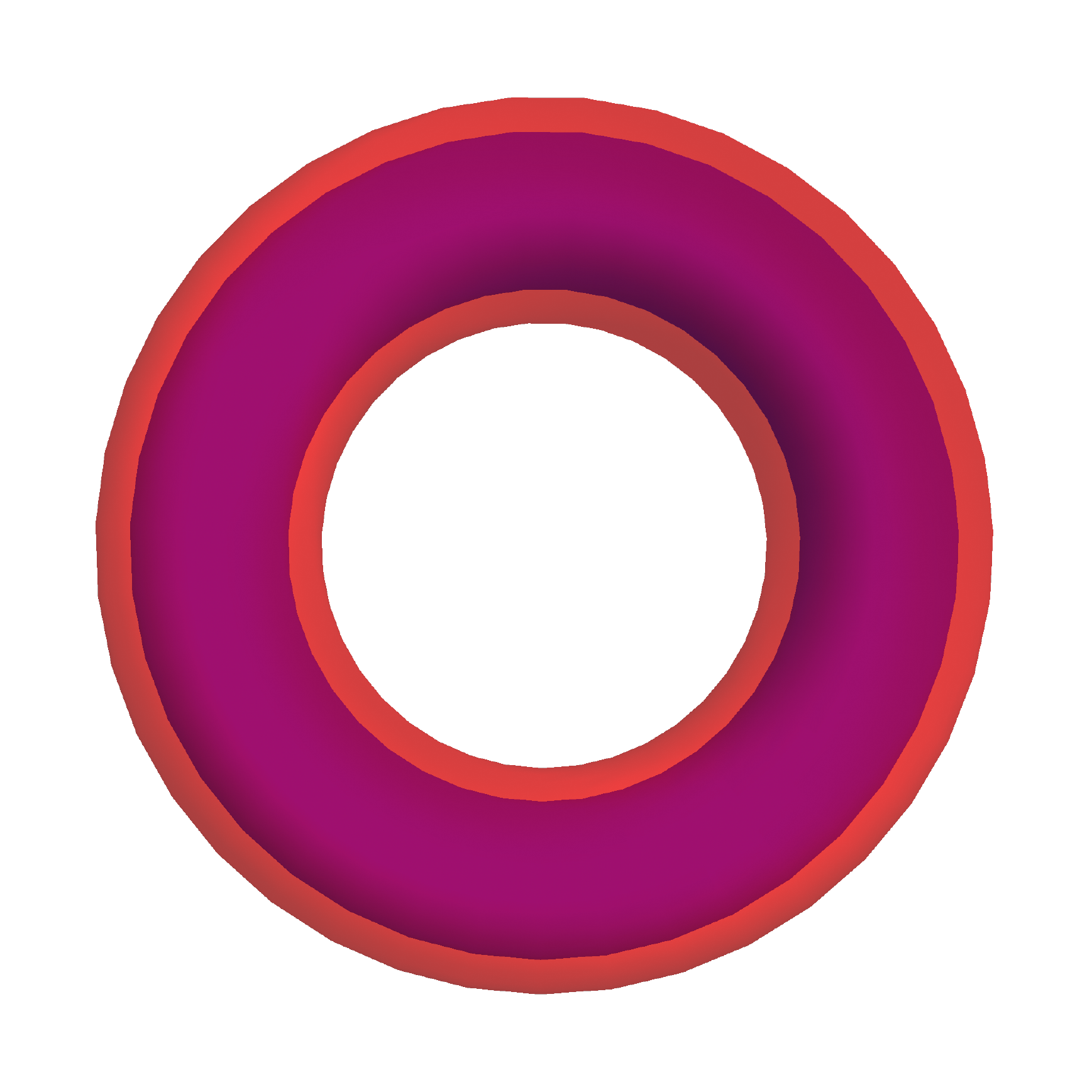}
\end{center}
\end{minipage}
\begin{minipage}{.49\textwidth}
\begin{center}
\includegraphics[width=\textwidth]{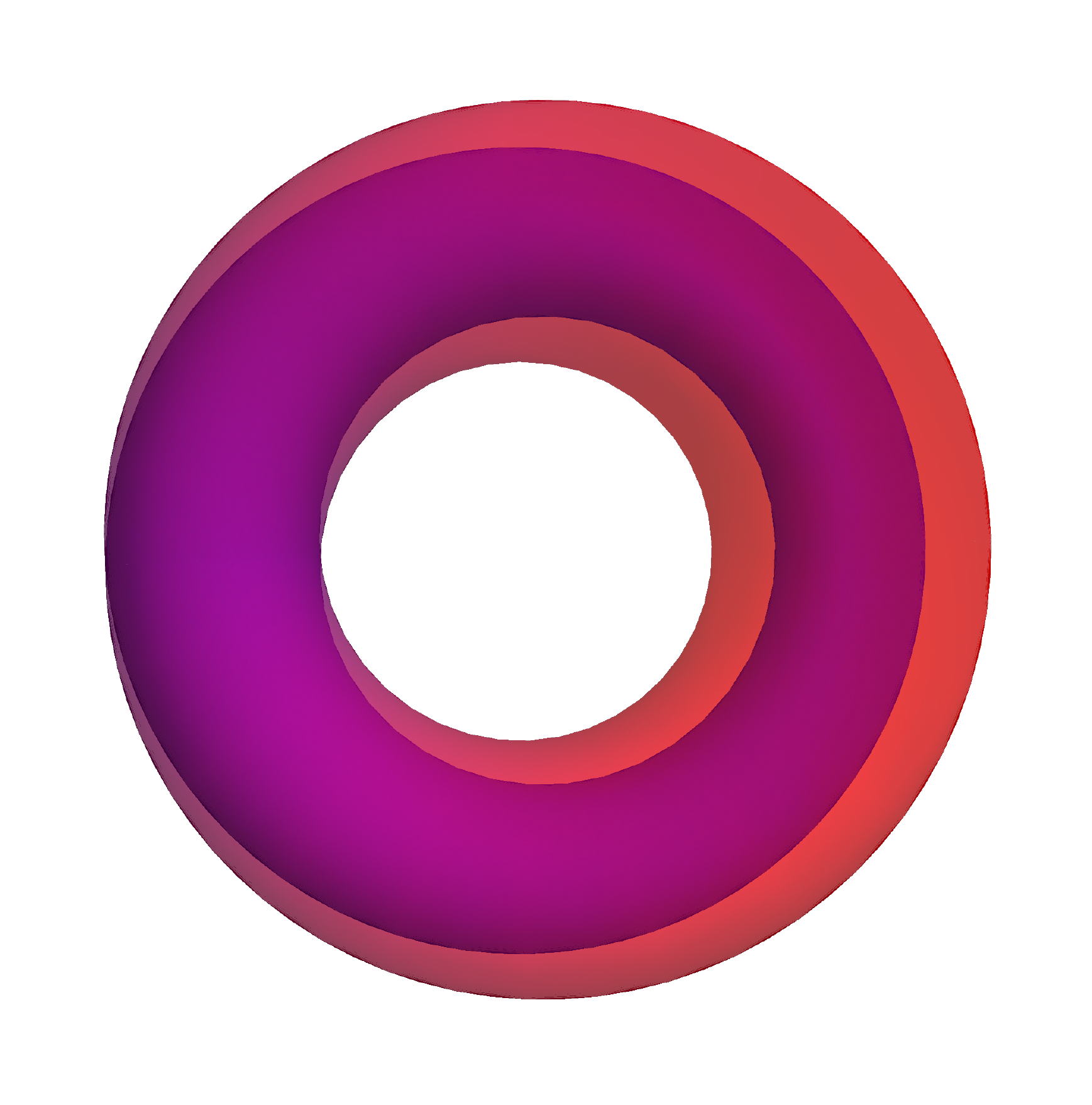}
\end{center}
\end{minipage}
\caption{The two averaged stringy geometries that condense to classical thermal $\AdS_3$ for a degree two covering surface. In this figure, the boundaries of the inner and outer torus are identified. } \label{fig:condensing N2}
\end{figure}

\begin{figure}
\begin{minipage}{.49\textwidth}
\begin{center}
\includegraphics[width=\textwidth]{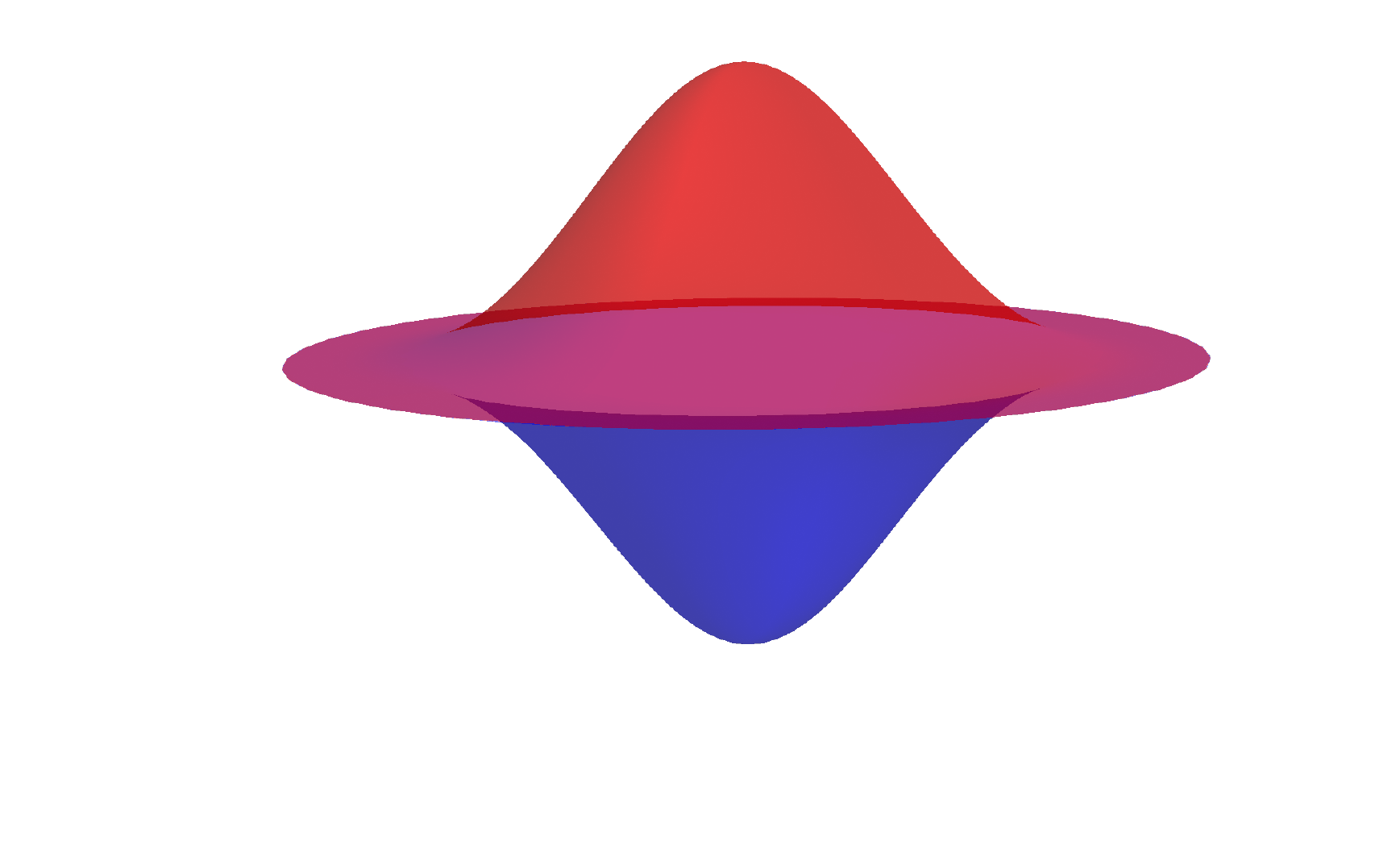}
\end{center}
\end{minipage}
\begin{minipage}{.49\textwidth}
\begin{center}
\includegraphics[width=\textwidth]{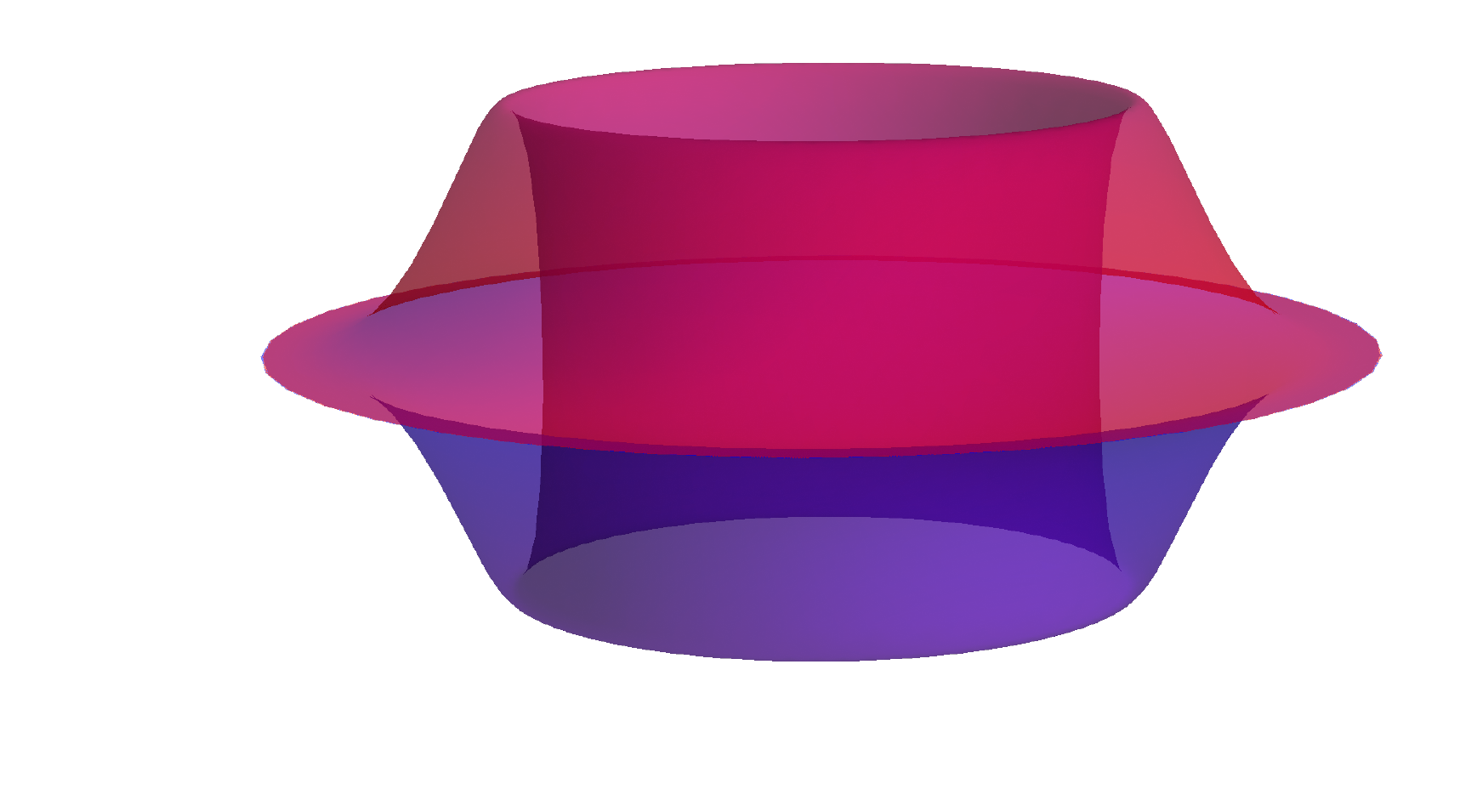}
\end{center}
\end{minipage}
\caption{The $t=0$ surfaces of some string geometries with two sheets. The boundary is the spatial boundary circle of global $\AdS_3$. } \label{fig:t0 surface}
\end{figure}

\begin{figure}
\begin{center}
\includegraphics[width=.5\textwidth]{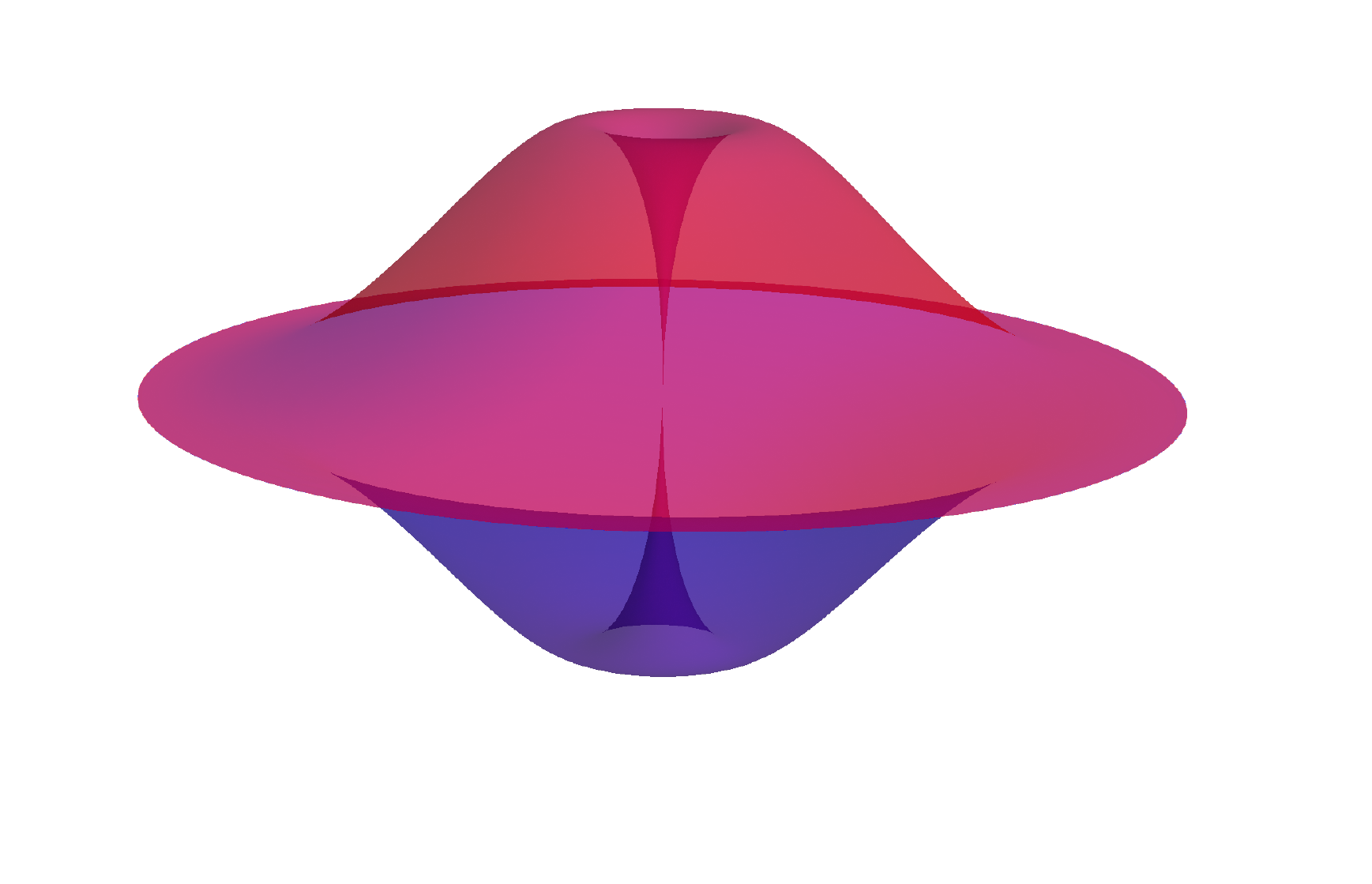}
\end{center}
\caption{The $t=0$ surface of the stringy conical deficit geometry with $M=2$. This geometry is obtained from the disc $\{z\in \CC\,|\, |z|\le 1\}$ by the identification $z \sim -z$ on the boundary. }  \label{fig:conical defect stringy geometry}
\end{figure}

In this averaged form, condensation of such averaged stringy geometries into classical geometries becomes much clearer. For example, \eqref{eq:thermal AdS3 boundary identification} is one geometry that for large $N$ resembles more and more thermal $\AdS_3$ itself, because the different copies of thermal $\AdS_3$ start to form a continuum. For example, the two condensing stringy geometries for a boundary torus and $N=2$ are depicted in Figure~\ref{fig:condensing N2}.
It might also help if we picture only the $t=0$ slices of the geometries in Figure~\ref{fig:condensing N2}. For this we cut the torus open and look only at the cut. This yields the first picture in Figure~\ref{fig:t0 surface}. The difference between the two geometries is invisible on the level of the $t=0$ slice, their difference lies in whether the blue and red sheets are interchanged when moving around the thermal circle or not. In the ensemble average picture, there is also a wormhole geometry that we did not picture in Figure~\ref{fig:condensing N2}, but whose $t=0$ surface is the second picture in Figure~\ref{fig:t0 surface}.
This by far does not exhaust the list of possible stringy geometries with two sheets. The conical defect geometry is pictured in Figure~\ref{fig:conical defect stringy geometry}. 
There are also other geometries that do not exhibit a clear Lorentzian interpretation. For example, $\text{thermal AdS}_3 \sqcup \text{Euclidean BTZ}$ with identified boundaries is such a case. This geometry can be pictured geometrically as two interlocking solid tori, which is in fact a description of the three-sphere as a Heegaard splitting.
We discuss some further speculations on these geometries in the Discussion~\ref{subsec:discussion}.

\subsection{Deforming the symmetric orbifold}
In the previous subsection, we explored ensemble averaging as a way  to let the semiclassical bulk geometry emerge semiclassically. The averaging process turns on an `interaction' between the different covering spaces, since it allows for wormholes connecting them. However, there is a more traditional way of turning on such an interaction, namely by deforming the symmetric orbifold away from its orbifold point. We shall see that the two procedures are in some ways similar.
A large deformation in the moduli space of the symmetric orbifold is expected to connect the tensionless string to the supergravity regime, where gravity can be treated semiclassically \cite{Seiberg:1999xz}. This means in particular that the deformation will need to connect the `stringy' gravity picture that we discussed smoothly with a semiclassical picture.

\paragraph{Marginal operator.} There is a marginal operator in the theory that turns on the deformation. Its construction is as follows. The twist field of a transposition has conformal weight $h=\frac{6}{24} \frac{2^2-1}{2}+\frac{4}{2} \cdot \frac{1}{16}=\frac{1}{2}$. The first contribution is the standard ground state energy of the twist 2 sector. The second term is the ground state energy of the Ramond sector of four free fermions, since the relevant spin structure in evenly twisted sectors is the Ramond sector. The ground states transform in the spinor representation of $\mathfrak{so}(4)$ that decomposes under the R-symmetry subgroup $\mathfrak{su}(2)$ as $\mathbf{2} \oplus 2 \cdot \mathbf{1}$. In particular, the highest weight state of the doublet is a BPS state of the $\mathcal{N}=4$ algebra. Thus it gives rise to a marginal operator (actually four marginal operators) by taking the level $\frac{1}{2}$ superconformal descendants. 
Let $\Psi^{\alpha\dot{\alpha}}$ be the BPS field (where $\alpha$ and $\dot{\alpha}$ are spinor indices of the left- and right-moving R-symmetry), then the marginal operator is given by
\be 
\Phi^{A\dot{A}}=\varepsilon_{\alpha\beta}\varepsilon_{\dot{\alpha}\dot{\beta}} G^{\beta A}_{-\frac{1}{2}}\bar{G}^{\dot{\beta} \dot{A}}_{-\frac{1}{2}}\Psi^{\alpha\dot{\alpha}}
\ee
Here, capital indices are spinor indices of the outer automorphism group of the $\mathcal{N}=4$ superconformal algebra.
\paragraph{Conformal perturbation theory.} We now imagine to compute the symmetric orbifold partition function perturbatively in the deformation parameter. The perturbed partition function takes the form\footnote{We evade the question on how to correctly normalize $\Phi$ in the grand canonical ensemble.}
\be 
\mathfrak{Z}_{\text{Sym}(\TT^4)}^{(\lambda)}=\left\langle \mathrm{e}^{-\lambda \int_\Sigma \Phi} \right \rangle=\sum_{n=0}^\infty \frac{(-\lambda)^n}{n!}\left \langle \left(\int_\Sigma \Phi \right)^n\right \rangle\ ,
\ee
where we take for definiteness $\Phi\equiv \varepsilon_{A \dot{A}}\Phi^{A\dot{A}}$ the singlet with respect to the diagonal automorphism group.\footnote{The following discussion is independent of this choice.} Thus, in perturbation theory, we need to compute correlation functions of twist 2 operators on the torus. They are as usual given by summing over all covering surfaces of the torus, but now these covering surfaces are branched over $n$ points at $n$-th order in perturbation theory. The branch points are simple, i.e.\ of order 2. Odd orders in the perturbation theory actually vanish. This can be seen geometrically from the Riemann-Hurwitz formula \eqref{eq:Riemann Hurwitz}, since for a branched cover the sum $\sum_i (w_i-1)$ over ramification indices has to be even. 

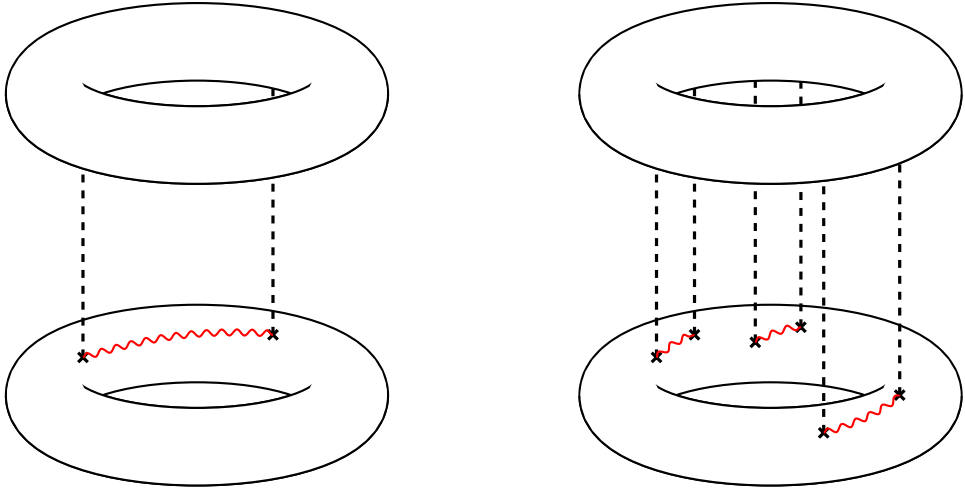
\begin{figure}
\begin{minipage}{.49\textwidth}
\begin{center}
\begin{tikzpicture}
\pic[thick] at (0,-2) {torus={2cm}{5mm}{70}};
\draw[very thick,dashed] (-1.5,-1.5)  to (-1.5,2.5);
\draw[very thick,dashed] (1,-1.2)  to (1,2.8);
\draw[very thick] (-1.5,-1.5) node[cross=3pt] {};
\draw[very thick] (1,-1.2) node[cross=3pt] {};
\draw[red,thick,decorate, decoration={snake,amplitude=.4mm,segment length=2mm},bend left=10] (-1.5,-1.5) to (1,-1.2);
\pic[thick] at (0,2) {torus={2cm}{5mm}{70}};
\end{tikzpicture}
\end{center}
\end{minipage}
\begin{minipage}{.49\textwidth}
\begin{center}
\begin{tikzpicture}
\pic[thick] at (0,-2) {torus={2cm}{5mm}{70}};
\draw[very thick,dashed] (-1.5,-1.5)  to (-1.5,2.5);
\draw[very thick,dashed] (-1,-1.2)  to (-1,2.8);
\draw[very thick,dashed] (-.2,-1.3)  to (-.2,2.7);
\draw[very thick,dashed] (.4,-1.1)  to (.4,2.9);
\draw[very thick,dashed] (.7,-2.5)  to (.7,1.5);
\draw[very thick,dashed] (1.7,-2)  to (1.7,2);
\draw[very thick] (-1.5,-1.5) node[cross=3pt] {};
\draw[very thick] (-1,-1.2) node[cross=3pt] {};
\draw[very thick] (-0.2,-1.3) node[cross=3pt] {};
\draw[very thick] (0.4,-1.1) node[cross=3pt] {};
\draw[very thick] (0.7,-2.5) node[cross=3pt] {};
\draw[very thick] (1.7,-2) node[cross=3pt] {};
\pic[thick] at (0,2) {torus={2cm}{5mm}{70}};
\draw[red,thick,decorate, decoration={snake,amplitude=.4mm,segment length=2mm},bend left=10] (-1.5,-1.5) to (-1,-1.2);
\draw[red,thick,decorate, decoration={snake,amplitude=.4mm,segment length=2mm},bend left=10] (-.2,-1.3) to (.4,-1.1);
\draw[red,thick,decorate, decoration={snake,amplitude=.4mm,segment length=2mm},bend right=10] (.7,-2.5) to (1.7,-2);
\end{tikzpicture}
\end{center}
\end{minipage}
\caption{Degree 2 branched covering maps appearing in conformal perturbation away from the symmetric product orbifold. The left figure shows the surface that appears at second order in conformal perturbation theory. The right surface appears at higher order in perturbation theory.} \label{fig:branched torus}
\end{figure}

\paragraph{Interactions between covering surfaces.} Inclusion of twist fields in correlators effectly turns on an interaction of the different covering spaces. Let us explain this at the simplest example of a symmetric orbifold with two copies. In the undeformed theory, there are four covering spaces in total of which one is disconnected. At second order in conformal perturbation theory, covering surfaces cover the torus holomorphically with two ramification points of order 2. Such a covering surface is of course connected. Schematically, we can write such a surface as in Figure~\ref{fig:branched torus}.
Here, the dashed lines correspond to the branch points of the covering and in between them are branch cuts.\footnote{It follows from the Burnside formula for covering maps that there are always four covering maps at every even order in perturbation theory. The surfaces depicted in Figure~\ref{fig:branched torus} are those originating from the disconnected covering surface in the undeformed symmetric orbifold. There are also surfaces originating from the connected covering surfaces.}

\paragraph{Emergence of semi-classical bulk geometry.} We now offer a speculative picture on how the semi-classical bulk geometry emerges at very high order in conformal perturbation theory.
As we just saw, the different sheets of the stringy geometry are independent in the symmetric orbifold point, but become more and more correlated as we deform the theory. Because of the independence of the covering surface at the tensionless point in moduli space, there are also independent bulk geometries for all the different covering surfaces. These are the multi-sheeted geometries that we constructed in the previous subsection. As we turn on the interaction between the covering surfaces, we should also turn on an interaction between the different sheets of the geometry.\footnote{Such an interaction is necessarily non-local from the point of view of the `micro-geometries', but can be local from the point of view of the emergent semi-classical geometry.} 
It is plausible that configurations where the sheets of the micro-geometries align have a much lower action once this interaction is turned on. Once we make the interaction strong enough, only those bulk geometries with aligning sheets survive and form semi-classical bulk geometries.

\section{Summary and discussion} \label{sec:summary discussion}
\subsection{Summary} \label{subsec:summary}
Let us summarize the most important points of the paper. We put the fine print in footnotes.
\begin{enumerate}
\item The worldsheet theory of the tensionless string in a background $\mathcal{M}_3 \times \text{S}^3 \times \TT^4$, where $\mathcal{M}_3$ is a hyperbolic 3-manifold, localizes on covering surfaces of the boundary Riemann surface(s) $\Sigma^{(1)} \sqcup \cdots \sqcup \Sigma^{(n)}$.\footnote{We have established that the worldsheet theory localizes on a discrete subset of the moduli space of Riemann surfaces. Covering surfaces are contained in this subset, but we have not been able to show equality of the two sets. In establishing these facts, we emphasized the role of a complex projective structure that emerges naturally from the worldsheet theory (branched complex projective structure in the case of correlation functions).}
\item We have argued that the string partition function on different hyperbolic 3-manifolds with the same boundary Riemann surface(s) $\Sigma^{(1)} \sqcup \cdots \sqcup \Sigma^{(n)}$ agree to all orders in string perturbation theory.\footnote{We realized $\mathcal{M}_3=\HH^3/\Gamma$ as a quotient of global hyperbolic space. Different bulk manifolds correspond to different orbifold groups $\Gamma$ and different choices of $\Gamma$ reorganize the sum over covering maps.}
\item The natural ensemble for string perturbation theory in $\text{AdS}_3/\text{CFT}_2$ is the grand canonical ensemble. Instead of fixing the number of fundamental strings in the background, an associated chemical potential is fixed, which on the string side is related to the string coupling constant.
\item For special values of the chemical potential, the grand canonical partition function can have poles. These poles can be associated to (possibly singular) on-shell classical bulk geometries. We argued that the classical bulk geometry should be understood as a condensate of covering maps. We proposed a set of (incomplete) rules how the classical geometry and the stringy geometries are related. The full grand canonical partition function can be recovered by a farey tail sum over classical geometries with a 1-loop determinant, see eq.~\eqref{eq:farey tail sum}.\footnote{We did this analysis very explicitly for a torus boundary, where the appearing geometries are $\SL(2,\ZZ)$ family of black holes and conical defects. For more complicated geometries, our assertion is much more speculative.}
\item The grand canonical partition function can depend on several chemical potentials -- one for each boundary in the geometry. The remaining chemical potentials can be introduced on the worldsheet via discrete torsion. While wormhole geometries cannot lead to poles in the full grand canonical partition function because it factorizes, we speculated that wormholes could become visible as singularities in the chemical potential once we integrate out the off-diagonal chemical potentials.
\item Averaging over the Narain moduli space of $\TT^4$ makes the relation between stringy geometries and classical geometries more manifest. Averaged stringy geometries can be interpreted as all possible fillings of the worldsheet, which are geometries with multiple sheets that meet at the boundary of $\AdS_3$. When tuning the chemical potential to the critical value, these multiple sheets can condensate to a single classical geometry. Deforming away from the symmetric orbifold has also the effect of introducing an interaction between the covering surfaces. We argued that also for strong interaction, the interaction favors those micro-geometries that align properly and leads to an emerging semi-classical bulk geometry.
\end{enumerate}
\subsection{Discussion}\label{subsec:discussion}
Let us discuss some open issues. The first points are technical and the later ones more conceptual.
\paragraph{The hybrid formalism in the free-field variables.} Our main technical goal in this paper was to compute the string partition function of the tensionless string on various hyperbolic manifolds. Even though the tensionless string enjoys a free field representation in the hybrid formalism, its BRST structure is quite complicated and we have not attempted to fully define the string partition function in these free field variables. In order to put this and similar computations on a firmer footing, it is indispensable to develop the hybrid formalism in these variables more rigorously.
\paragraph{Localization to covering maps.} We have not succeeded to demonstrate fully that the string partition function localizes only on covering maps. We gave a simple counterexample in Section~\ref{subsec:special cases}: the worldsheet torus partition function seems in general to be non-zero also in geometries with higher genus boundaries. However, since the orbifold is infinite, the inclusion of the order of the orbifold group is rather delicate and can suppress these contributions. We do not know whether the `fake' covering surfaces that appear have a deeper meaning or are just a complication of the formalism.
\paragraph{Worldsheet theory from the boundary CFT.} The complications of perturbative string theory are striking, when compared to the simplicity of the result from a dual CFT point of view. String theory has to deal with lots of additional structures that are not visible in the final result, such as BRST cohomologies, integrals over $\mathcal{M}_{g,n} \times \Jac(\Sigma_{g})$ etc. One could hope to construct a version of string perturbation theory that does not have all this additional structure. The idea of reconstructing the worldsheet from the boundary CFT is an old one, see \cite{Gopakumar:2003ns,Gopakumar:2004qb, Gopakumar:2005fx, Razamat:2008zr} and was recently made concrete for symmetric orbifolds in \cite{Gaberdiel:2020ycd}.
\paragraph{Backreaction.} One might question the validity of string perturbation theory in our work. We have trusted string perturbation theory far beyond the regime where it is usually applicable, since we have considered arbitrary numbers of highly winding strings. We would expect such heavy string configurations to backreact on the geometry and change it. This is in fact precisely what we saw, since we can interpret condensates of string configurations as new classical geometries. Nonetheless, the string perturbative description seems to work well in that it succeeds to describe the full boundary theory correctly. One potential reason why this happens is perhaps the fact that the worldsheets (except for the sphere) seem to stay close to the boundary of the bulk. Thus even though the interior of the bulk geometry changes, perturbative strings do not get affected by this change.
\paragraph{Non-saddle geometries.} Recently, the importance of including non-saddle geometries in the gravitational path integral has been emphasized \cite{Saad:2019lba, Maxfield:2020ale, Cotler:2020ugk}. If these appear in the grand canonical partition function, they are hard to isolate. Let us discuss this for the boundary torus partition function. The reason is that they are only expected to have an order 1 action and thus would lead to poles in the grand canonical partition function at $p=1$, i.e.\ at $\sigma=0$. However, $\sigma=0$ is already an accumulation point of poles of the partition function. Thus, non-saddle geometries do not seem to naturally appear in the grand canonical farey tail sum \eqref{eq:farey tail sum}.

\paragraph{Black hole-string and wormhole-string transition.} It has  been proposed that black holes of string size transition into a very long string winding around the horizon \cite{Susskind:1993ws, Horowitz:1996nw, Giveon:2005mi}. Our results of this paper that this is true very generally in our setting. While all computations have been done in Euclidean signature, the qualititative picture of winding strings should be preserved under Wick rotation. In some sense or analysis thus leads to more general wormhole-string transitions etc.

\paragraph{Bulk emergence.} While the condensation of covering maps to classical geometries seems to work quantiatively, it does not explain how the bulk emerges from the two-dimensional worldsheets. Ultimately, we should not only see the bulk action emerge, but also more local notions like bulk reconstruction from entanglement wedges. It remains to be seen whether this toy model can be used to obtain more fine-grained information.
\paragraph{Ensemble average and chaos.} The tensionless string lacks any chaotic behaviour. Even after ensemble averaging any member of the ensemble is a solvable CFT. In particular, indicators such as the spectral form factor and out of time order correlators diagnose a clearly non-chaotic behaviour of the symmetric orbifold \cite{Balasubramanian:2016ids, Belin:2017jli}.
This is in sharp contrast with the situation in theories of pure gravity, such as JT-gravity \cite{Saad:2019lba}. Thus our analysis corroborates that the ensemble is a feature of low-energy effective descriptions of the theory.

\paragraph{Baby universes and third quantization.} We have seen in Section~\ref{subsec:ensemble average} that one can introduce an ensemble average in the symmetric orbifold that seems to make the nature of the stringy geometries somewhat more manifest.
It is interesting that the average of a single partition function already contains many geometries that feature disconnected wormholes, even though the asymptotic boundaries of the wormhole are in fact identified, see Figure~\ref{fig:t0 surface}. 
Perhaps these `micro geometries' can be viewed as baby universes \cite{Coleman:1988cy, Giddings:1988cx, Giddings:1988wv, Marolf:2020xie}.\footnote{In the examples of Figures~\ref{fig:t0 surface} and \ref{fig:conical defect stringy geometry}, the Cauchy slice is such that the spatial surface has no compact disconnected region. However, one can see that one can arrange the Cauchy slice in more complicated examples such that all the sheets are disconnected universes.}

\paragraph{Deformation away from symmetric orbifold.} We have started to analyze the behaviour of the symmetric orbifold partition function when deforming away from the symmetric orbifold and shown that it turns on an interaction of the covering surfaces. It would be very interesting to make our speculations about the emergence of the bulk geometry more concrete.

\paragraph{Effective spacetime theory.} The description of the tensionless string is formulated on the worldsheet. It would be far more enlightening to understand the corresponding \emph{spacetime} theory. At the tensionless point, such a theory would be highly non-local. Constructing the spacetime theory explicitly would presumably shed further light on the emergence of a local bulk dual when deforming away from the symmetric product. For vector-models, the corresponding higher spin spacetime theory was recently constructed in \cite{Aharony:2020omh}. 

\paragraph{$\AdS_5 \times \S^5$.} There is the obvious question how this generalizes to higher dimensional examples of the AdS/CFT correspondence, most notably $\AdS_5 \times \S^5$. The dual $\mathcal{N}=4$ SYM theory possesses a free point that corresponds to tensionless strings in the bulk. While there is not yet a complete formulation of the corresponding worldsheet theory (see however \cite{Berkovits:2007rj, Berkovits:2019ulm} for attempts and progress), the question can be studied from the boundary. This has a long history \cite{Sundborg:1999ue, Sundborg:2000wp}. We hope that the insights of this paper can be useful also in higher dimensions.

\section*{Acknowledgements}
I would like to thank Alexandre Belin, Andrea Dei, Bob Knighton, Shota Komatsu, Juan Maldacena, Tom\'a\v{s} Proch\'azka, Bo Sundborg and Edward Witten for discussions. I would also like thank Andrea Dei, Bob Knighton, Shota Komatsu and especially Bo Sundborg for very helpful comments on the manuscript. I thank Alessandro Sfondrini for organizing the workshop ``Correlation functions in low-dimensional AdS/CFT'' in Castasegna, where this work was started. This work is supported by the IBM Einstein Fellowship at the Institute for Advanced Study.
\appendix

\section{Hyperbolic 3-manifolds} \label{app:hyperbolic 3 manifolds}
In this Appendix, we collect some facts about hyperbolic 3-manifolds. These are locally Euclidean $\text{AdS}_3$ spaces and serve as bulk manifolds in our investigation. Every hyperbolic 3-manifold can be written as a quotient $\HH^3/\Gamma$ for a discrete subgroup $\Gamma \subset \PSL(2,\CC)$, where $\HH^3$ is hyperbolic 3-space. Thus the study of hyperbolic 3-manifolds is equivalent to studying discrete subgroups $\Gamma \subset \PSL(2,\CC)$ -- so-called \emph{Kleinian groups}.
\subsection{Hyperbolic 3-space}
We work with the Poincar\'e ball model, where $\HH^3$ is identified with the unit ball inside $\RR^3$, equipped with the metric
\be 
4\frac{\sum_{i=1}^3 \mathrm{d}x_i^2}{(1-\abs{x}^2)^2}\ .
\ee
The isometry group is the conformal group $\PSL(2,\CC)$.
The boundary of $\HH^3$ is the Riemann sphere $\mathbb{CP}^1$ and $\PSL(2,\CC)$ acts by M\"obius transformations on it. 

$\HH^2\subset\HH^3$ can then be thought of as the equatorial plane $x_3=0$, which is preserved by a subgroup $\PSL(2,\RR) \subset \PSL(2,\CC)$. The boundary of $\HH^2$ inside $\HH^3$ can be identified with the equatorial circle $\mathbb{R} \cup \{\infty\}$ of $\mathbb{CP}^1$, which is indeed preserved by $\PSL(2,\RR)$. This equatorial plane corresponds to the $t=0$ spacelike surface in global \emph{Lorentzian} $\text{AdS}_3$.

\subsection{General properties} 
While a Kleinian group $\Gamma$ acts properly discontinuously on $\HH^3$, it typically does not on the boundary $\mathbb{CP}^1$. Let $\Omega$ be the maximal open set in $\mathbb{CP}^1$ on which it does act properly discontinuously. We shall in the following assume that $\Omega \ne \varnothing$. The complement $\Lambda=\mathbb{CP}^1 \setminus \Omega$ is called the domain of discontinuity or the limit set. By Ahlfohr's measure theorem, $\Lambda$ has measure zero. The set $\Lambda$ is however typically very discontinuous and fractal.

Let us furthermore assume that $\Gamma$ is finitely generated (which is the case in all examples of interest to us). Then Ahlfohr's finiteness theorem states that $\Omega/\Gamma$ is a finite union of punctured Riemann surfaces. $\Omega$ itself can have either 1, 2 or infinitely many connected components.

Hyperbolic 3-manifolds are rigid, which means that for each boundary geometry and topological type, there is at most one hyperbolic structure. 
As a consequence, the sum over geometries in 3d gravity is indeed a sum and thus not contain continuous pieces.
\subsection{(Co)homology of 3-hyperbolic manifolds} \label{subapp:cohomology of hyperbolic manifolds}
In the main text, we need some basic aspects of the topology of hyperbolic manifolds. In this discussion, we assume the manifold $\mathcal{M}=\HH^3/\Gamma$ to be smooth and do not allow singularities. We determine the integer (co)homology groups in terms of the Kleinian group $\Gamma$ and basic geometric invariants of the boundary. We have of course the basic groups
\be 
\mathrm{H}_0(\mathcal{M},\ZZ) \cong  \mathrm{H}^0(\mathcal{M},\ZZ) \cong \ZZ
\ee
and $\mathrm{H}_3(\mathcal{M},\ZZ) \cong \mathrm{H}^3(\mathcal{M},\ZZ) \cong 0$, since the manifold is non-compact (we will assume in the following that it has at least one boundary). Higher (co)homology groups also vanish.
By the Hurewicz theorem, we have
\be 
\mathrm{H}_1(\mathcal{M},\ZZ) \cong \pi_1(\mathcal{M})^\text{ab} \cong \ZZ^r \oplus \Gamma_\text{tor}\ .
\ee
Here, $\Gamma_\text{tor}$ is the torsion subgroup of the abelianization of $\pi_1(\mathcal{M}) \cong \Gamma$ (that is not necessarily the torsion subgroup of $\Gamma$). $r$ is the rank of the abelianization of $\Gamma$. By the universal coefficients theorem, we have
\be 
\mathrm{H}^1(\mathcal{M},\ZZ) \cong \ZZ^r\ , \qquad \mathrm{H}^2(\mathcal{M},\ZZ) \cong \ZZ^R \oplus \Gamma_\text{tor}\ , \qquad  \mathrm{H}_2(\mathcal{M},\ZZ) \cong \ZZ^R\ .
\ee
We will determine the rank $R$ below. $\mathrm{H}_2(\mathcal{M},\ZZ)$ has no torsion, since otherwise by the universal coefficients theorem, the torsion would also appear in $\mathrm{H}^3(\mathcal{M},\ZZ) $.
Next, we determine the rank $R$ of $\mathrm{H}_2(\mathcal{M},\RR)$.\footnote{It is more convenient to use real coefficients here, since we already determined all the torsion parts.} We have the long exact sequence of (relative) homology groups
\begin{align}
0 &\longrightarrow\mathrm{H}_3(\mathcal{M},\partial \mathcal{M};\RR)  \cong \RR & &\hspace{-.4cm}\longrightarrow\mathrm{H}_2(\partial \mathcal{M},\RR)  \cong \RR^n & &\hspace{-.4cm}\longrightarrow \mathrm{H}_2(\mathcal{M},\RR) \cong \RR^R\nonumber\\
&\longrightarrow \mathrm{H}_2(\mathcal{M},\partial\mathcal{M};\RR) \cong \RR^r &  &\hspace{-.4cm}\longrightarrow \mathrm{H}_1(\partial\mathcal{M},\RR) \cong \RR^{\sum_i \! 2G_i} & &\hspace{-.4cm}\longrightarrow \mathrm{H}_1(\mathcal{M},\RR) \cong \RR^r\nonumber\\
&\longrightarrow \mathrm{H}_1(\mathcal{M},\partial \mathcal{M},\RR)\cong \RR^R  & &\hspace{-.4cm} \longrightarrow \mathrm{H}_0(\partial \mathcal{M},\RR) \cong \RR^n & & \hspace{-.4cm} \longrightarrow \mathrm{H}_0(\mathcal{M},\RR)\cong \RR \longrightarrow 0\ .
\end{align}
Here, $n$ is the number of boundaries. We used Poincar\'e-Lefschetz duality, which is the generalization of Poincar\'e duality to orientable manifolds with boundary,
\be 
\mathrm{H}_k(\mathcal{M},\RR) \cong \mathrm{H}^{3-k}(\mathcal{M},\partial \mathcal{M},\RR) \ , \qquad \mathrm{H}^k(\mathcal{M},\RR) \cong \mathrm{H}_{3-k}(\mathcal{M},\partial \mathcal{M},\RR) 
\ee
to express the relative homology groups in term of known quantities. Because the alternating sum of the ranks of the group has to vanish in any exact sequence, we can calculate $R$ from this and obtain 
\be 
R=n+r-\sum_i G_i-1=r-1-\sum_i (G_i-1)\ . \label{eq:rank second homology group}
\ee
Thus, everything is expressible in terms of the group $\Gamma$ and the Euler characteristics of the boundary surfaces. 

For our discussion, we also need $\mathrm{H}^2(\mathcal{M},\U(1))$. From the universal coefficients theorem, it follows that
\be 
\mathrm{H}^2(\mathcal{M},\U(1)) \cong \U(1)^R\ .
\ee

\subsection{Examples} \label{subapp:Kleinian groups examples}
Here we discuss examples of Kleinian groups. These are essentially the bulk manifolds for the uniformizations of the boundary surfaces that we reviewed in Appendix~\ref{app:uniformization}.
\paragraph{Schottky groups.} Since the boundary of $\HH^3$ is $\CP^1$, it is natural to consider the quotient space $\HH^3/\Gamma_G^\text{S}$, where $\Gamma_G^\text{S}$ is a genus $G$ Schottky group. The resulting 3-manifold is a handlebody and can be thought of as the interior of a Riemann surface when embedded in 3-space.

The characteristic feature of handlebodies is that some homology cycles of Riemann surface become contractible in the bulk. The $\alpha_I$ cycles of the boundary can be identified with the circles $C_I$ bounding the fundamental domain. They can be seen to be contractible in the bulk. Topologically, this yields a classification of handlebodies: for every Lagrangian sublattice of $\mathrm{H}_1(\Sigma_G,\ZZ)$ (i.e.~a $G$-dimensional sublattice with trivial induced intersection form), there is a corresponding handlebody where these cycles can be contracted. 
However for genus $G \ge 2$ surfaces, there are also other bulk manifolds.

Handlebodies can be interpreted as Euclidean continuations of multiboundary \emph{Lorentzian} wormholes. This analytic continuation can be performed in different ways leading to different physical interpretations \cite{Krasnov:2000zq}.

\paragraph{Fuchsian groups.} We can also act with a Fuchsian group on $\HH^3$. Since the Fuchsian group $\Gamma_G^\text{F}$ fixes the extended real line $\RR \cup \{\infty\} \subset \CP^1$, the domain of discontinuity $\Omega$ equals the union of the upper and lower half-plane. Consequently, the boundary $\Omega/\Gamma_G^\text{F}$ of the resulting hyperbolic 3-manifold coincides with \emph{two} copies of the Riemann surface $\Sigma_G=\HH^2/\Gamma_G^\text{F}$. Thus the resulting geometry is a Euclidean wormhole connecting the two copies of the surface. The metric can be written down very explicitly \cite{Maldacena:2004rf, Maxfield:2016mwh}
\be 
\mathrm{d}s^2_{\HH^3/\Gamma_G^\text{F}}=\mathrm{d}\chi^2+\cosh^2\chi \, \mathrm{d}s_{\Sigma_G}^2\ , \label{eq:Fuchsian wormhole metric}
\ee
where $\mathrm{d}s_{\Sigma_G}^2$ is the hyperbolic metric on the Riemann surface. $\chi \in \RR$ parametrizes the location along the throat of the wormhole. The thinnest part of the throat is located at $\chi=0$. The solution is reflection symmetric $\chi \mapsto -\chi$. We could further quotient by this isometry to obtain a geometry with a single boundary. It is singular, because the $\{\chi=0\}$ surface is fixed, but we can get a non-singular geometry by assuming that the Riemann surface $\Sigma_G$ has a non-trivial involution (automorphism of order 2) without fixed points and by composing it with the reflection $\chi \mapsto -\chi$. The resulting geometry has only a single boundary, but is not a handlebody. In fact, this geometry is very simple from a topological point of view. The wormhole is homotopy equivalent to the Riemann surface $\Sigma_G$ and thus the geometry obtained by further dividing out the $\ZZ_2$ action is homotopy equivalent to $\Sigma_G/\ZZ_2$. Depending on whether the involution is orientation preserving or reversing, $\Sigma_G/\ZZ_2$ is either a Riemann surface of genus $\frac{G+1}{2}$ (for this $G$ has to be odd), or it is the non-orientable Klein bottle of genus $G$. In the two cases, the non-trivial homology groups are\footnote{Since $\chi \mapsto -\chi$ also reverses the orientation, the whole manifold is orientable when the involution is orientation-reversing and vice-versa.}
\begin{subequations}
\begin{align}
&\text{orientation preserving involution:}& \!\mathrm{H}_1(\mathcal{M},\ZZ) &\cong \ZZ^{G+1}\ , &\!\!\!\mathrm{H}_2(\mathcal{M},\ZZ) &\cong \ZZ\ ,\! \\
&\text{orientation reversing involution:}& \!\mathrm{H}_1(\mathcal{M},\ZZ) &\cong \ZZ^{G}\oplus \ZZ_2\ , &\!\!\!\mathrm{H}_2(\mathcal{M},\ZZ) &\cong 0\ ,
\end{align}
\end{subequations}
respectively. This shows that the resulting manifold is not equivalent to a handle-body. The reader can easily verify that the formula  \eqref{eq:rank second homology group} holds in both cases. Moreover, the first case gives an example with $\mathrm{H}_2(\mathcal{M},\ZZ) \ne 0$, which means that an additional chemical potential can be introduced in the theory using discrete torsion, see Section~\ref{subsec:discrete torsion and more chemical potentials}. However, in this case $\mathcal{M}$ is also non-orientable and it is not clear to us whether one can formulate the worldsheet theory on a non-orientable target space.
\paragraph{Quasi-Fuchsian groups.} We can similarly consider the manifold $\HH^3/\Gamma_G^\text{QF}$ for $\Gamma_G^\text{QF}$ a quasi-Fuchsian group, see also Appendix~\ref{subapp:simultaneous uniformization}. In this case the resulting hyperbolic manifold can be thought of as a Euclidean wormhole connecting two genus $G$ surfaces with possibly different moduli. 

\section{Some facts about Riemann surfaces}\label{app:Riemann surfaces}
In this Appendix, we will briefly recall some facts about Riemann surfaces. This Appendix is mostly meant for reference and to fix conventions. We shall denote by $\Sigma_g$ a Riemann surface (with no restrictions on the genus). 
\subsection{Differentials} \label{subapp:differentials}
We fix canonical generators of $\pi_1(\Sigma_g)$, $\alpha_1,\dots,\alpha_g$, $\beta_1,\dots,\beta_g$, satisfying $\prod_{I=1}^g [\alpha_I,\beta_I]=1$. We often view these generators as elements of $\mathrm{H}_1(\Sigma_g,\ZZ)$, where they have canonical intersection products:
\be 
\alpha_I \cap \alpha_J=\beta_I \cap \beta_J=0\ , \qquad\alpha_I \cap \beta_J=\delta_{IJ}\ , \label{eq:canonical homology basis}
\ee
for $I=1,\dots,g$. Let $\omega_1,\dots,\omega_g$ be the corresponding dual basis of $\mathrm{H}^{(1,0)}(\Sigma_g,\mathbb{C})$:
\be 
\int_{\alpha_I} \omega_J=\delta_{IJ}\ , \qquad \int_{\beta_I} \omega_J=\Omega_{IJ}\ ,
\ee
where $\Omega_{IJ}$ is the period matrix. It satisfies
\be 
\boldsymbol{\Omega}=\boldsymbol{\Omega}\tran\, \qquad \Im \boldsymbol{\Omega}>0\ .
\ee
A change of generators $\alpha_1,\dots,\alpha_g$, $\beta_1,\dots,\beta_g$ is an element of $\Aut(\pi_1(\Sigma_g))$ and induces an automorphism of $\mathrm{H}_1(\Sigma_g,\ZZ)$. Such an automorphism preserves the intersection product and can thus be identified with an element of $\Sp(2g,\ZZ)$. It acts on the period matrix by fractional linear transformations:
\be 
\boldsymbol{\Omega} \mapsto (A\, \boldsymbol{\Omega}+B)(C\, \boldsymbol{\Omega}+D)^{-1}\ , \qquad \begin{pmatrix}
A & B \\ C & D
\end{pmatrix}\in \Sp(2g,\ZZ)\ .
\ee
\subsection{Divisors}
A divisor is a formal finite sum of points on the surface:
\be 
D=\sum_i n_i z_i\ , \qquad n_i \in \mathbb{Z}\ , \quad z_i \in \Sigma\ .
\ee
The group of divisors is the free abelian group on the points of the surface. For a meromorphic function $f$ on $\Sigma$, define the principal divisor as $(f)=\sum_z \text{ord}_z(f)z$, where $\text{ord}_z(f)$ is the order of vanishing of $f$ at $z$ (negative when $f$ has a pole at $z$). Since $(fg)=(f)+(g)$, principal divisors form a subgroup of all divisors and we can form equivalence classes $D \sim D+(f)$, which are the divisor classes. The degree of a divisor is 
\be 
|D|=\sum_i n_i\in \mathbb{Z}\ ,
\ee
which is well-defined also on divisor classes since $|(f)|=0$. One can define divisors also for 1-forms etc., since the order of vanishing is always a well-defined concept for any tensorfield.

A particularly important divisor is the canonical divisor $K$, given by the divisor of any meromophic one-form. The equivalence class is well-defined, since the ratio $\omega_1/\omega_2$ of any two one-forms is a meromorphic function. The degree of the canonical divisor is $|K|=2g-2$.

A divisor $D$ on a Riemann surface determines a line bundle $\mathcal{O}(D)$, whose sections are the space of meromorphic functions on $\Sigma$ which vanish at least as fast at $z_i$ as prescribed by the divisor $D$.

\subsection{Classification of line bundles} \label{subapp:classification of line bundles}
Here, we sketch the classification of line bundle on Riemann surfaces, since this plays a role in Section~\ref{sec:localization}. Line bundles form a group with respect to the tensor product.
Topologically, line bundles are just classified by their first Chern class (which is a group homomorphism from the line bundles to the $\mathrm{H}^2(\Sigma_g,\ZZ) \cong \ZZ$), which coincides with the degree of the corresponding divisor. We are however interested in the analytic classification. We can restrict ourselves to line bundles with vanishing first Chern class, since any line bundle of degree $d$ can be obtained by tensoring the flat line bundle with a fixed line bundle of degree $d$. Line bundles with vanishing first Chern class carry a connection with vanishing curvature and are hence flat. They can be characterized by a homomorphism
\be 
\rho:\pi_1(\Sigma_g) \longrightarrow \mathbb{C}^\times\ ,
\ee
that describes the holonomy of the bundle. Since the target group is abelian, such a homomorphism can be also understood as a homomorphism from $\mathrm{H}_1(\Sigma_g,\mathbb{Z})$ into $\mathbb{C}^\times$. We can in turn identify $\mathrm{H}_1(\Sigma_g,\mathbb{Z})$ with $\mathbb{Z}^{2g}$ by employing the canonical basis $\alpha_1,\dots,\alpha_g$, $\beta_1,\dots,\beta_g$. Thus, we can set $\rho(\alpha_I)=\mathrm{e}^{2\pi i s_I}$ and $\rho(\beta_I)=\mathrm{e}^{2\pi i t_I}$. Most of these parameters are actually redundant. Consider the following quasiperiodic function in $z$:
\be 
\mathrm{exp}\left(-2\pi i \sum_I s_I \int^z_{z_0}\omega_I \right)\ .
\ee
This function can be viewed as a holomorphic transformation between line bundles with different multipliers around the cycles.
By definition, its holonomies around the $\alpha$-cycles remove the phases given by $s_I$ completely. Thus, we can set $s_I=0$. There are also some identifications on the parameters $t_I$. We have of course $t_I\sim t_I+1$ because of the exponential map. We can also use the same quasiperiodic function with $s_I \in \mathbb{Z}$ (so that its holonomies around the $\alpha_I$-cycles are still trivial). This has the effect of changing $t_I \mapsto t_I-\Omega_{IJ}s_J$. In summary, the parameters $t_I$ form the space
\be 
\Jac(\Sigma_g)=\mathbb{C}^g/(\mathbb{Z}^g \oplus \boldsymbol{\Omega} \mathbb{Z}^g)\ .
\ee
This is a $g$-dimensional complex torus -- the so-called Jacobian. This is the moduli space of flat line bundles on $\Sigma_g$. It carries a natural complex metric and has volume
\be 
\vol(\Jac(\Sigma_g))=\det \Im \boldsymbol{\Omega}\ .
\ee

The same can also be seen by using the Abel-Jacobi map, that maps injectively (for $g \ge 1$)  $\Sigma_g \longrightarrow \Jac(\Sigma_g)$:
\begin{align}
\mathbf{u}:\Sigma_g &\longrightarrow \Jac(\Sigma_g)\ , \\
z &\longmapsto \int_{z_0}^z \boldsymbol{\omega} \ .
\end{align}
Note that since the integration path is unspecified, the result is only well-defined as an element of the Jacobian. $z_0$ is an arbitrary reference point on $\Sigma_g$. This map extends naturally to divisors by linearity. For a flat line bundle, the associated divisor has degree 0 and is hence independent of the choice of $z_0$. The Abel-Jacobi map now yields a isomorphism between line bundles (or their divisors) and the Jacobian. A divisor $D$ is hence principal if and only if
\be 
|D|=0 \qquad \text{and}\qquad \mathbf{u}(D)=0\ .
\ee
\subsection{Spin structures}
In this work, we consider fields with half-integer spin on the Riemann surface $\Sigma_g$. They are sections of a spin bundle $S$. $S$ satisfies $S^2=K$. The degree of any spin structure is $|S|=g-1$. Consequently, $2\, \mathbf{u}(S)=\mathbf{u}(K)$. Since the Jacobian is a real $2g$-dimensional torus, there are $2^{2g}$ ways to choose $\mathbf{u}(S)\in \Jac(\Sigma_g)$ such that $2\,\mathbf{u}(S)=\mathbf{u}(K)$, corresponding to the $2^{2g}$ spin structures on the Riemann surface. 
Notice that for two spin bundles $S_1$ and $S_2$, $S_1\otimes S_2^{-1}$ is a flat line bundle on $\Sigma_g$ and satisfies $2\, \mathbf{u}(S_1\otimes S_2^{-1})=0 \in \Jac(\Sigma_g)$. Thus given a fixed spin structure, we obtain any other spin structure by tensoring with such a special flat line bundle with structure group $\ZZ_2 \subset \CC^\times$.

Spin structures can be further divided into even and odd spin structures according to whether they have an even or odd number of holomorphic sections (or alternatively via theta-characteristics). There are $2^{g-1}(2^g+1)$ even spin structures and $2^{g-1}(2^g-1)$ odd spin structures.

\subsection{Riemann-Roch theorem}
The Riemann-Roch theorem is a statement about the number of holomorphic sections of a line bundle $L$. Denoting the space of sections by $\mathrm{H}^0(\Sigma_g,L)$, the theorem states that
\be 
\dim \mathrm{H}^0(\Sigma_g,L) -\dim\mathrm{H}^0(\Sigma_g,K \otimes L^{-1})=|L|+1-g\ .
\ee
For example, we frequently make use of the following well-known statements (valid for $g \ge 2$):
\begin{subequations}
\begin{align} 
\dim\mathrm{H}^0(\Sigma_g,K) &=g\ , \label{eq:dimension holomorphic 1 forms}\\
\dim\mathrm{H}^0(\Sigma_g,K^2) &=3g-3\ . \label{eq:dimension quadratic differentials}
\end{align}
\end{subequations}
For a spin bundle $S$, the Riemann-Roch theorem makes no predictions. The dimension of sections of $S$ is generically 1 for an odd spin structure and 0 for an even spin structure, but can jump on subloci of $\mathcal{M}_g$ \cite{Hitchin}.
\section{Subgroups of the fundamental group and covering spaces}\label{app:subgroups}
This Appendix is more algebraic in nature and describes how to efficiently list the subgroups of the fundamental group of Riemann surfaces. The fundamental group is generated by $\alpha_1,\dots,\alpha_g$, $\beta_1,\dots,\beta_g$ with $\prod_I [\alpha_I,\beta_I]=1$, as described in Appendix~\ref{subapp:differentials}.
\subsection{Regular covering spaces}
Connected covering spaces of degree $N_\text{c}$ of the genus $g$ Riemann surface are in one-to-one correspondence to subgroups of $\pi_1(\Sigma_g)$ up to conjugacy. (We are not interested in covering spaces of punctured spaces and hence consider conjugate subgroups to be equivalent).

Regular (or normal) covering spaces are those for which the corresponding subgroup $\mathrm{H}$ of $\pi_1(\Sigma_g)$ is normal. In this case $\mathrm{G}=\pi_1(\Sigma_g)/\mathrm{H}$ is a finite group of order $N_\text{c}$. The covering space $\Sigma_{N_\text{c}(g-1)+1}$ enjoys a group action of $\mathrm{G}$ and $\Sigma_{N_\text{c}(g-1)+1}/\mathrm{G}=\Sigma_{g}$.
Starting from degree 3, not all covering spaces are regular.

\subsection{Relation to homomorphisms to \texorpdfstring{$S_N$}{SN}} \label{subapp:homomorphisms SN}
A different useful perspective to think about covering spaces is to label them by group homomorphisms
\be 
\phi: \pi_1(\Sigma_g) \longrightarrow S_{\text{N}}\ ,
\ee
whose image acts transitively on $\{1,\dots,N\}$. If we drop the transitivity condition than we also get disconnected covering spaces. This is because a covering space can be understood as a fibration with fibre $\{1,\dots,N\}$ and structure group $S_N$ that permutes the different sheets. A covering space that is obtained by relabelling $\{1,\dots,N\}$ is equivalent and thus we are again interested in group homomorphisms up to conjugacy.

Such group homomorphisms are much more managable since we only have to specify them on the generators. We only have to ensure that the relation
\be 
\prod_{I=1}^g [\phi(\alpha_I),\phi(\beta_I)]=1 \label{eq:homomorphism constraint}
\ee
is satisfied. 
The relation to subgroups of $\pi_1(\Sigma_g)$ is obtained by setting
\be 
\mathrm{H}_\phi=\Stab(\{1\})\ ,
\ee
which by the orbit-stabilizer theorem is an index $N$ subgroup (which is not necessarily normal). 

\subsection{Number of subgroups}
Let us next discuss the number of subgroups of a fixed degree. This question is answered by Hurwitz theory. The number of \emph{disconnected} covering surfaces of degree $N$ of a genus $N$ surfaces is given by Burnside's formula:\footnote{This is just the genus $g$ partition function of two-dimensional gauge theory with gauge group $S_N$. This gauge theory counts covering spaces, because every covering space can be viewed as a fibre bundle where the fiber consists of $N$ points, as discussed in~\ref{subapp:homomorphisms SN}. Since gauge theory counts the number of $S_N$-bundles, it counts the number of covering spaces.}
\be 
H^\text{disc}(N,g)=\sum_{R} \left(\frac{\dim R}{N!}\right)^{2-2g}\ ,
\ee
where the sum extends over all irreducible representations of the symmetric group $S_N$. Here, $H$ stands for Hurwitz. In this formula, covering surfaces with non-trivial automorphism group are weighted by the inverse order of their automorphism group. To find the number of connected coverings one can form the Hurwitz potential:
\be 
\mathcal{H}_g^\text{disc}(q)=1+\sum_{d=1}^\infty H^\text{disc}(N,g) q^d\ .
\ee
Then the connected covering surfaces can be obtained by taking a logarithm, which passes to the connected Hurwitz potential:
\be 
\mathcal{H}_g^\text{conn}(q)=\log \mathcal{H}_g^\text{disc}(q)=\sum_{N_\text{c}=1}^\infty H^{\text{conn}}(N_\text{c},g)q^{N_\text{c}}\ .
\ee
The actual number of connected covering spaces is then given by $N_\text{c} H^{\text{conn}}(\text{c},g)$. The factor of the automorphism group also appears in the symmetric orbifold partition function \eqref{eq:symmetric orbifold partition function}.
For low values of $N_\text{c}$ , these values are listed in Table~\ref{eq:number of subgroups fundamental group}. 
\begin{table}
\begin{center}
\begin{tabular}{c|l}
\text{index} & \text{number of subgroups up to conjugacy} \\
\hline
2 & $4 \cdot 2^{x}-1$ \\
3 & $6 \cdot 6^{x}+3 \cdot 3^{x}-6 \cdot 2^{x}+1$ \\
4 & $8 \cdot 24^x+4 \cdot 12^x+8 \cdot 8^x-8 \cdot 6^x-8 \cdot 4^x-4 \cdot 3^x+8 \cdot 2^x-1$
\end{tabular}
\end{center}
\caption{The number of subgroups of the fundamental group of a Riemann surface of a given index. Here, $x=2g-2$. The structure $\sum_{a \, | \, d!} n_a a^x$ for $n_a$ integers persists also at higher index orders.}\label{eq:number of subgroups fundamental group}
\end{table}
The asymptotics $d \to \infty$ is easy to describe (for genus $g \ge 2$), since the sum in Burnside's formula is completely dominated by the smallest representations -- the trivial and alternating representation. Thus, we get
\be 
N^{\text{disc}}(d,g) \sim N^{\text{conn}}(d,g) \sim 2\,(d!)^{2g-2}\ .
\ee
Passing to the connected Hurwitz numbers does not change the asymptotics.

\subsection{Enumerating subgroups}
Actually enumerating subgroups is difficult. We do so for illustration for low indices.
\paragraph{Index 2.} For index 2, the situation is simple, because every subgroup is normal. Consequently, every index 2 subgroup is obtained as the kernel of a surjective homomorphism $\phi: \pi_1(\Sigma_g) \longrightarrow \ZZ_2$. Such a homomorphism is entirely determined by specifying it on its generators. For every generator, there are 2 choices and in total there are hence $2^{2g}-1$ such homomorphisms. We need to subtract one to account for the fact that the trivial homomorphism is not surjective. 
\paragraph{Index 3.} For index 3 subgroups, the situation is much more difficult, because most subgroups are actually not normal. We proceed by constructing all homomorphisms $\phi: \pi_1(\Sigma_g) \longrightarrow S_3$. Let us first disregards the transitivity condition.  If the image of the homomorphism lies actually in $\ZZ_3$, then the covering space is regular. Let us now look at a general homomorphism into $S_3$ and discuss the constraint \eqref{eq:homomorphism constraint}. For any choice of $\phi$ on the generators, the product in \eqref{eq:homomorphism constraint} is an even permutation. If the image of the homomorphism lies not in $\ZZ_3$, then there is at least one generator that is mapped to a transposition. By changing the transposition, one changes the resulting permutation in the product of \eqref{eq:homomorphism constraint}. Thus one sees that when the image of the homomorphism lies not in $\ZZ_3$, exactly $\frac{1}{3}$ of the choices satisfy the constraint \eqref{eq:homomorphism constraint}. If the image lies in $\ZZ_3$, the constraint is trivially satisfied. Hence the number of homomorphisms $\phi:\pi_1(\Sigma_g) \longrightarrow S_3$ is given by
\be 
\frac{1}{3}(6^{2g}-3^{2g})+3^{2g}\ .
\ee
Relabeling would divide this number further by $6$.
The second term corresponds to the regular covering spaces. This way of counting quickly becomes complicated.
\paragraph{Index 4.} Let us only mention here that there are different regular covering spaces for degree 4 coverings, corresponding to the cases where the image of $\phi$ lies in $\ZZ_2\times\ZZ_2$ or in $\ZZ_4$.

\section{Uniformization} \label{app:uniformization}
In this appendix, we survey the existing (simultaneous) uniformizations of Riemann surfaces. In the main text, we use these sometimes for the worldsheet and sometimes for the boundary surfaces. In the main text, we use $g$ as the worldsheet genus and $G$ as the boundary genus. To keep notation in this appendix uniform, we use $g$ in the following.
\subsection{Teichm\"uller space and the mapping class group} \label{subapp:Teichmuller space}
 We start by recalling some facts about the structure of the moduli space of Riemann surfaces.
The dimension of the moduli space of Riemann surfaces $\mathcal{M}_g$ is $\dim(\mathcal{M}_g)=3g-3$ for $g \ge 2$ (and $=1$ for $g=1$). Deformations of the complex structure are parametrized by Beltrami-differentials $\tensor{\mu}{^z_{\bar{z}}}$, which form the tangent space $T_{\Sigma_g}\mathcal{M}_g$. The cotangent space is in turn naturally identified with the space of quadratic differentials (i.e.~holomorphic sections of $K^2$).
 The structure of the moduli space $\mathcal{M}_g$ itself is quite complicated:
\begin{enumerate}
\item $\mathcal{M}_g$ is not compact, but can be compactified in a canonical way by including nodal Riemann surfaces (the Deligne-Mumford compactification $\overline{\mathcal{M}}_g$). In this work, the integrands of string path integrals are supported only in the interior of the moduli space and all our statements are independent of the compactification.
\item $\mathcal{M}_g$ has the structure of an orbifold that keeps track of the automorphism groups of the Riemann surfaces. The moduli spaces $\mathcal{M}_0$ and $\mathcal{M}_1$ are unstable because they have a continuous automorphism group. All Riemann surfaces with $g \ge 2$ have finite automorphism group (whose order is bounded by $84(g-1)$). Riemann surfaces with non-trivial automorphism groups lead to orbifold singularities in $\mathcal{M}_g$.
\item $\mathcal{M}_g$ naturally carries a K\"ahler metric, the so-called Weil-Petersson metric. It can be described as follows. For two tangent-vectors (Beltrami differentials) at $\Sigma_g \in \mathcal{M}_g$, define an inner product as follows:
\be 
\langle \tensor{\mu}{^z_{\bar{z}}} | \tensor{\nu}{^z_{\bar{z}}} \rangle=\int_{\Sigma_g} \mathrm{d}^2 z \ \sqrt{\gamma} \ \overline{\tensor{\mu}{^z_{\bar{z}}}} \tensor{\nu}{^z_{\bar{z}}}\ .
\ee
This inner product depends on the choice of the metric on the surface $\gamma_{ab}$. The Weil-Petersson metric is defined by choosing the unique metric with constant negative curvature on $\Sigma_g$ (see uniformization theorems below).
\item $\mathcal{M}_g$ is not simply connected. Its universal covering space is Teichm\"uller space $\mathcal{T}_g$. On $\mathcal{T}_g$, the mapping class group $\MCG(\Sigma_g)$ acts and $\mathcal{T}_g/\MCG(\Sigma_g) \cong \mathcal{M}_g$. The mapping class group consists of all the orientation-preserving homeomorphisms of $\Sigma_g$ modulo those that are continuously connected to the identity and is isomorphic to the outer automorphism group of the fundamental group of a genus $g$ surface (by the Dehn-Nielsen-Baer theorem):
\be 
\MCG(\Sigma_g) \cong \Out(\pi_1(\Sigma_g))\ .
\ee
The mapping class group for genera $g \ge 2$ is not well-understood. Any outer automorphism on $\pi_1(\Sigma_g)$ induces an outer automorphism on its abelianization $\mathrm{H}_1(\Sigma_g,\mathbb{Z})$. Generators of the mapping class group preserve the intersection product on $\mathrm{H}_1(\Sigma_g,\mathbb{Z})$ and one obtains a surjection
\be 
\MCG(\Sigma_g) \longrightarrow \Sp(2g,\mathbb{Z})\ .
\ee
The kernel of this morphism is non-trivial for genera $g \ge 2$ and is called the Torelli subgroup $\Tor(\Sigma_g)$. It is sometimes useful to consider the Torelli space $\mathcal{U}_g=\mathcal{T}_g/\Tor(\Sigma_g)$, that can be described as the space of Riemann surfaces together with a choice of canonical homology cycles.
\end{enumerate}
For $g=1$, most of these assertions become trivial. We have $\mathcal{T}_1=\HH^2$, the upper half-plane. The Weil-Petersson metric on $\mathcal{T}_1$ coincides with the Poincar\'e upper half plane metric.
The mapping class group for genera $g=1$ is the well-known modular group $\SL(2,\ZZ)\cong \Sp(2,\mathbb{Z})$ and since the fundamental group is abelian, the Torelli subgroup is trivial. Consequently, $\mathcal{T}_1$ and $\mathcal{U}_1$ coincide.

For genus 2 and 3, the situation is similar. While the Torelli subgroup is non-trivial, the Torelli space $\mathcal{U}_g$ is still simple to describe. The period mapping $\Sigma \mapsto \boldsymbol{\Omega}_\Sigma$ embeds the Torelli space in the Siegel upper half plane
\be 
\mathcal{H}_g=\{\boldsymbol{\Omega} \in \mathbb{C}^{g \times g}\,|\, \boldsymbol{\Omega}\tran=\boldsymbol{\Omega}\,,\ \Im \boldsymbol{\Omega}>0\}\ .
\ee
For $g=2$ and $3$, this is actually an isomorphism and thus $\mathcal{U}_g\cong\mathcal{H}_g$. The period map ceases to be surjective for $g=4$ and the image becomes much harder to describe.
 The moduli space of Riemann surfaces is then given by $\mathcal{M}_g\cong \mathcal{H}_g/\Sp(2g,\ZZ)$ in these genera. 

\subsection{Fuchsian uniformization}\label{subapp:Fuchsian uniformization}
For genera $g \ge 2$, the universal covering space $\Sigma_g$ is the upper half plane $\HH^2$. Hence 
we can write $\Sigma_g=\HH^2/\Gamma_g^\text{F}$ for some discrete subgroup $\Gamma_g^\text{F} \subset \PSL(2,\mathbb{R})$ that acts properly discontinuously on the upper half-plane. Such a group is called a Fuchsian group. 
Since $\Gamma_g^\text{F} \cong \pi_1(\Sigma_g)$, $\Gamma_g^\text{F}$ is represented by $2g$ matrices satisfying
\be 
\left\{A_1,\dots,A_g,B_1,\dots,B_g \in \PSL(2,\RR)\,\bigg|\, \prod_{I=1}^g [A_I,B_I]=\id \right\} \ . \label{eq:Fuchsian group presentation}
\ee
Such a collection of matrices depends on $6g-3$ \emph{real} parameters. Since collections of matrices that are related by an overall conjugation lead to the same Riemann surface, a Fuchsian group depends on $6g-6$ real parameters, which coincides with the $3g-3$ complex parameters of moduli space. The fundamental domain of the Fuchsian uniformization is given by the $4g$-gon that is obtain by cutting the Riemann surface $\Sigma_g$ along the cycles $\alpha_I$ and $\beta_I$. The Fuchsian uniformization does not make the complex structure of moduli space manifest. However, it naturally realizes the unique metric with constant negative curvature on $\Sigma_g$. Since the upper half-plane metric on $\HH^2$ is preserved by the action of $\PSL(2,\mathbb{R})$, it descends to a well-defined metric on $\Sigma_g$ with constant negative curvature. 

\paragraph{Group cohomology.} For our application, the cohomology of Fuchsian groups plays some role, since we are computing an orbifold with Fuchsian group. $\Sigma_g$ is an Eilenberg MacLane space $K(\Gamma_g^\text{F},1)$.\footnote{This means that the only non-trivial homotopy group $\pi_n$ occurs for $n=1$ and $\pi_1(\Sigma_g)=\Gamma_g^\text{F}$ by construction. The reason for this is that $\pi_n$ for $n \ge 2$ coincides with the same hopotopy group of its universal covering space. In this case the universal covering space $\HH^2$ is contractible, so $\pi_n(\Sigma_g)$ for $n \ge 2$ vanishes.} Since the group cohomology can be computed in terms of the (singular) cohomology of the corresponding Eilenberg MacLane space, we have
\begin{align} 
\text{H}^n(\Gamma_g^\text{F},M)&=\text{H}^n (\Sigma_g,M)=\begin{cases}
M\ , & n=0,2\ , \\
M^{2g} \ , & n =1\ , \\
0 \ , & n>2\ .
\end{cases}
\end{align}
for any abelian group $M$.
In particular, $\mathrm{H}^2(\Gamma_g^\text{F},\U(1))=\U(1)$ and thus, there should be a phase that we can freely choose when orbifolding by a Fuchsian group. The corresponding cocycle can be written down very explicitly. Let $[g]_{A_I}$ and $[g]_{B_I}$ denote the number of $A_I$'s (or $B_I$'s) in $g$ when $g$ is written as a word in the generators. Since the constraint in \eqref{eq:Fuchsian group presentation} satisfies $[\cdots]_{A_I}=[\cdots]_{B_I}=0$, this remains also well-defined in the Fuchsian group. The homomorphisms $[\, \cdot\,]_{A_I}$ and $[\,\cdot\, ]_{B_I}$ can be taken as the generators of $\mathrm{H}^1(\Gamma_g^\text{F},\ZZ)$. The generator of $\mathrm{H}^2(\Gamma_g^\text{F},\ZZ)$ can be chosen to take the form
\be 
\varphi(g,h)=\sum_{I=1}^g \left([g]_{A_I} [h]_{B_I}-[g]_{B_I} [h]_{A_I}\right)\ .
\ee
An element for the $\U(1)$ cohomology is then given by $\exp\left(2\pi i \theta \varphi(g,h)\right)$ for $\theta \in [0,1)$.

\paragraph{Spin structure.} We can describe spin structures very naturally using Fuchsian uniformization. The idea is to identify spin structures with lifts $\widetilde{\Gamma}_g^\text{F}$ of $\Gamma_g^\text{F}$ to $\SL(2,\RR)$. There are $2^{2g}$ such lifts, since we may choose the sign for every generator freely. Any such lift is compatible with the relation $\prod_I [A_I,B_I]=0$. Group elements $\gamma$ acts by M\"obius transformations on the upper half plane. Using the lift to $\SL(2,\RR)$ we can define $\sqrt{\partial \gamma(z)}$ consistently for every group element. If 
\be 
\gamma=\begin{pmatrix}
a & b \\ c & d
\end{pmatrix}\ ,
\ee
then $\sqrt{\partial \gamma(z)} \equiv (cz+d)^{-1}$. We can then define spinors to be automorphic forms $\psi(z)$ satisfying
\be 
\sqrt{\partial \gamma(z)} \psi(\gamma(z))=\psi(z)
\ee
on the upper half plane. This hence defines a spin bundle $S$.
\subsection{Schottky uniformization}
Another type of uniformization is Schottky uniformization. It applies to all Riemann surfaces. Here, we let a subgroup of $\Gamma_g^\text{S} \subset \PSL(2,\CC)$ act properly discontinuously on an open subset $\Omega \subset \CP^1$. This is a natural uniformization for the context of $\text{AdS}_3$, since the boundary of global Euclidean $\text{AdS}_3$ is $\CP^1$. A Schottky group $\Gamma_g^\text{S}$ is more precisely characterized by the following properties:
\begin{enumerate}
\item $\Gamma_g^\text{S}$ is isomorphic to a free group in $g$ generators, whose elements are all loxodromic $\PSL(2,\CC)$ transformations (meaning that the corresponding M\"obius transformation has two fixed points).
\item A fundamental domain of the Schottky group can be described as follows. Let $C_{-g},\dots,C_{-1}$, $C_1,\dots,C_g$ be $2g$ circles in the complex plane bounding the discs $D_{-g},\dots,D_{-1}$, $D_1,\dots, D_g$ such that the discs are all disjoint. Then the generators for the Schottky group are taken to be $B_1,\dots,B_g$, where $B_I(C_{-I})=C_I$. Moreover, $B_I$ maps the interior of $C_{-I}$ to the exterior of $C_I$. The fundamental domain for the Schottky group may be taken to be
\be 
\mathcal{F}=\mathbb{CP}^1\setminus (D_{-g} \cup\cdots \cup D_{-1} \cup D_1 \cup \cdots \cup D_g)\ .
\ee
\end{enumerate}
$\Gamma_g^\text{S}$ does not act properly discontinuously on all of $\CP^1$. To have a properly discontinuous action, one has to excise a limit set $\Lambda \subset \CP^1$. $\Lambda$ consists of infinitely many points for $g \ge 2$ and has a complicated fractal structure. We define the open set $\Omega=\CP^1 \setminus \Lambda$. The Riemann surface is then obtained as $\Omega/\Gamma_g^\text{S}$.

Every Riemann surface admits a Schottky uniformization (this is the Koebe retrosection theorem). The Schottky group naturally depends on $3g$ complex parameters corresponding to the matrices $B_1,\dots,B_g$. Overall conjugation of the collection $B_1,\dots,B_g$ again leads to the same Riemann surface and thus the parametrization really depends on $3g-3$ complex parameters, in agreement with the dimension of $\mathcal{M}_g$. While Fuchsian uniformization gives a natural description of the Teichm\"uller space $\mathcal{T}_g$, Schottky uniformization gives a description of an intermediate cover, the so called Schottky space $\mathcal{S}_g$. For $g=1$, we have
\be 
\mathcal{S}_1=\HH^2/(\tau \sim \tau+1)\ .
\ee
Schottky uniformization makes the complex structure of the moduli space manifest. However, since the action of $\PSL(2,\CC)$ on $\CP^1$ does not preserve a metric, it does not lead to a metric on the Riemann surface $\Sigma_g$.

$\Gamma_g^\text{S}$ is isomorphic to a free group and as such the group cohomology of $\Gamma_g^\text{S}$ is well-known. It follows in particular that its group cohomology $\mathrm{H}^n(\Gamma_g^\text{S},M)$ is trivial for $n \ge 2$. Here, $M$ is any abelian group.$\mathrm{H}^1(\Gamma_g^\text{S},M)$ is isomorphic to the abelianization $(\Gamma_g^\text{S})_{\text{ab}} \otimes_\ZZ M \cong M^g$.

\paragraph{Spin structure.} Using Schottky uniformization, we can describe $2^g$ out of the $2^{2g}$ spin structures on the Riemann surface naturally. They again correspond to lifts of the Schottky group $\Gamma_g^\text{S} \subset \PSL(2,\CC)$ to $\SL(2,\CC)$. The remaining spin structures are harder to define, since its sections involve branch cuts running between the circles $C_I$ and $C_{-I}$. We will not have need of these additional spin structures.
\subsection{Simultaneous uniformization} \label{subapp:simultaneous uniformization}
Finally, we discuss simultaneous uniformization of two Riemann surfaces by quasi-Fuchsian groups. This is relevant for the context of $\text{AdS}_3$ for the Euclidean wormhole. The simplest case occurs when a Fuchsian group acts on $\CP^1$. In this case, the limit set $\Lambda$ is $\mathbb{R} \cup \{\infty\}$. Thus, $\Omega=\CP^1\setminus \Lambda=\HH^2 \cup \overline{\HH}^2$, where $\overline{\HH}^2$ is the lower half-plane. Consequently, $\Omega/\Gamma_g^\text{F}=\Sigma_g^{(1)} \cup \Sigma_g^{(2)}$ is the union of two Riemann surfaces with identical moduli.\footnote{This depends slightly on the orientation we choose. If we want the two Riemann surfaces to have the induced orientation of the wormhole geometry, i.e.~that $\partial (\HH^3/\Gamma_g^\text{F})=\Sigma_g^{(1)} +\Sigma_g^{(2)}$ in homology, then the moduli are actually complex conjugate to each other.} 

The statement of simultaneous uniformization generalizes this statement. A quasi-Fuchsian group $\Gamma_g^\text{QF}$ is a discrete subgroup of $\PSL(2,\CC)$, whose limit set equals a Jordan-curve in $\CP^1$ (i.e.~a non-intersecting loop on $\CP^1$).\footnote{Sometimes this is called a quasi-Fuchsian group of the first kind.} The case of a Fuchsian group is a special case, since it preserves the Jordan curve $\mathbb{R} \cup \{\infty\}$. In this case, $\Omega$ has always two components and so $\Omega/\Gamma_g^\text{QF}\cong \Sigma_g \cup \Sigma_g'$ consists of two Riemann surfaces of the same genus, but not necessarily the same moduli. The simultaneous uniformization theorem \cite{Bers:1960} states that any two Riemann surfaces can always be uniformized in this way and hence the space of quasi-Fuchsian groups can be identified with two copies of Teichm\"uller space $\mathcal{T}_g$.

\section{(Branched) complex projective structures}\label{app:branched complex projective structures}
In this Appendix, we review some facts about complex projective structures on Riemann surfaces. There are several equivalent definition of this. We will follow the exposition of \cite{Gallo, Gunning}.
\subsection{Complex projective structures}
\paragraph{Definition.}
Let $\Sigma_g$ be a Riemann surface of genus $g \ge 2$. We choose a Fuchsian uniformization of $\Sigma_g$, $\Sigma_g=\HH^2/\Gamma_g^\text{F}$, where $\HH^2$ is the upper half plane and $\Gamma_g^\text{F}$ is a Fuchsian group, see Appendix~\ref{subapp:Fuchsian uniformization}. A complex projective structure is a holomorphic function $\gamma:\HH^2 \longrightarrow \mathbb{CP}^1$ such that $\partial \gamma(z) \ne 0$ for all $z \in \HH^2$. We also require that $\partial (\gamma(z)^{-1})\ne 0$. Equivalently, $\gamma(z)$ is locally injective. Moreover, $\gamma(z)$ has the following automorphic property:
\be 
\gamma(g(z))=\rho(g)(\gamma(z))
\ee 
for some homomorphism $\rho:\Gamma_g^\text{F} \longrightarrow \Gamma\subset \PSL(2,\mathbb{C})$ and all $g \in \Gamma_g^\text{F}$. The map $\gamma$ is also often called the developing map. We consider two complex projective structures to be equivalent if they differ only by an overall composition with a M\"obius transformation,
\be 
\widetilde{\gamma}=M \circ \gamma\ , \qquad \tilde{\rho}(g)=M \circ \rho(g) \circ M^{-1}
\ee
for some matrix $M \in \PSL(2,\CC)$.
\paragraph{Equivalence to a projective atlas.}
Often, a complex projective structure is defined differently as follows. We first choose a coordinate covering $\{U_\alpha,z_\alpha\}$ of the Riemann surface $\Sigma_g$. $z_\alpha$ are the coordinate maps $z_\alpha: U_\alpha \longrightarrow V_\alpha \subset \mathbb{CP}^1$ ($z_\alpha$ as usual are biholomorphisms).\footnote{We use here $\mathbb{CP}^1$ instead of the usual $\mathbb{C}$ because it allows us to treat cases more uniformly.} \footnote{Alternatively, we can define a complex projective structure on a real surface, since a complex projective structure in particular induces a a complex structure. We take the point of view that the complex projective structure is subordinate to the complex structure.} On intersections, we have transition maps
\be 
f_{\alpha\beta}=z_\alpha \circ z_\beta^{-1}: z_\beta(U_\alpha \cap U_\beta) \longrightarrow z_\alpha(U_\alpha \cap U_\beta)\ .
\ee
A complex projective structure is such an atlas for which all the transition functions are projective maps. 

The transition maps satisfy the obvious consistency condition
\be 
f_{\alpha\beta} \circ f_{\beta\gamma}=f_{\alpha\gamma}\ ,
\ee
whereever these maps are defined. Thus, $f_{\alpha\beta}$ define the \emph{coordinate bundle} over $\Sigma$. The group of the bundle is $\PSL(2,\mathbb{C})$ and the fibre is $\mathbb{CP}^1$. Since the transition maps are constant (when considered as a mapping from $U_\alpha \cap U_\beta$ into $\PSL(2,\mathbb{C})$) and thus the coordinate bundle is flat. Specifying a flat coordinate bundle is equivalent to specifying the complex projective structure.

This definition of complex projective structure is equivalent to the previous one. To see this, we cover a fundamental domain of the Fuchsian realization by open subsets $U_\alpha$. We then simply identify
\be 
z_\alpha=\Gamma|_{U_\alpha}\ .
\ee
thus defining the coordinate maps. 
One can easily check that this identifies the two structures.

\paragraph{Examples from uniformization.} Uniformizing the Riemann surface $\Sigma_g$ either via Fuchsian, Schottky or some other uniformization leads to a complex projective structure (or in the case of Fuchsian uniformization even to a real projective structure). As we shall see, the space of complex projective structures is however much bigger.
\paragraph{The Schwarzian derivative.}
It is useful to look at the Schwarzian derivative of the developing map,
\be 
S(\gamma)(z)=\frac{\partial^3 \gamma(z)}{\partial \gamma(z)}-\frac{3(\partial^2 \gamma(z))^2}{2(\partial \gamma(z))^2}\ .
\ee
Here, we view $\gamma$ as map from the upper half-plane and so $\partial$ is the usual derivative, not a covariant derivative.
Let us recall some crucial properties of the Schwarzian derivative:
\begin{enumerate}
\item Invariance under postcomposition with M\"obius transformations:
\be 
S(g(\gamma(z)))=S(\gamma(z))
\ee
for $g$ a M\"obius transformation.
\item Covariance under precomposition with M\"obius transformations:
\be 
S(\gamma(g ( z)))=S(\gamma(z))(\partial g(z))^2
\ee
for $g$ a M\"obius transformation.
\item Relation to a second order differential equation. Let $f_1$ and $f_2$ be two linearly independent solutions to the differential equation (viewed on $\HH^2$)
\be 
\partial^2 f(z)+\frac{1}{2}\phi(z) f(z)=0\ .
\ee
Then the ratio $\gamma(z)=f_1(z)/f_2(z)$ satisfies $S(\gamma(z))=\phi(z)$. In fact, we may take
\be 
f_1(z)=\frac{\gamma(z)}{\sqrt{\partial\gamma(z)}}\ , \qquad f_1(z)=\frac{1}{\sqrt{\partial\gamma(z)}}
\ee
for a choice of square root.\footnote{This is well-defined because $\partial \gamma(z)$ does not have zeros. Poles of $\partial \gamma(z)$ are double poles, since also $\partial (\gamma(z)^{-1})\ne 0$.} The pair $\boldsymbol{f}(z)=(f_1(z),f_2(z))$ is a section of a rank 2 vector bundle and satisfies
\be 
(\partial g(z))^{-\frac{1}{2}}\boldsymbol{f}(g(z))=\rho(g)(\boldsymbol{f}(z))\ ,
\ee
where $\rho$ is a lift to $\SL(2,\mathbb{C})$ of the original homomorphism. 
\end{enumerate}
Coming back to the Schwarzian derivative of the developing map, we see that $\phi(z)=S(\gamma)(z)$ is a well-defined quadratic differential on the Riemann surface $\Sigma$, thanks to the first two properties. $\phi(z)$ has no poles, because $\partial \gamma(z) \ne 0$ for all $z \in \HH^2$.

There is thus a map 
\be 
\{\text{complex projective structures}\} \longrightarrow \{\text{holomorphic quadratic differentials}\}
\ee
given by taking the Schwarzian of the developing map.

Conversely, given a holomorphic quadratic differential $\phi(z)$ on $\Sigma_g$, we can find a corresponding complex projective structure as follows. Essentially, one has to solve the differential equation $S(\gamma)(z)=\phi(z)$ on $\HH^2$, which leads to the developing map $\gamma(z)$ (unique up to composition with M\"obius transformation). The properties above imply that $\gamma(z)$ has the automorphic property for some homomorphism $\rho: \Gamma_g^\text{F} \longmapsto\PSL(2,\mathbb{C})$ and $\partial \gamma(z) \ne 0$, $\partial (\gamma(z)^{-1}) \ne 0$. One has to work harder to show that the differential equation indeed always admits a solution. We refer to \cite{Hehjal} for this.
This shows that the relation between complex projective structures and quadratic differentials on $\Sigma_g$ is 1--to--1.

Property 3 shows that the homomorphism $\rho$ for a complex projective structure always lifts to a homomorphism
\be 
\rho: \Gamma_g^\text{F} \longrightarrow \SL(2,\mathbb{C})\ .
\ee

\paragraph{Parameter counting.} It is useful to count the number of complex parameters that enter these definitions. It is well-known that the dimension of quadratic differentials is $3g-3$, see \eqref{eq:dimension quadratic differentials}. Thus, also the dimension of complex projective structures is $(3g-3)$-dimensional. However, the definition of the homomorphism $\rho$ involves $6g-6$ complex parameters: we choose $2g$ complex matrices for the generators $\alpha_1,\dots,\alpha_g,\beta_1,\dots,\beta_g$ of $\pi_1(\Sigma)$, but they have to satisfy the relation
\be 
\prod_{I=1}^g [\rho(\alpha_I),\rho(\beta_I)]=\id\ .
\ee
Moreover, two homomorphisms differing by an overall conjugation are considered equivalent which accounts for another 3 parameters.

Thus, for most homomorphisms $\rho$, there will not be a developing map and so they do not define a complex projective structure. From the parameter counting, we see however that if we allow both the complex structure and the complex projective structure on $\Sigma_g$ to vary, then we get $6g-6$ complex parameters. See property~\ref{item:Gallo theorem} below.

\paragraph{Further properties.}
Here we list further useful properties of complex projective structures. 
\begin{enumerate}
\item If two developing maps $\gamma_1$ and $\gamma_2$ lead to the same homomorphism $\rho$, then $\gamma_1=\gamma_2$.  See \cite[Theorem 3]{Gunning}.
\item The following conditions for the developing map $\gamma$ are equivalent \cite[Theorem 7]{Gunning}:
\begin{enumerate}
\item $\gamma(\HH^2)\ne \mathbb{CP}^1$.
\item $\gamma:\HH^2 \longrightarrow \gamma(\HH^2)$ is a covering map.
\item $\rho(\Gamma_g^\text{F})$ acts properly discontinuously on the image $\gamma(\HH^2)$.
\end{enumerate}
If these conditions are satisfied then, $\gamma$ descends to a well-defined covering map
\be 
\tilde{\gamma}:\Sigma_g \longrightarrow \gamma(\HH^2)/\rho(\Gamma_g^\text{F}) \ .
\ee
These are precisely the projective structures in which we are interested in this paper.
\item \label{item:Gallo theorem} A group of M\"obius transformations is \emph{elementary} if its action on $\HH^3$ has either one fixed point in $\HH^3$ or one fixed in $\partial \HH^3\cong \CP^1$. In the first case, it is conjugate to a group of unitary transformations, whereas in the second case, it is conjugate to a group of affine transformations. 

Any homomorphism $\rho:\pi_1(\Sigma) \longrightarrow \PSL(2,\CC)$ that is liftable to $\SL(2,\CC)$ and whose image is not an elementary group is realized by a complex structure on $\Sigma_g$ subordinate to some complex structure \cite{Gallo}.
\end{enumerate}
\subsection{Branched complex projective structures}
While complex projective structures are the relevant structure on the worldsheet for computing partition functions, we have to turn to branched complex projective structures for correlation functions.
\paragraph{Definition.} For branched complex structures, we do not require the developing map to be locally injective. Thus, a branched complex structure is a map $\gamma:\HH^2 \longrightarrow\CP^1$ satisfying
\be 
\gamma(g(z))=\rho(g)(\gamma(z))
\ee
for some homomorphism $\rho: \Gamma_g^\text{F} \longrightarrow \Gamma \subset \PSL(2,\CC)$ and all $g \in \Gamma_g^\text{F}$. $\gamma$ is branched over finitely many points $z_1, \dots, z_n$ on the Riemann surface with ramification indices $w_1,\dots, w_n$, meaning that
\be 
\partial^{n} \gamma(z_i)=0\ , \qquad 0<n< w_i
\ee
and $\partial^{w_i} \gamma(z_i)\ne 0$.
Note that the ramification index is preserved under actions of M\"obius transformations, so that this is well-defined.

The definition in terms of a complex atlas is much less useful in this case and hence we will work only with the developing map.
\paragraph{Relation to meromorphic quadratic differentials.} 
We can still define the Schwarzian in this case. The resulting quadratic differential has now quadratic differentials at $z_i$. A straightforward computation gives 
\be 
S(\gamma)(z) \sim \frac{1-w_i^2}{2(z-z_i)^2}+\mathcal{O}\left((z-z_i)^{-1}\right)\ .
\ee
Hence the Schwarzian gives us a map
\begin{multline} 
\{\text{branched complex projective structures with branch locus } (z_1,w_1),\dots,(z_n,w_n)\} \\
\longrightarrow \left\{\text{meromorphic quadratic differentials with }\QRes_{z=z_i} \phi(z)=\frac{1-w_i^2}{2}\right\}\ ,
\end{multline}
where $\QRes$ is the quadratic residue. Contrary to the unbranched case, this map is not 1--to--1. We can easily understand this via an example. Consider the case $g=0$ with four punctures, each having $w_i=2$. Let us take $z_1=0$, $z_2=1$, $z_3=2$ and $z_4=3$. In this case, a generic quadratic differential with the required poles takes the form
\be 
\phi(z)=\frac{-24z^2+72z-54}{z^2(z-1)^2(z-2)^2(z-3)^2}+\frac{q}{z(z-1)(z-2)(z-3)}\ .
\ee
However, we do not expect a familiy of maps $\gamma(z)$ with simple ramifications at $z_i$. Such a map is necessarily a branched cover of $\CP^1$ and as such has degree 3 by the Riemann Hurwitz formula. Hence we can write it as $\gamma(z)=P_3(z)/Q_3(z)$ for two third order polynomials $P_3(z)$ and $Q_3(z)$. Counting parameters, such a map depends on 7 parameters. Overall M\"obius transformations make 3 parameters redundant, the other 4 are fixed by requiring the ramifications at $z_i$. Hence we do not get a family of solutions for the Schwarzian derivative. Instead we have only two discrete solutions $q=-10\pm 2 \sqrt{13}$. 

In general, one needs to impose $n$ additional constraints on the quadratic differential in order for a solution $\gamma$ to exist. As one can see from this example, this constraint is quite complicated. The general statement is that there exist $n$ polynomials on the affine vectorspace of quadratic differentials $P_1,\dots,P_n$ with the required poles such that the joint zero locus can be identified with space of \emph{integrable} quadratic differentials, i.e.~those that give rise to a developing map $\gamma$ \cite{Hehjal, Mandelbaum}. 

Thus, the correct statement is in this case that branched complex projective structures with given branch locus are in 1--to--1 correspondence with integrable meromorphic quadratic differentials with the correct poles. For genus $g \ge 2$, the space of branched complex projective structures with given ramification locus has hence again generic dimension $3g-3$.

\section{Topologically twisted partition function} \label{app:topologically twisted partition function}
In this Appendix, we explain the form of topologically twisted partition functions. Our main example is the sigma-model on $\TT^4$ that is relevant for the present paper.
\subsection{Topological twist}
The $\mathcal{N}=2$ superconformal algebra possesses chiral supercurrents $G^+$ and $G^-$ of weight $\frac{3}{2}$. It also possesses an R-symmetry current $J=i\partial H$ with defining OPE
\be 
J(z) J(w) \sim \frac{c}{3(z-w)^2}\ ,
\ee
where $c$ is the central charge of the theory. 

We now topologically twist the theory, which corresponds to a redefinition of the Virasoro tensor
\be 
\hat{T}=T-\frac{1}{2}\partial J\ .
\ee
With respect to this new Virasoro tensor, the supercharges have weight $1$ ($G^+$) and $2$ ($G^-$). Moreover, the new central charge vanishes. This ensures that the correlation functions in the twisted theory are Weyl-anomaly free.
The topological twist leads to an anomalous $\U(1)$-current. In a correlator, the total charge with respect to the current $J$ has to be $\frac{c}{3}(1-g)$ in order to get a non-vanishing result. Thus, the topologically twisted partition function that we would like to compute is
\be 
\left\langle \prod_{j=1}^{g-1} \mathrm{e}^{-i H}(z_j) \right\rangle\ .
\ee
We suppress here as usual the right-movers. 
This is still not quite what we want, since this partition function vanishes identically. One way to see this is that the path integral still has a zero mode: the insertion of $\mathrm{e}^{-i H}(z) $in the correlation function is a Grassmannian 1-form that should have an a zero at $z=z_i$. Such a one-form does exist and its presence leads to a vanishing result. 

This is exactly the same issue that is present in the naive definition of the correlators of the $\mathcal{N}=4$ topological string \cite{Berkovits:1994vy, Berkovits:1999im}. It is remedied in \eqref{eq:N4 correlation function} by the inclusion of the current $J$. In our context, this means that we consider the following topologically twisted partition function:
\be 
\left\langle \prod_{j=1}^{g-1} \mathrm{e}^{-i H}(z_j) \partial H(z)\right\rangle\ .
\ee
\subsection{Change of variables}
We want to relate this partition function to the untwisted partition function. To do so, we follow the strategy of \cite{Gerasimov:1990fi, Iengo}. Let us consider a complex fermion $\psi$ and $\bar{\psi}$ (that are both spinors) and the topologically twisted versions $\Psi$ (that is a function) and $\bar{\Psi}$ (that is a 1-form). We relate the two as follows:
\be 
\bar{\Psi}=\Omega \bar{\psi}\ , \qquad \Psi=\Omega^{-1} \psi\ .
\ee
Here $\Omega$ is a holomorphic spinor. For $\Omega$ to exist, we need it to be an element of an odd spin bundle $S$. $\Omega$ has $g-1$ zeros which are located at $z_1^*$,\dots, $z_{g-1}^*$ and for simplicity, we are computing the correlator
\be 
\langle \bar{\Psi}(z_1^*) \cdots \bar{\Psi}(z_{g-1}^*) (\Psi\bar{\Psi})(z)\rangle\ ,
\ee
which satisfies the anomalous charge conservation. When computing this correlator via a path integral integral approach, we simply have to analyze how the measure and the action changes under this change of variables. In the $(\Psi,\bar{\Psi})$ variables, $\Psi(z)$ needs to have a first order pole at $z_i^*$ and $\bar{\Psi}(z)$ has a first order zero at $z_i^*$ because of the operator insertions. Since also $\Omega(z)$ has a zero at these points, this means that $(\psi,\bar{\psi})$ is regular at these points. Since $(\Psi\bar{\Psi})=(\psi\bar{\psi})$, nothing changes at the additional point.
The action in the $(\Psi,\bar{\Psi})$ variables translates correctly to the action in the $(\psi,\bar{\psi})$ variables. Finally, we have to analyze the relation between the measures. The norms are related according to
\be 
\lVert \delta \bar{\Psi} \rVert^2=\lVert \delta \bar{\psi} \rVert^2 |\Omega|^2\ , \qquad \lVert \delta \Psi \rVert^2=\lVert \delta \psi \rVert^2 |\Omega|^{-2}\ .
\ee
In order for this measure to coincide with the standard measure on differentials, we need to specify the metric according to
\be 
g_{z\bar{z}}=|\Omega|^4\ .
\ee
Thus, we see that with this choice of metric
\be 
\langle \bar{\Psi}(z_1^*) \cdots \bar{\Psi}(z_{g-1}^*) (\Psi\bar{\Psi})(z) \rangle_{\text{twisted}}=\langle (\psi\bar{\psi})(z) \rangle_{\text{untwisted}}
\ee
The right-hand side is evaluated with the fixed odd spin structure $S$. We see again that the additional insertion of $(\psi\bar{\psi})(z)$ is necessary to get a non-vanishing result.

\bibliographystyle{JHEP}
\bibliography{bib}
\end{document}